\documentclass[jphysa]{iopart}

\usepackage{graphicx}
\usepackage{amsmathred}
\usepackage{amsfonts}
\usepackage{amssymb}
\usepackage{braket}
\usepackage{times,txfonts}
\usepackage{subfigure}
\usepackage{hyperref}
\usepackage[mathscr]{euscript}
\usepackage[section]{placeins}
\usepackage{enumitem}

\usepackage{color}

\newcounter{desccount}
\newcommand{\descitem}[1]{%
  \item[#1] \refstepcounter{desccount}\label{#1}
}
\newcommand{\descref}[1]{\hyperref[#1]{#1}}

\newcounter{propcount}
\newcommand{\propitem}[1]{%
  \item[#1] \refstepcounter{propcount}\label{#1}
}
\newcommand{\propref}[1]{\hyperref[#1]{#1}}

\newcounter{reqcount}

\newcommand{\reqitem}[1]{%
  \item[#1] \refstepcounter{reqcount}%\label{#1}
}

\newcommand{\reqref}[1]{\hyperref[#1]{#1}}

\begin{document}

\topical{}
\title[Measures and applications of quantum correlations]{\sf \bfseries Measures and applications of quantum correlations}

\author{ {\sf \bfseries Gerardo Adesso}, {\sf \bfseries Thomas R. Bromley},   {\sf \bfseries Marco Cianciaruso}}
\address{Centre for the Mathematics and Theoretical Physics of Quantum Non-Equilibrium Systems, School of Mathematical Sciences, The University of Nottingham, University Park, Nottingham NG7 2RD, United Kingdom}

\begin{abstract}
Quantum information theory is built upon the realisation that quantum
resources like coherence and entanglement can be exploited for
novel or enhanced ways of transmitting and manipulating
information, such as quantum cryptography, teleportation, and quantum
computing. We now know that there is potentially much more than
entanglement behind the power of quantum information processing. There exist more
general forms of non-classical correlations, stemming from fundamental principles such as the necessary disturbance induced by a local measurement, or the persistence of quantum coherence in all possible local bases. These signatures can be identified and are resilient in
almost all quantum states, and have been linked to the enhanced performance of certain quantum
protocols over classical ones in noisy conditions. Their presence
represents, among other things, one of the most essential manifestations of quantumness in cooperative systems, from the subatomic to the macroscopic domain. In this work we give an overview of the current quest for a proper understanding and characterisation of the
frontier between classical and quantum correlations in composite states.
We  focus  on various approaches to
define and quantify general quantum correlations, based on different yet interlinked physical perspectives, and comment on the operational
significance of the ensuing measures for quantum technology tasks such as information encoding, distribution, discrimination and
metrology. We then provide a broader outlook of a few applications in which
quantumness beyond entanglement looks fit to play a key role.
\end{abstract}


\clearpage

\tableofcontents
\title[Measures and applications of quantum correlations]
\clearpage
\section{Introduction}

Quantum theory has been astonishingly successful for roughly a century now. Beyond its explanatory power, it
has enabled us to break new grounds in technology. Lasers, semiconductor devices, solar panels, and magnetic resonance imaging, are
just examples of everyday technologies based on quantum theory, nowadays classified as ``Quantum Technologies 1.0''.
The ultimate exploitation of the quantum laws applied to information processing is now promising to revolutionise the information and communication
technology sector as well, unfolding the era of ``Quantum Technologies 2.0'' \cite{dowling2003quantum}.
Secure quantum communication, enhanced quantum sensing, and the prospects of quantum computation,
have been ignited by the realisation that quintessential quantum features, like superposition and entanglement, can be exploited as resources for data encoding and transmission in ways which are substantially more efficient or radically novel compared to those allowed by classical resources alone \cite{nielsen2010quantum}. It is then clear that the deep understanding of the most genuine traits of quantum mechanics bears a strong promise for application into disruptive technologies.

Remarkably, the question of defining when a system behaves in a {\it quantum} way, rather than as an effectively {\it classical} one, still lacks a universal answer. This statement holds in particular when analysing composite systems and the nature of the correlations between their subsystems.
Until recently, theoretical and experimental attention has been mainly devoted to entanglement among different subsystems \cite{horodecki2009quantum}: entangled states clearly display a non-classical nature, and some of them can exhibit even stronger deviations from classicality such as steering \cite{cavalcanti2016quantum} and nonlocality \cite{brunner2014bell}. However, even unentangled states are not, in most cases, amenable to a fully classical description. More general forms of quantum correlations, exemplified by the so-called quantum discord \cite{ollivier2001quantum,henderson2001classical}, capture basic aspects of quantum theory, such as the fact that local measurements on parts of a composite system necessarily induce an overall disturbance in the state.
%Quantum correlations can  then be identified, for example, with those destroyed by the act of minimally disturbing measurements on one or more subsystems of a multipartite system. Classical correlations are consequently characterised as those that can be extracted by minimally disturbing local measurements, and are left in the state afterwards.
If one recognises genuine ``quantumness'' according to this paradigm, then almost all generally mixed quantum states of two or more subsystems display quantum correlations \cite{ferraro2010almost}, even in the absence of entanglement.

The present topical review  focuses precisely on this type of quantum correlations (QCs). In the last one-and-a-half decades, an increasing interest has been devoted by the international community to the study and the characterisation of QCs beyond entanglement, and notable progress has been achieved. Pursuing such an investigation further is important for two main reasons. On the fundamental side, it shines light on the ultimate frontiers of the quantum world, that is, on the most elemental (and consequently elusive) traits that distinguish the behaviour of a quantumly correlated system from one fully ascribed to a joint classical probability distribution.  On the practical side, it may reveal operational tasks where QCs even in absence of entanglement can still translate into a quantum enhancement, thus yielding more resilient and more accessible resources for future quantum technologies.

Given the large extent of research devoted to various aspects of QCs in recent years, there have been already a few introductory and review articles discussing the main concepts and presenting some of the most important findings in this research area. In particular, the reader may consult \cite{modi2014pedagogical} for a pedagogical introduction, \cite{horodecki2013quantumness} for a discussion of QCs in the context of resource theories, \cite{streltsov2014quantum} for a self-contained exposition of QCs and their role in quantum information theory, and \cite{modi2012classical} for a longer and more encyclopaedic coverage of relevant results until 2012. Nevertheless, it is fair to say that the topic of general QCs has still not reached the level of comprehension and appreciation by the broader mathematical and physical community that is instead claimed by entanglement theory. The reasons for this are varied. On the one hand, the topic is certainly not mature yet, and many open questions remain at the foundational level which are still in need of deeper insights and original solutions. On the other hand, this is also partly due to the excessive dispersion of certain research on QCs towards studies with little physical content, such as the mere calculation of some QCs quantifier in a particular system or model. Such a fragmentation has made it more difficult for some landmark advances stemming from the study of QCs to stand out and be clearly recognised by the international pool of non-specialists.

This review aims to give the ``{\it ABC}'' of QCs,\footnote{``{\em ABC}'' also stands for the acronym of the authors' initials.} with the intention of highlighting the physical understanding of the concepts involved without compromising their mathematical rigour, and with a specific focus on the following three basic questions:
\begin{enumerate}
\item[({\it A})] What are the signature traits of QCs from a fundamental point of view?
\item[({\it B})] How can we meaningfully quantify QCs in arbitrary quantum states?
\item[({\it C})] What are QCs good for in practical applications?
\end{enumerate}
The review is particularly targeting beginners who are willing to embark in fascinating research at the quantum-classical border --- and who may hopefully find here the right motivation and a set of problems to start tackling --- as well as expert colleagues who may have already addressed a particular aspect of QCs research --- and who may be looking for further inspiration from the bigger picture to move the next step forward.
To keep it close to its focus while maintaining an acceptable size, however, the review is also necessarily incomplete (meaning that several topics which even witnessed intense research are excluded, such as the one dealing with the dynamics of QCs in open quantum systems, or their potential role in quantum computing schemes with mixed states, among all), and many relevant references may have been omitted from our already comprehensive bibliography. In those cases, while issuing apologies in advance, we invite the interested reader to consult e.g.~\cite{modi2012classical} as well as the original literature for further information.

The review  is organised as follows. In Section~\ref{Section:QCs} we introduce the hierarchy of correlations in composite quantum systems, briefly highlighting the classification of QCs with respect to entanglement, steering and nonlocality. We then  give the formal definition of QCs (or lack thereof) and present in an original fashion various defining traits that pinpoint QCs as opposed to classical correlations. In Section~\ref{Section:Measures} we first introduce the progress achieved so far in the formalisation of a resource theory of QCs, collecting some necessary requirements and desiderata that any valid measure of QCs should satisfy. We then review a plethora of recently introduced QCs measures, which are shown to capture quantitatively the different defining traits introduced in the previous section. Particular effort is devoted to highlighting interlinks and dependency relations between the various types of measures, including the most recent insights not covered in other existing reviews. Section~\ref{Section:Applications} contains an overview of various important applications of QCs in the contexts of quantum information theory, thermodynamics, and many-body physics. Emphasis is placed on those settings which provide direct operational interpretations for some of the measures defined in the previous section. We conclude in Section~\ref{Section:Conclusions} with a summary of covered and uncovered topics accompanied by an outlook of a few currently open questions in QCs research.

%Includes a nice motivating opening and the plan of the review article. Including history.~\cite{ollivier2001quantum,henderson2001classical}

\section{Quantum correlations}\label{Section:QCs}

\subsection{The many shades of quantumness of correlations}\label{Section:Shades}

\begin{figure}[t!]
    \centering
    \includegraphics[width=0.45\textwidth]{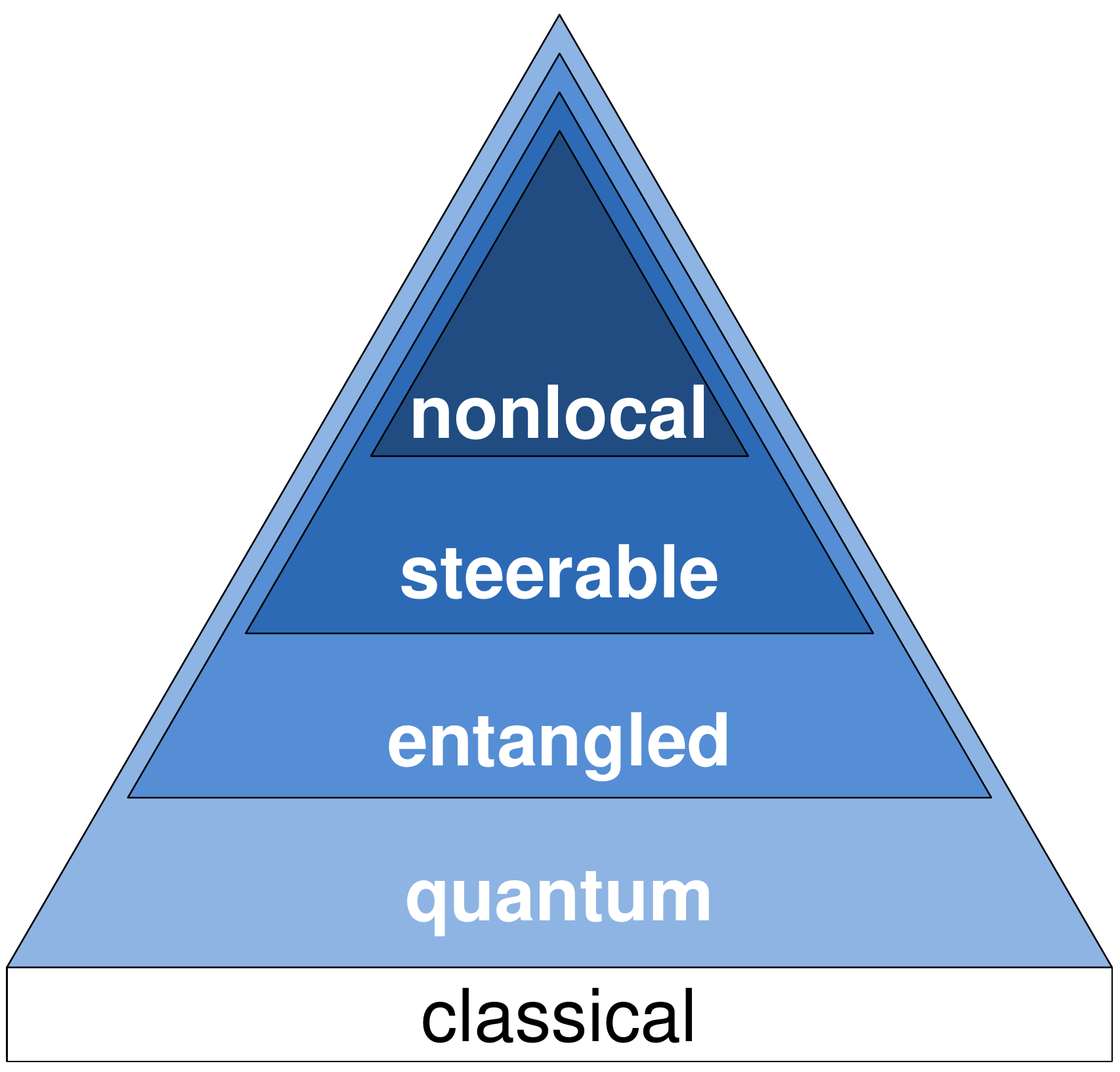}
    \caption{Hierarchy of correlations in states of composite quantum systems. Pure states can be either uncorrelated or just entangled. For mixed states, several layers of non-classical correlations have been identified, going significantly beyond the seminal  paradigm of \cite{werner1989quantum}. In order of decreasing strength, these can be classified as: nonlocality $\Rightarrow$  steering $\Rightarrow$  entanglement $\Rightarrow$  general quantum correlations. All of these forms of non-classical correlations can enable classically impossible tasks.
For instance, device-independent quantum cryptography requires nonlocality \cite{brunner2014bell}, entanglement-assisted subchannel discrimination depends on steering \cite{piani2015necessary}, while quantum teleportation and dense coding exploit plain entanglement \cite{horodecki2009quantum}. In this review we shall focus on the lower end of the spectrum, i.e., quantum correlations (QCs) beyond entanglement. They incarnate the most general yet arguably the least understood manifestation of non-classical correlations in composite quantum systems. Their fundamental and operational value will be illustrated in various physical settings throughout the review.}
    \label{Fig:CorrHier}
\end{figure}

The simplest testbed for the study of correlations is that of a composite quantum system made of two subsystems $A$ and $B$, each associated with a (finite dimensional) Hilbert space $\mathscr{H}_A$ and $\mathscr{H}_B$, respectively. If the system is prepared in a pure quantum state $\ket{\psi}_{AB} \in \mathscr{H}_{AB}$, where the Hilbert space of the composite system is defined as the tensor product $\mathscr{H}_{AB} := \mathscr{H}_A \otimes \mathscr{H}_B$ of the marginal Hilbert spaces, then essentially two possibilities can occur. The first is that the two subsystems are completely independent, in which case the state takes the form of a tensor product state $\ket{\psi}_{AB} = \ket{\alpha}_A \otimes \ket{\beta}_B$, with $\ket{\alpha}_A \in \mathscr{H}_A$ and $\ket{\beta}_B \in \mathscr{H}_B$. In this case there are no correlations of any form (classical or quantum) between the two parts of the composite system. The second possibility is that, instead, there exists no local state for $A$ and $B$ such that the global state can be written in tensor product form,
\begin{equation}
\ket{\psi}_{AB} \neq \ket{\alpha}_A \otimes \ket{\beta}_B\,. %, \quad \forall \ket{\alpha}_A \in \mathscr{H}_A, \ket{\beta}_B \in \mathscr{H}_B\,.
\end{equation}
In this case, $\ket{\psi}_{AB}$ describes an {\it entangled} state of the two subsystems $A$ and $B$. Entanglement encompasses any possible form of correlations in pure bipartite states, and can manifest in different yet equivalent ways. For instance, every pure entangled state is nonlocal, meaning that it can violate a Bell inequality \cite{gisin1991bell}. Similarly, every pure entangled state is necessarily disturbed by the action of any possible local measurement \cite{ollivier2001quantum,henderson2001classical}. Therefore, entanglement, nonlocality, and QCs are generally synonymous for pure bipartite states.

As illustrated in Fig.~\ref{Fig:CorrHier}, the situation is subtler and richer in case $A$ and $B$ are globally prepared in a mixed state, described by a density matrix $\rho_{AB} \in \mathscr{D}_{AB}$, where $\mathscr{D}_{AB}$ denotes the convex set of all density operators acting on $\mathscr{H}_{AB}$. The state $\rho_{AB}$ is {\it separable}, or unentangled, if it can be prepared by means of local operations and classical communication (LOCC), i.e., if it takes the form
\begin{equation}
\label{Equation:Separable}
\rho_{AB} = \sum_i p_i \varsigma_A^{(i)} \otimes \tau_B^{(i)}\,,
\end{equation}
with $\{p_i\}$ a probability distribution, and quantum states $\{\varsigma_A^{(i)}\}$ of $A$ and $\{\tau_B^{(i)}\}$ of $B$.
The set $\mathscr{S}_{AB} \subset \mathscr{D}_{AB}$ of separable states is constituted therefore by all states $\rho_{AB}$ of the form given by Eq.~(\ref{Equation:Separable}),
\begin{equation}\label{Equation:SSet}
\mathscr{S}_{AB} := \Big\lbrace\ \rho_{AB} \quad | \quad \rho_{AB} = \sum_i p_i \varsigma_A^{(i)} \otimes \tau_B^{(i)} \Big\rbrace\,.
\end{equation}
Any other state $\rho_{AB} \not\in \mathscr{S}_{AB}$ is {\it entangled}. Mixed entangled states are hence defined as those which cannot be decomposed as a convex mixture of product states. Notice that, unlike the special case of pure states, the set of separable states is in general strictly larger than the set of product states, $\mathscr{S}_{AB} \supset \mathscr{P}_{AB}$, where
\begin{equation}\label{Equation:PSet}
\mathscr{P}_{AB} := \Big\lbrace\ \rho_{AB} \quad | \quad \rho_{AB} = \rho_A \otimes \rho_B \Big\rbrace\,.
\end{equation}

Entanglement, one of the most fundamental resources of quantum information theory, can be then recognised as a direct consequence of two key ingredients of quantum mechanics: the superposition principle and the tensorial structure of the Hilbert space.  We defer the reader to \cite{horodecki2009quantum} for a comprehensive review on entanglement, and to \cite{plenio2007an} for a compendium of the most widely adopted entanglement measures.
Within the set of entangled states, one can further distinguish some layers of more stringent forms of non-classicality. In particular, some but not all entangled states are {\it steerable}, and some but not all steerable states are {\it nonlocal}.

Steering, i.e.~the possibility of manipulating the state of one subsystem by making measurements on the other, captures the original essence of inseparability adversed by Einstein, Podolsky and Rosen (EPR) \cite{einstein1935can} and appreciated by Schr\"odinger \cite{schrodinger1935discussion}, and has been recently formalised in the modern language of quantum information theory \cite{wiseman2007steering}. It is an asymmetric form of correlations, which means that some states can be steered from $A$ to $B$ but not the other way around. The reader may refer to \cite{cavalcanti2016quantum} for a  recent review on EPR steering.

On the other hand, nonlocality, intended as a violation of EPR local realism \cite{einstein1935can},
represents the most radical departure from a classical description of the world, and has received considerable
attention in the last half century since Bell's 1964 theorem \cite{bell1964on}. Recently, a triplet of experiments demonstrating Bell nonlocality free of traditional loopholes have been accomplished \cite{loopholefree1,loopholefree2,loopholefree3}, confirming the predictions of quantum theory. Nonlocality, like entanglement, is a symmetric type of correlations, invariant under the swap of parties $A$ and $B$. The reader is referred to \cite{brunner2014bell} for a review on nonlocality and its applications.

As remarked, here we are mainly interested in signatures of quantumness beyond entanglement. Therefore, an important question we should consider is: Are the correlations in separable states completely classical? In the following we argue that, in general, this is not the case. The only states which may be regarded as classically correlated form a negligible corner of the subset of separable states, and will be formally defined in the next subsection.

% Not sure if we want these but they looked useful!
%- Links to QCs: Minimum distillable entanglement equals one-way information defecit~\cite{streltsov2011linking}. Relative entropy of entanglement links with minimum distillable entanglement~\cite{piani2011all,gharibian2011characterizing,piani2012quantumness}. Entanglement irreversibility and quantum discord~\cite{cornelio2011entanglement}. Link between entanglement of formation and discord~\cite{fanchini2011conservation,wu2012correlation}.

\subsection{Classically correlated states}\label{Section:Facts}

Consider a composite system consisting of a classical bit and a quantum bit (qubit). For convenience, we shall adopt the same formalism for both. The classical bit can either be in the state $\ket{0}$ or in the state $\ket{1}$, representing e.g.~``off" or ``on" in modern binary electronics and communications. If the classical bit is in $\ket{0}$, one can write the composite state as $\ket{0}\bra{0}_{A}\otimes \rho_{B}$, where we have identified the classical bit as subsystem $A$ and the qubit as subsystem $B$, with  $\rho_{B}$ a quantum state. However, if the state of the classical bit is unknown, then the composite state $\rho_{AB}$ is a statistical mixture, i.e.~\begin{equation}
\rho_{AB} = p_{0} \ket{0}\bra{0}_{A}\otimes \rho^{(0)}_{B} + p_{1}\ket{1}\bra{1}_{A}\otimes \rho^{(1)}_{B},
\end{equation}
where $p_{0}$ and $p_{1}$ are probabilities adding up to one, while $\rho^{(0)}_{B}$ and $\rho^{(1)}_{B}$ denote the state of the quantum bit $B$ when the classical bit $A$ is in $\ket{0}$ and $\ket{1}$, respectively. This is an example of what we call a \emph{classical-quantum} state, or classical on $A$, since subsystem $A$ is classical. Equivalently, we can say that $\rho_{AB}$ is {\it classically correlated} with respect to subsystem $A$, the motivation for this terminology becoming clear soon.

Going beyond just bits, one can think about any classical system as consisting of a collection of distinct states, which can be represented using an orthonormal basis $\{\ket{i}\}$. Any classical-quantum state can then be written as a statistical mixture in the following way \cite{piani2008no},
\begin{equation}\label{Equation:CQ}
\rho_{AB} = \sum_{i} p_{i} \ket{i}\bra{i}_{A} \otimes \rho_{B}^{(i)},
\end{equation}
where $\rho_{B}^{(i)}$ is the quantum state of subsystem $B$ (now in general a $d$-dimensional quantum system, or qudit) when $A$ is in the state $\ket{i}_{A}$, and $\{p_{i}\}$ is a probability distribution. The set $\mathscr{C}_{A}$ of classical-quantum states  is then formed by any state that can be written as in Eq.~(\ref{Equation:CQ}),
\begin{equation}\label{Equation:CQSet}
\mathscr{C}_{A} := \Big\lbrace\ \rho_{AB} \quad | \quad \rho_{AB} = \sum_{i} p_{i} \ket{i}\bra{i}_{A} \otimes \rho_{B}^{(i)} \Big\rbrace\,,
\end{equation}
where $\{\ket{i}_{A}\}$ is any orthonormal basis of subsystem $A$ and $\{\rho_B^{(i)}\}$ are any quantum states of subsystem $B$. We stress that the orthonormal basis $\{\ket{i}_{A}\}$ appearing in Eq.~(\ref{Equation:CQSet}) is not fixed but rather can be chosen from all the orthonormal bases of subsystem $A$. The set $\mathscr{C}_A$ may look similar to the set $\mathscr{S}_{AB}$ of separable states, defined in Eq.~(\ref{Equation:SSet}), but there is an important difference: in Eq.~(\ref{Equation:SSet}), any state of subsystem $A$ is allowed in the ensemble, while in Eq.~(\ref{Equation:CQSet}) only projectors corresponding to an orthonormal basis can be considered. This reveals that classical-quantum states form a significantly smaller subset of the set of separable states, $\mathscr{C}_A \subset \mathscr{S}_{AB}$.

Swapping the roles of $A$ and $B$, one can define the {\it quantum-classical} states along analogous lines,
\begin{equation}\label{Equation:QuantumClassical}
\rho_{AB} = \sum_{j} p_{j} \rho_{A}^{(j)} \otimes \ket{j}\bra{j}_{B},
\end{equation}
and the corresponding set $\mathscr{C}_{B}$,
%\begin{equation}\label{Equation:QCSet}
%\mathscr{C}_{B} \equiv \left\lbrace \rho_{AB} \,\,\,\,\,\, | %\,\,\,\,\,\, \rho_{AB} = \sum_{j} p_{j} \rho_{A}^{(j)} \otimes \ket{j}\bra{j}_{B} \right\rbrace
%\end{equation}
%for any orthonormal basis $\{\ket{j}_{B}\}$.
\begin{equation}\label{Equation:QCSet}
\mathscr{C}_{B} := \Big\lbrace\ \rho_{AB} \quad | \quad \rho_{AB} = \sum_{j} p_{j} \rho_{A}^{(j)} \otimes \ket{j}\bra{j}_{B}  \Big\rbrace\,,
\end{equation}
where $\{\ket{j}_{B}\}$ is any orthonormal basis of subsystem $B$ and $\{\rho_A^{(j)}\}$ are any quantum states of subsystem $A$. These states can be equivalently described as {\it classically correlated} with respect to subsystem $B$.

Finally, if we consider the composition of two classical objects, we can introduce the set of \emph{classical-classical} states, or classical on $A$ and $B$, which are {\it classically correlated} with respect to both subsystems. A state is classical-classical if it can be written as
\begin{equation}\label{Equation:CC}
\rho_{AB} = \sum_{ij} p_{ij} \ket{i}\bra{i}_{A} \otimes  \ket{j}\bra{j}_{B},
\end{equation}
where we now have a joint probability distribution $\{p_{ij}\}$ and orthonormal bases for both subsystems $A$ and $B$.  The set $\mathscr{C}_{AB}$ of classical-classical states  is then formed by any state that can be written as in Eq.~(\ref{Equation:CC}),
\begin{equation}\label{Equation:CCSet}
\mathscr{C}_{AB} := \Big\lbrace\ \rho_{AB} \quad | \quad \rho_{AB}=\sum_{ij} p_{ij} \ket{i}\bra{i}_{A} \otimes  \ket{j}\bra{j}_{B} \Big\rbrace\,,
\end{equation}
%\begin{equation}\label{Equation:CCSet}
%\mathscr{C}_{AB} \equiv \left\lbrace \rho_{AB} \,\,\,\,\,\, | \,\,\,\,\,\, \rho_{AB} = \sum_{ij} p_{ij} \ket{i}\bra{i}_{A} \otimes  \ket{j}\bra{j}_{B} \right\rbrace
%\end{equation}
where $\{\ket{i}_{A}\}$ and $\{\ket{j}_{B}\}$ are any orthonormal bases of subsystem $A$ and $B$, respectively.
%The description of these states is completely specified by the joint classical probability distribution $\{p_{ij}\}$, which is just formally embedded into a density matrix formalism labelled by orthonormal index vectors on each subsystem.
These states can be thought of as the embedding of a joint classical probability distribution $\{p_{ij}\}$ into a density matrix formalism labelled by orthonormal index vectors on each subsystem.

\begin{figure}[t!]
    \centering
    \begin{subfigure}[]
        \centering
        \includegraphics[height=5.5cm]{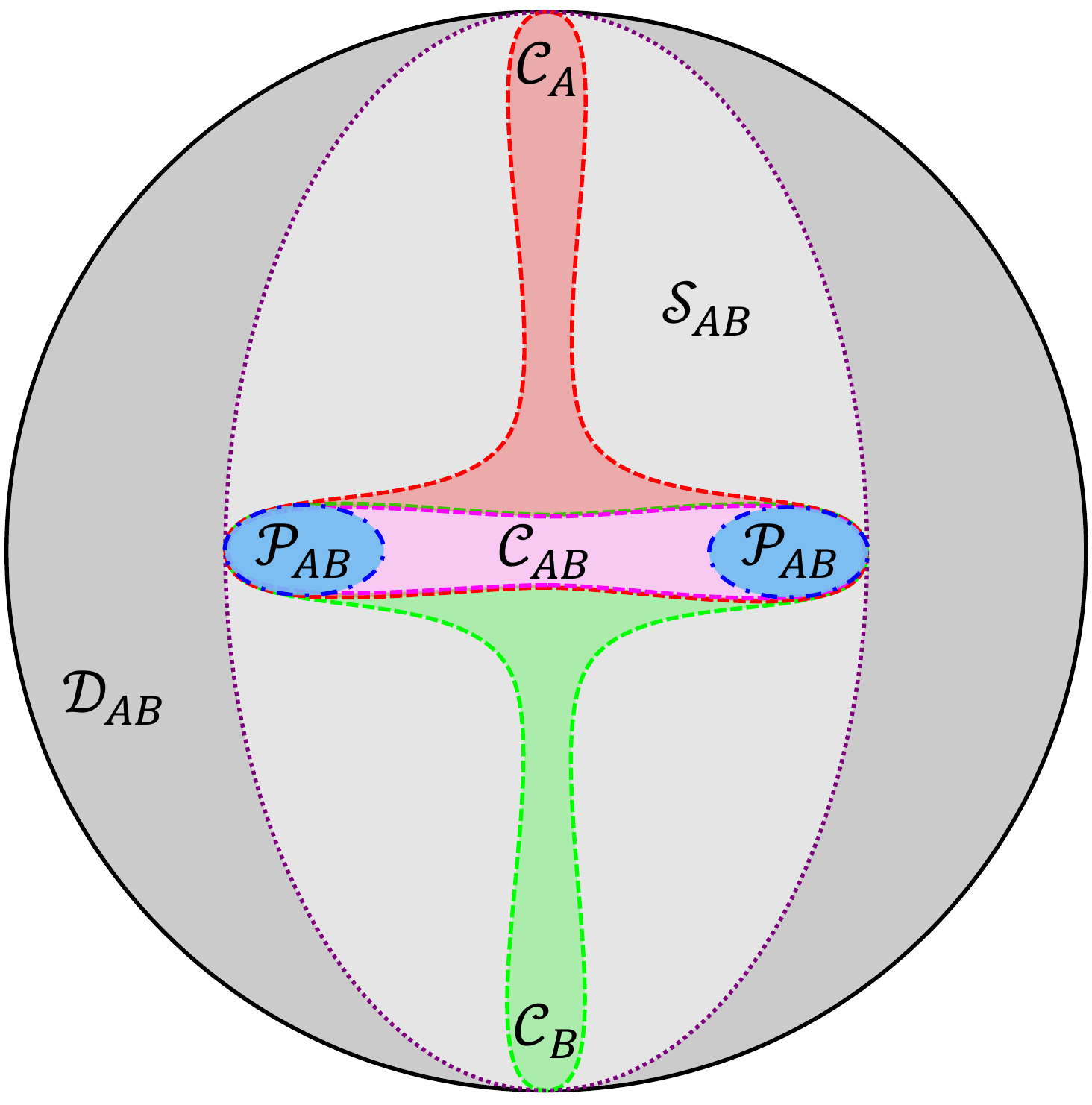}
    \end{subfigure}%
    \hspace*{.5cm}
    \begin{subfigure}[]
        \centering
        \includegraphics[height=5.5cm]{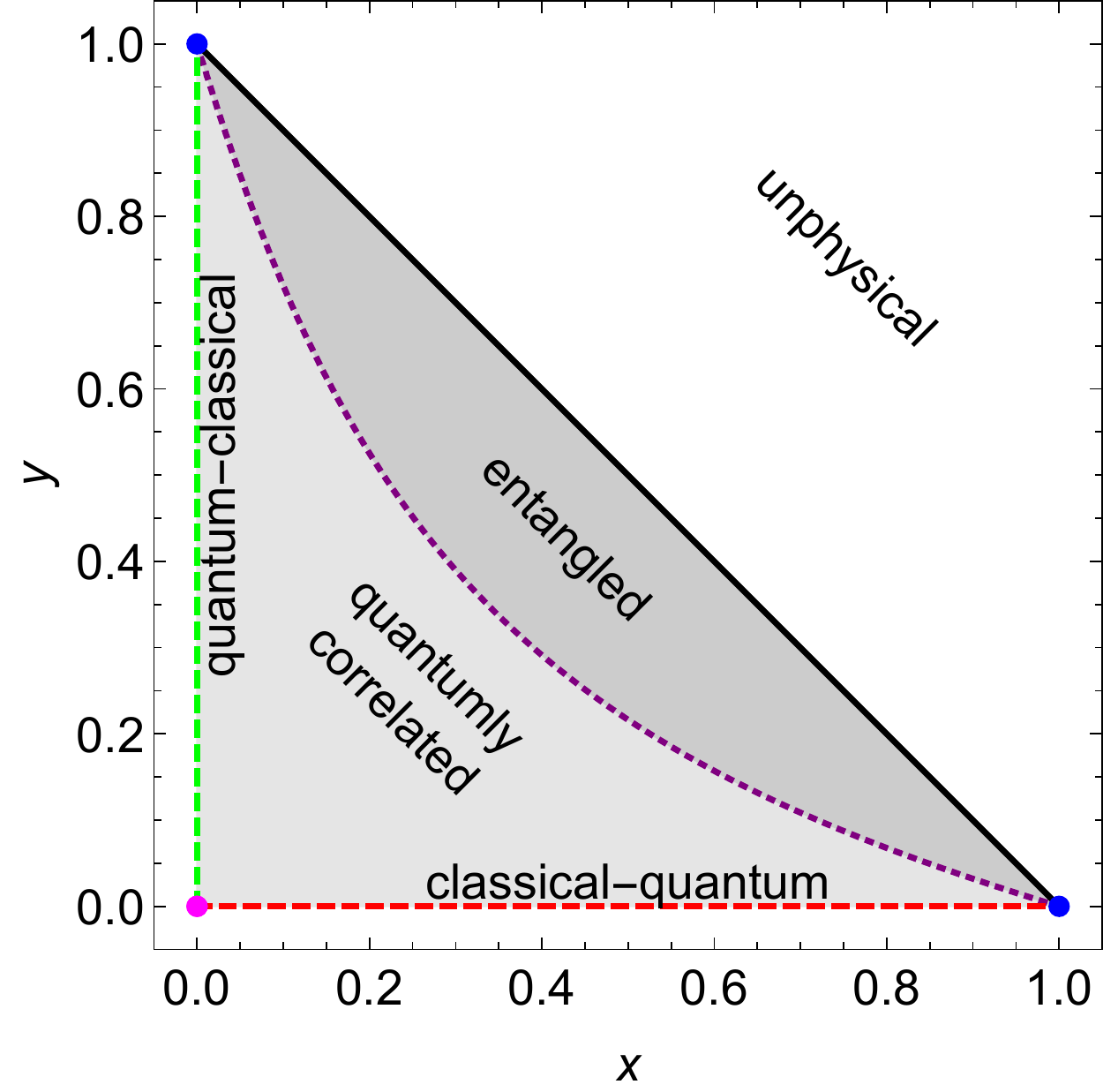}
    \end{subfigure}%
    \caption{Visualisation of different subsets of correlated states in bipartite quantum systems. In both panels, we depict in blue the (uncorrelated) product states belonging to the subset $\mathscr{P}_{AB}$, in magenta the classical-classical states belonging to the subset $\mathscr{C}_{AB}$, in dashed red the classical-quantum states belonging to the subset $\mathscr{C}_A$, in dashed green the quantum-classical states belonging to the subset $\mathscr{C}_B$, in light grey (with dotted purple boundary) the separable states belonging to the subset $\mathscr{S}_{AB}$, and in dark grey (with solid black boundary) all the remaining (entangled) states, completing the whole set $\mathscr{D}_{AB}$. Panel (a) is an artistic impression which does not faithfully reflect the  topology of the various subsets (in particular, the sets $\mathscr{C}_{AB}$, $\mathscr{C}_{A}$, $\mathscr{C}_{B}$, and $\mathscr{P}_{AB}$ are all of null measure and nowhere dense). Panel (b) depicts the actual subsets for the  two-qubit states $\rho_{AB}(x,y)$ of Eq.~(\ref{Equation:ExSets}), with $0 \leq x+y \leq 1$. For $y=0$ (horizontal axis, dashed red), we have $\rho_{AB}(x,0) = \frac{1-x}{2} \ket{0}\bra{0}_A \otimes \ket{0}\bra{0}_B + \frac{1+x}{2} \ket{1}\bra{1}_A \otimes \rho^{(1)}_B \in \mathscr{C}_A$, with $\rho^{(1)}_B = (1+x)^{-1}(x \ket{0}\bra{0}_B + x \ket{0}\bra{1}_B + x \ket{1}\bra{0}_B + \ket{1}\bra{1}_B)$. Analogously, for $x=0$ (vertical axis, dashed green), we have $\rho_{AB}(0,y) \in \mathscr{C}_B$, obtained from the previous case upon swapping $A \leftrightarrow B$ and $x \leftrightarrow y$. In particular, for $x=y=0$ (magenta dot), we have $\rho_{AB}(0,0)=\frac12 (\ket{00}\bra{00} + \ket{11}\bra{11}) \in \mathscr{C}_{AB}$. Furthermore, $\rho_{AB}(1,0) = \ket{1}\bra{1}_A \otimes \ket{+}\bra{+}_B \in \mathscr{P}_{AB}$ (blue dot), with $\ket{+}=(\ket{0} +  \ket{1})/\sqrt2$, and similarly  $\rho_{AB}(0,1)=\ket{+}\bra{+}_A \otimes \ket{1}\bra{1}_B \in \mathscr{P}_{AB}$ (blue dot). Finally, the state $\rho_{AB}(x,y)$ has QCs for all the other admissible values of $x$ and $y$, and is further entangled if ${1+10 x y-16 x^2 y^2 (4 x+1) (4 y+1)-7(x^2+y^2)+2 (x+y)\big(1+2 (x^2-4 x y+y^2)\big)<0}$ (the threshold is indicated by the dotted purple curve), as it can be verified by applying the partial transposition criterion \cite{horodecki2009quantum}.}
    \label{Fig:Sets}
\end{figure}

It holds by definition that classical-classical states amount to those which are both classical-quantum and quantum-classical, that is, $\mathscr{C}_{AB} = \mathscr{C}_A \cap \mathscr{C}_B$. More generally, we have
\begin{equation}\label{Equation:Subsets}
\mathscr{P}_{AB} \subset \mathscr{C}_{AB} \subset \{\mathscr{C}_A, \mathscr{C}_B\} \subset \mathscr{S}_{AB} \subset \mathscr{D}_{AB}\,.
\end{equation}
In this hierarchy, only the two rightmost sets  (containing separable states, and all states, respectively) are convex, while all the remaining ones are not; that is, mixing two classically correlated states one may obtain a state which is not classically correlated anymore. Another interesting fact is that, while separable states span a finite volume in the space of all quantum states, the sets $\mathscr{C}_A$, $\mathscr{C}_B$, and consequently $\mathscr{C}_{AB}$ are of null measure and nowhere dense within $\mathscr{S}_{AB}$ \cite{ferraro2010almost}. Their topology is therefore difficult to visualise, and we can only present an artistic impression in Fig.~\ref{Fig:Sets}(a). Nevertheless, to achieve a more specific rendition, in Fig.~\ref{Fig:Sets}(b) we discuss a particular example of a family of two-qubit states defined as
\begin{equation}\label{Equation:ExSets}
\rho_{AB}(x,y) =\frac{1}{2+8 x y} \left( \begin{array}{cccc}  1-x-y & 0 & 0 & 0 \\  0 & 4 x y+y & 4 x y & y \\  0 & 4 x y & 4 x y+x & x \\  0 & y & x & 1 \\ \end{array}\right)\,,\end{equation}
 dependent on two real parameters $x$ and $y$ with $0 \leq x+y \leq 1$, and spanning the various types of correlations discussed in this section (from no correlations up to entanglement). Another frequently studied instance where the geometry of correlations is particularly appealing is that of so-called Bell diagonal states of two qubits, defined as arbitrary mixtures of the four maximally entangled Bell states, which (up to local unitaries) represent all two-qubit states with maximally mixed marginals, see e.g.~\cite{dakic2010necessary,lang2010quantum,aaronson2013comparative}.

Throughout the review, we will consider without loss of generality the settings in which classicality is either attributed to subsystem $A$ alone, or to both subsystems, and we will sometimes use the simple terminology ``classical states'' to refer either to classical-quantum or classical-classical states, depending on the context.

%\subsection{Defining properties of quantum correlations}

%Any state of a composite system $AB$ which is not a classical-classical

%Correspondingly, any state which is not a classical state will be regarded as a state displaying quantumness of correlations, or in short QCs. Clearly, the notion of QCs embodies a generally nonsymmetric type of correlations, similar to steering:
%any state $\rho_{AB} \neq \mathscr{C}_{A}$ has QCs with respect to subsystem $A$, or one-way QCs.
%The terminology will be further clarified in the following.

\subsection{Characterising classically correlated states}\label{Section:Facts-C}

\begin{figure*}[t]
    \centering
    \includegraphics[width=7.3cm]{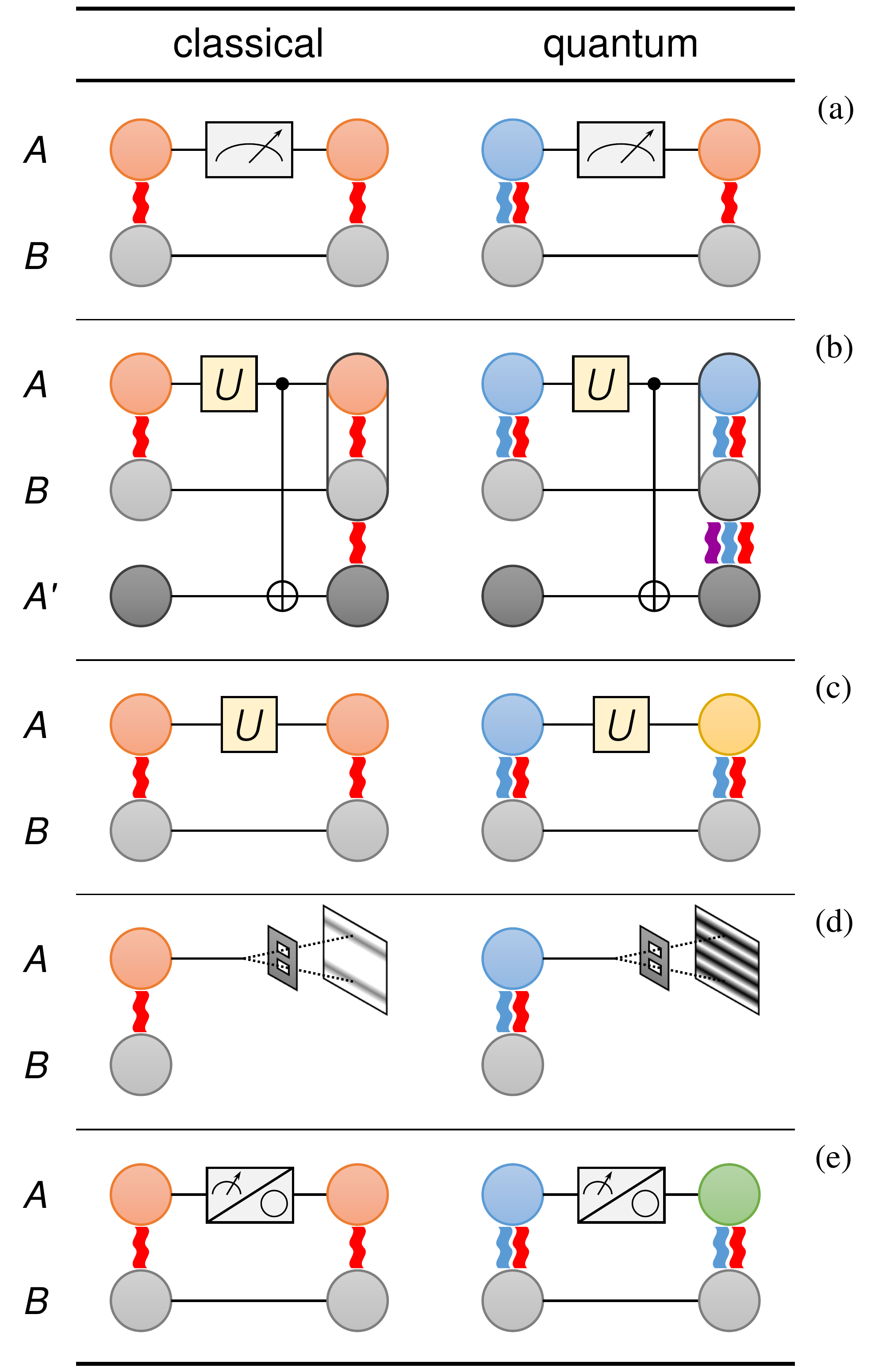}
    \caption{Schematics of the {\it defining properties} of classical-quantum states (left) versus states with one-sided QCs with respect to subsystem $A$ (right) in a bipartite system $AB$, as detailed in Sections~\ref{Section:Facts-C} and \ref{Section:Facts-Q}. Analogous schemes can be drawn to depict classical-classical states versus states with two-sided QCs, with the exception of row (c). In the drawings, single, double, and triple ribbon connectors denote respectively classical correlations, QCs, and entanglement. Row (a): Classical states may remain invariant under a complete local projective measurement (Property~\ref{Property 2:}), while states with nonzero QCs are always altered by any such measurement (Definition~\ref{Definition 2:}).
    Row (b): Classical states may not get entangled with an apparatus $A'$ during a local measurement (Property~\ref{Property 3:}), while states with nonzero QCs always lead to creation of entanglement with an apparatus at the pre-measurement stage (Definition~\ref{Definition 3:}); the pre-measurement interaction is represented as a local unitary followed by a generalised control-\textsc{not} gate. Row (c): Classical states may remain invariant under a local non-degenerate unitary (Property~\ref{Property 4:}), while states with nonzero QCs are always altered by any such operation (Definition~\ref{Definition 4:}). Row (d): Classical states may be incoherent with respect to a local basis (Property~\ref{Property 5:}), while states with nonzero QCs exhibit coherence (rendered as ability to display interference in a double slit experiment) in all local bases (Definition~\ref{Definition 5:}).
    Row (e): Classical states may remain invariant under an entanglement-breaking channel (Property~\ref{Property 6:}), while states with nonzero QCs are always altered by any such channel (Definition~\ref{Definition 6:}); the entanglement-breaking channel is depicted as a local measurement followed by a preparation map.
    These characterisations form the basis to {\it quantify} QCs in an arbitrary bipartite quantum state as well, as discussed further in Section~\ref{Section:Types} (see Table~\ref{Table:Types}).
 }
    \label{Fig:Defs}
\end{figure*}

We now aim to illustrate several alternative characterisations of classical states in (finite dimensional) composite quantum systems, in order to aid with their physical understanding. This will accordingly be  useful to appreciate the various defining traits of QCs in Section~\ref{Section:Facts-Q}, and guide the possible approaches to their quantification in Section~\ref{Section:Measures}.

Here we first succinctly introduce the reader to such physical properties, which are illustrated schematically in the left column of Fig.~\ref{Fig:Defs}. We will then go into further details for each property (including the definition and explanation of the specialised concepts mentioned below) in the following subsections.

Specifically, in a bipartite system, a state $\rho_{AB}$ is classical if and only if it complies with either one of the following defining properties, which are all equivalent to each other.
\begin{description}
\propitem{Property 1:}{$\rho_{AB}$ is of the form given in Eq.~(\ref{Equation:CQ}) or Eq.~(\ref{Equation:CC}) (referring to elements of $\mathscr{C}_A$ or $\mathscr{C}_{AB}$, respectively).}
\propitem{Property 2:}{$\rho_{AB}$  is  invariant under at least one local complete projective measurement on one or both subsystems (referring to elements of $\mathscr{C}_A$ or $\mathscr{C}_{AB}$, respectively).}
\propitem{Property 3:}{$\rho_{AB}$ does not become entangled with an apparatus during the pre-measurement stage of at least one local complete projective measurement of one or both subsystems (referring to elements of $\mathscr{C}_A$ or $\mathscr{C}_{AB}$, respectively).}
\propitem{Property 4:}{$\rho_{AB}$  is   invariant under at least one local unitary operation with non-degenerate spectrum on  subsystem $A$ (referring to elements of $\mathscr{C}_A$).}
\propitem{Property 5:}{$\rho_{AB}$  is incoherent with respect to at least one local orthonormal basis for one or both subsystems (referring to elements of $\mathscr{C}_A$ or $\mathscr{C}_{AB}$, respectively).}
\propitem{Property 6:}{$\rho_{AB}$  is   invariant under at least one entanglement-breaking channel acting on one or both subsystems (referring to elements of $\mathscr{C}_A$ or $\mathscr{C}_{AB}$, respectively).}
\end{description}

Most of these notions can be extended to introduce a hierarchy of partly classical states in multipartite systems \cite{piani2012quantumness}, however we shall focus on the bipartite case in the present review.
%The following subsections provide details on the above  properties, which are illustrated schematically in the left column of Fig.~\ref{Fig:Defs}.
Notice further that additional necessary and sufficient characterisations of classical states have been provided e.g.~in \cite{piani2008no,dakic2010necessary,datta2010a,brodutch2010quantum,coles2012unification,modi2012classical}.
%\FloatBarrier

\subsubsection{Invariance under a complete local projective measurement}\label{Section:Facts-Measuring}

We begin this section with a brief refresh of measurements in quantum mechanics \cite{vonneumann1932mathematical,nielsen2010quantum}, in particular focusing on how local measurements can be used to define and motivate the notion of classically correlated states. A detailed introduction to measurements within the context of QCs can be found in~\cite{spehner2014quantum}.

A generalised quantum measurement can be mathematically described by resorting to the notion of a {\it quantum instrument}, which is a set $\mathscr{M}=\{\mu_i\}$ of subnormalised (i.e., trace nonincreasing) completely positive maps $\mu_i$, %known as the measurement operators,
that sum up to a completely positive trace preserving (CPTP) map, i.e., to a quantum channel $\mu:=\sum_i \mu_i$.
%A generalised measurement $M$, or quantum instrument, is represented generally by a positive operator valued measure (POVM), which consists of a set of positive operators $\{M_{i}\}$ satisfying $\sum_{i} M_{i} = \mathbb{I}$, and by the specification of a family of measurement operators $\{\mu_i\}$ such that $M_{i} = \mu_{i}^{\dagger}\mu_{i}$. The set of POVMs is convex, i.e.~considering any two POVMs $\{M_{i}^{(1)}\}$ and $\{M_{i}^{(2)}\}$, a convex combination $M_{i}^{(3)} = q M_{i}^{(1)} + (1-q) M_{i}^{(2)}$ for some probability $q$ leads to another POVM $\{M_{i}^{(3)}\}$.  In particular, extremal POVMs are  defined as those that cannot be written as a convex combination of other POVMs. Furthermore, rank one POVMs are those that consist only of rank one operators $\{M_{i}\}$.
We can think of each subnormalised map $\mu_{i}$ as corresponding to a measurement outcome $i$.
Given a state $\rho$, the probability of obtaining the outcome $i$ when applying the quantum instrument $\mathscr{M}$ to $\rho$ is
\begin{equation*}
p_{i} = \mbox{Tr}\left( \mu_i[ \rho] \right),
\end{equation*}
and the corresponding state after the measurement, also called `subselected' post-measurement state, is
\begin{equation*}
\rho_{i} = \mu_i[\rho]/p_{i}.
\end{equation*}
If the outcome of the measurement is not known, then the resultant state, simply called post-measurement state, is the statistical mixture
\begin{equation}
\sum_{i}p_{i}\rho_{i} =\sum_{i} \mu_i[\rho] =: \mu[\rho].
\end{equation}
It is also useful to think about how the result of this quantum measurement is stored in a classical register. If we think of an orthonormal basis $\{\ket{i}\}$ corresponding to the measurement outcomes $\{i\}$, the classical register will read $\ket{i}\bra{i}$ when the outcome $i$ occurs. Furthermore, if the outcome is not known, then the system composed of the output quantum system and the output classical register will be overall in the state
\begin{equation*}
\sum_{i} p_{i} \rho_i \otimes \ket{i}\bra{i} = \sum_{i}\mu_i[\rho] \otimes \ket{i}\bra{i}.
\end{equation*}
Consequently,
%by focusing only on what is recorded by the classical register,
by ignoring the measured quantum system, i.e., by tracing it out, we get the following classical output
\begin{equation}
\sum_{i} p_{i} \ket{i}\bra{i} = \sum_{i}\mbox{Tr}\left( \mu_i[\rho] \right)\ket{i}\bra{i} =: M[\rho].
\end{equation}
The map $M$ transforming an arbitrary quantum state $\rho$ into the classical state $M[\rho]$ is referred to as the \textit{quantum-to-classical map} (or {\it measurement map}) corresponding to the quantum instrument $\mathscr{M}$ with subnormalised maps $\{\mu_i\}$.

We note that in order to know the probabilities $p_i$ of each outcome $i$
%and thus the form of the quantum-to-classical map corresponding to a quantum instrument $I$,
we do not need to know all the details of $\mathscr{M}$. Conversely, we just need to know how the dual $\mu^*_i$ of every subnormalised map acts on the identity.
%where $\mu^*_i$ is defined as follows:
%\begin{equation}
%\mbox{Tr}\left( \mu_i[A] X \right) = \mbox{Tr}\left( A \mu_i^*[X] \right), \ \ \forall A\in \mathscr{B}, X \in %\mathscr{B}'.
%\end{equation}
Indeed,
\begin{equation*}
p_i = \mbox{Tr}\left( \mu_i[\rho] \right) = \mbox{Tr}\left( \mu_i[\rho] \mathbb{I} \right) = \mbox{Tr}\left( \rho \mu_i^*[\mathbb{I}] \right).
\end{equation*}
The operators $M_i:=\mu_i^*[\mathbb{I}]$ form a so-called positive operator valued measure (POVM), i.e., a set of positive semidefinite operators that sum up to the identity, $\sum_i M_i = \mathbb{I}$.

On the other hand, we must stress that, in general, given only a POVM $\{M_i\}$ we can just determine the probabilities of the outcomes
%and the quantum-to-classical map corresponding to a quantum measurement
but we do not know how the quantum system transforms. For example, one can easily see that a given POVM $\{M_i\}$ is compatible with infinitely many quantum instruments, such as all the ones that can be defined as follows: $\mu_i[\rho]=m_i \rho m_i^\dagger$, where the $m_i$'s are such that $M_i = m_i^\dagger m_i$. Indeed, it is clear that such $m_i$'s are not uniquely defined, but rather also any  $m_i'=U_i m_i$, with $U_i$ being arbitrary unitaries, are such that $M_i={m_i'}^\dagger m_i'$.

A special type of measurement is a projective measurement, for which the maps $\mu_{i}$ are Hermitian orthogonal projectors, i.e.,
%they act on an arbitrary state $\rho$ as follows:
%\begin{equation*}
%\mu_i[\rho] = \Pi_i \rho \Pi_i,
%\end{equation*}
they satisfy the constraint $\mu_{i}\mu_{j} = \delta_{ij}\mu_{i}$. Furthermore, rank one measurements are those that consist only of rank one subnormalised maps $\mu_i$.

% In this case, if the result of the projective measurement is unknown, the state becomes
%\begin{equation}
%\sum_{i}p_{i}\rho_{i} = \sum_{i} m_{i} \rho m_{i}^{\dagger}.
%\end{equation}

%Here we are concerned with the correlations of a bipartite quantum state $\rho_{AB}$, which naturally involves investigating the subsystems $A$ and $B$ through local measurements.
%Let us now consider a {\it local generalised measurement} (herein called an LGM) consisting of generalised measurements $\tilde{\Pi}_A$ and/or $\tilde{\Pi}_B$ acting separately on subsystems $A$ and/or $B$ in a bipartite state $\rho_{AB}$, with measurement operators $\{(\tilde{\pi}_{A})_{a}\}$ and/or $\{(\tilde{\pi}_{B})_{b}\}$, respectively. We shall use a tilde from now on to denote an LGM, whereas the lack of a tilde will be reserved for the particular case of a {\it local projective measurement} (LPM). We will also use the notation $\tilde{\Pi}_{AB}$ to denote in particular a joint LGM on both subsystems, $\tilde{\Pi}_{AB} = \tilde{\Pi}_A \otimes \tilde{\Pi}_B$, and equivalently for a joint LPM without the tilde.
Let us now consider a quantum system composed of two subsystems $A$ and $B$. A {\it local generalised measurement} (herein called an LGM) is a generalised measurement acting either only on $A$ or only on $B$ or separately on both $A$ and $B$, and whose subnormalised maps are, respectively, either $\{(\tilde{\pi}_{A})_{a}\otimes \mathbb{I}_B\}$ or $\{\mathbb{I}_A\otimes (\tilde{\pi}_{B})_{b}\}$ or $\{(\tilde{\pi}_{A})_{a}\otimes (\tilde{\pi}_{B})_{b}\}$. We shall use a tilde from now on to denote an LGM, whereas the lack of a tilde will be reserved for the particular case of a {\it local projective measurement} (LPM).
%We will also use the notation $\tilde{\Pi}_{AB}$ to denote in particular a joint LGM on both subsystems, $\tilde{\Pi}_{AB} = \tilde{\Pi}_A \otimes \tilde{\Pi}_B$, and equivalently for a joint LPM without the tilde.

The probability of obtaining the outcome $a$ when performing an LGM only on $A$ is %with the corresponding state and state of the register when the result is unknown, is
\begin{equation*}
p_{a} = \mbox{Tr}\left\lbrace\left((\tilde{\pi}_{A})_{a} \otimes \mathbb{I}_{B}\right) [\rho_{AB}]\right\rbrace ,
\end{equation*}
with corresponding subselected post-measurement state
\begin{equation*}
\rho_{AB|a} = \big[\left((\tilde{\pi}_{A})_{a} \otimes \mathbb{I}_{B}\right) [\rho_{AB}]\big]/p_{a}.
\end{equation*}
For each outcome $a$, one can associate a vector in an orthonormal basis $\{\ket{a}_{A'}\}$ representing a classical register $A'$ (note that this basis may live in a different Hilbert space to that of subsystem $A$). By ignoring the state of the measured subsystem $A$, the resulting state of the classical register $A'$ and of the unmeasured subsystem $B$ is then $\ket{a}\bra{a}_{A'} \otimes \rho_{B|a}$, where $\rho_{B|a} = \mbox{Tr}_{A}(\rho_{AB|a})$. Hence, if the result is not known, this composite system will be in the classical-quantum state
\begin{equation}\label{Equation:PostLGMA}
\tilde{\Pi}_{A}[\rho_{AB}]:=\sum_{a} p_{a} \ket{a}\bra{a}_{A'} \otimes \rho_{B|a}.
\end{equation}

Likewise, for an LGM on both subsystems $A$ and $B$, the probability of getting respectively the outcomes $a$ and $b$ is
\begin{equation*}
p_{ab} = \mbox{Tr}\left\lbrace\left((\tilde{\pi}_{A})_{a} \otimes (\tilde{\pi}_{B})_{b}\right) [\rho_{AB}]\right\rbrace
\end{equation*}
with corresponding subselected post-measurement state
\begin{equation*}
\rho_{AB|ab} = \big[\left((\tilde{\pi}_{A})_{a} \otimes (\tilde{\pi}_{B})_{b}\right) [\rho_{AB}]\big]/p_{ab}.
\end{equation*}
Again, associating to each outcome $a$ one vector of the orthonormal basis $\{\ket{a}_{A'}\}$ corresponding to a classical register $A'$ and to each outcome $b$ one vector of the orthonormal basis $\{\ket{b}_{B'}\}$ corresponding to a classical register $B'$, the composite classical register reads $\ket{a}\bra{a}_{A'} \otimes \ket{b}\bra{b}_{B'}$ when both outcomes $a$ and $b$ are recorded. By ignoring the state of the measured subsystems $A$ and $B$, the resulting state of the classical registers $A'$ and $B'$ is then $\ket{a}\bra{a}_{A'} \otimes \ket{b}\bra{b}_{B'} $. Hence, if the result of the LGM is unknown, the system composed of the two classical registers will be in the classical-classical state
\begin{equation}\label{Equation:PostLGMAB}
\tilde{\Pi}_{AB}[\rho_{AB}]:=\sum_{ab} p_{ab} \ket{a}\bra{a}_{A'} \otimes \ket{b}\bra{b}_{B'}.
\end{equation}

Analogous definitions can be obtained for LPMs by just removing the tilde. Notice that, according to Naimark's theorem, every generalised measurement
%described by a POVM
can be represented as a projective measurement on a larger system \cite{peres1993quantum}. In particular, this means that we can think of the action of an LGM
%$\tilde{\Pi}_{A}[\rho_{AB}]$
applied only to subsystem $A$ as equivalent to the action of an LPM
%$\Pi_{A'}[\rho_{A'B}]$
applied only to a subsystem $A'$ whose Hilbert space is obtained by extending the Hilbert space of $A$ into a larger one. Similarly, we can think of the action of an LGM
%$\tilde{\Pi}_{A}[\rho_{AB}]$
applied separately to both subsystems $A$ and $B$ as equivalent to the action of an LPM
%$\Pi_{A'}[\rho_{A'B}]$
applied separately to both subsystem $A'$ and $B'$ whose Hilbert spaces are obtained by extending, respectively, the Hilbert spaces of $A$ and $B$ into larger ones. See e.g.~\cite{spehner2014quantum} for a more detailed account.

% to $\tilde{\Pi}_{AB}[\rho_{AB}]$ as equivalent to $\Pi_{A'B'}[\rho_{A'B'}]$ considering an embedding into a larger space extended on both subsystems, see e.g.~\cite{spehner2014quantum}.

Let us then consider in detail the relevant case of measuring  $\rho_{AB}$ via a so-called complete rank one LPM, also known as a local von Neumann measurement, whose subnormalised maps form a complete set of orthogonal rank one projectors, i.e.,~$(\pi_{A})_{a}[\rho_A] = \ket{a}\bra{a}_{A} \rho_A \ket{a}\bra{a}_{A}$ and $(\pi_{B})_{b}[\rho_B] = \ket{b}\bra{b}_{B} \rho_B \ket{b}\bra{b}_{B}$, with $\{\ket{a}_A\}$ and $\{\ket{b}_B\}$ orthonormal bases and $\rho_A$ and $\rho_B$ arbitrary states for subsystem $A$ and $B$ respectively. Then,  it is immediate to see that the post-measurement states can be written as
\begin{eqnarray}
\pi_{A}[\rho_{AB}] &:=& \sum_{a}\left(\ket{a}\bra{a}_{A} \otimes \mathbb{I}_{B}\right) \rho_{AB}\left(\ket{a}\bra{a}_{A} \otimes \mathbb{I}_{B}\right) \nonumber \\ &=& \sum_{a} p_{a} \ket{a}\bra{a}_{A} \otimes \rho_{B|a} \,,  \label{Equation:PostLPMA}\\
\pi_{AB} [\rho_{AB}] &:=& \sum_{ab}\left(\ket{a}\bra{a}_{A} \otimes \ket{b}\bra{b}_{B}\right) \rho_{AB}\left(\ket{a}\bra{a}_{A} \otimes \ket{b}\bra{b}_{B}\right) \nonumber \\ &=& \sum_{ab} p_{ab} \ket{a}\bra{a}_{A} \otimes \ket{b}\bra{b}_{B}\,, \label{Equation:PostLPMAB}
\end{eqnarray}
i.e., any state $\rho_{AB}$ (even if initially entangled or generally non-classical) is {\em mapped} into a classical-quantum or a classical-classical state, after such a complete rank one LPM acting on subsystem $A$ or on both subsystems $A$ and $B$, respectively. This simple observation captures the fundamental perturbing role of local (von Neumann) quantum measurements on general states of composite systems.

%It is clear that if we perform again the same complete rank one LPM on the output state of Eq.~(\ref{Equation:PostLPMA}), then the resulting post-measurement state will not be further affected, that is, $\pi_A\big[\pi_A[\rho_{AB}]\big]=\pi_A[\rho_{AB}]$.

Indeed, since the action of any complete LPM $\pi_A$ on any input state $\rho_{AB}$ results in a classical-quantum post-measurement state of the form given in Eq.~(\ref{Equation:PostLPMA}),  if a state  $\rho_{AB}$ is invariant under such an operation it must be necessarily classical-quantum. This means that the only states left invariant by a complete rank one LPM on subsystem $A$ are classical-quantum states. Conversely, for every classical-quantum state $\rho_{AB} \in \mathscr{C}_A$ as defined in Eq.~(\ref{Equation:CQ}), there exists a complete rank one LPM on subsystem $A$ which leaves such a state invariant, defined precisely as the one with a complete set of orthogonal rank one projectors given by $(\pi_{A})_{i}[\rho_A] = \ket{i}\bra{i}_{A} \rho_A \ket{i}\bra{i}_{A}$.
It thus holds that
\begin{equation}\label{Equation:ClassicalStateProjectiveDefinitionA}
\rho_{AB} \in \mathscr{C}_{A}\quad \Leftrightarrow \quad \exists \mbox{ a complete rank one LPM on $A$  such that} \ \pi_{A}[\rho_{AB}] = \rho_{AB}.
\end{equation}
A similar argument for classical-classical states and LPMs on both subsystems implies that
\begin{equation}\label{Equation:ClassicalStateProjectiveDefinitionAB}
\rho_{AB} \in \mathscr{C}_{AB} \quad \Leftrightarrow \quad \exists \mbox{ a complete rank one LPM on $A$ and $B$ such that} \ \pi_{AB}[\rho_{AB}] = \rho_{AB}.
\end{equation}
These two statements formalise the defining Property~\ref{Property 2:} given earlier for classical states, and illustrated in Fig.~\ref{Fig:Defs}(a). Moreover, they also motivate why they are called classically correlated states: because it is possible to perform a local von Neumann measurement on them that leaves such states unperturbed.

Even more, since the state of a measured subsystem becomes itself diagonal in the basis in which the complete rank one LPM is performed, we have that the corresponding post-measurement state $\pi_A[\rho_{AB}]$ (resp., $\pi_{AB}[\rho_{AB}]$) and output of the quantum-to-classical map $\Pi_A[\rho_{AB}]$ (resp., $\Pi_{AB}[\rho_{AB}]$) can be considered equivalent.

Overall, we can thus locally access the information about a system in a classical state  by probing the subsystem via a von Neumann measurement without inducing a global disturbance (in contrast to the general quantum mechanical scenario). This latter feature reveals that the correlations that are left after a complete rank one LPM are akin to those described by classical probability theory, hence justifying the terminology for the corresponding output states.

%The fact that the information about these states can be accessed locally by probing the subsystems without inducing a disturbance (in contrast to the general quantum mechanical scenario) reveals that their correlations are akin to those described by classical probability theory, hence justifying the terminology.

\subsubsection{Non-creation of entaglement with an apparatus}\label{Section:Facts-Entanglement}

Let us proceed by analysing in more detail the workings of an LPM according to von Neumann's model \cite{vonneumann1932mathematical}. Given a bipartite system in the initial state $\rho_{AB}$, any LPM on subsystem $A$ with a complete set of orthogonal rank one projectors $(\pi_{A})_{a}[\rho_A] = \ket{a}\bra{a}_{A}\rho_A\ket{a}\bra{a}_{A}$ can be realised by letting subsystem $A$ interact via a unitary operation $V^{\{\ket{a}\}}_{AA'}$ with an ancillary system $A'$ initialised in a reference pure state, say $\ket{0}_{A'}$ in its computational basis; in this case the ancillary system $A'$ has the same dimension as $A$ and plays the role of a measurement apparatus \cite{streltsov2011linking}. The state of the three parties $A,B,A'$ after the interaction is known as the {\it pre-measurement} state
\begin{equation}\label{Equation:Premeasurement}
\rho'^{\{\ket{a}_A\}}_{ABA'} := (V^{\{\ket{a}_A\}}_{AA'} \otimes \mathbb{I}_B) (\rho_{AB} \otimes \ket{0}\bra{0}_{A'}) (V^{\{\ket{a}_A\}}_{AA'} \otimes \mathbb{I}_B)^{\dagger}\,.
\end{equation}
The LPM is completed by a readout of the apparatus $A'$ in its eigenbasis, in such a way that
\begin{equation}\label{Equation:Readout}
\mbox{Tr}_{A'} \rho'^{\{\ket{a}_A\}}_{ABA'} = \sum_{a}\left(\ket{a}\bra{a}_{A} \otimes \mathbb{I}_{B}\right) \rho_{AB}\left(\ket{a}\bra{a}_{A} \otimes \mathbb{I}_{B}\right) = \pi_{A}[\rho_{AB}]\,.
\end{equation}
Imposing Eq.~(\ref{Equation:Readout}), one finds that the pre-measurement interaction between $A$ and $A'$ has to be of a very specific form (up to a local unitary on $A'$), namely that of an isometry % controlled unitary
\begin{equation}\label{Equation:PremeasurementUnitary}
V^{\{\ket{a}_A\}}_{AA'} \ket{a}_A \ket{0}_{A'} = \ket{a}_A \ket{a}_{A'}\,.
\end{equation}
We can always think of $V^{\{\ket{a}_A\}}_{AA'}$ as the combination $V^{\{\ket{a}_A\}}_{AA'} = C_{AA'}  (U^{\{\ket{a}_A\}}_{A} \otimes \mathbb{I}_{A'})$ of a local unitary $U^{\{\ket{a}_A\}}_{A}$ on $A$, which serves the purpose of determining in which basis $\{\ket{a}_A\}$ the subsystem is going to be measured (i.e., which observable is being probed), followed by a generalised controlled-\textsc{not} gate $C_{AA'}$, whose action on the computational basis $\ket{i}_A \ket{i'}_{A'}$ of $\mathbb{C}^d \otimes \mathbb{C}^d$ is  $C_{AA'} \ket{i}_A \ket{i'}_{A'} = \ket{i}_A \ket{i \oplus i'}_{A'}$, with  $\oplus$ denoting addition modulo $d$ \cite{piani2011all,piani2012quantumness}.

In case of a joint LPM on both subsystems $A$ and $B$, the pre-measurement state reads
\begin{equation}\label{Equation:PremeasurementAB}
\rho'^{\{\ket{a}_A, \ket{b}_B\}}_{ABA'B'} := (V^{\{\ket{a}_A\}}_{AA'} \otimes V^{\{\ket{b}_B\}}_{BB'} ) (\rho_{AB} \otimes \ket{0}\bra{0}_{A'} \otimes \ket{0}\bra{0}_{B'}) (V^{\{\ket{a}_A\}}_{AA'} \otimes V^{\{\ket{b}_B\}}_{BB'})^{\dagger}\,,
\end{equation}
where $B'$ is an additional ancilla of the same dimension as $B$, which plays the role of an apparatus measuring $B$ in the basis $\{\ket{b}_B\}$.

In general, the pre-measurement state $\rho'^{\{\ket{a}\}}_{ABA'}$ ($\rho'^{\{\ket{a}_A, \ket{b}_B\}}_{ABA'B'}$) is entangled across the split $AB:A'$ ($AB:A'B'$), meaning that the ancilla(e) become entangled with the whole system $AB$ due to the pre-measurement interaction(s); it is indeed because of the presence of such entanglement that the system $AB$ is mapped to a classical state upon tracing over the ancillary system(s), an act which generally amounts to an irreversible information loss similar to a decoherence phenomenon, perceived as a disturbance on the state of the system due to the local measurement(s) \cite{zurek2003decoherence}. However, we know that classical states can be locally measured without perturbation. In the present context, this means that for classical states there is at least one local measurement basis such that the corresponding pre-measurement state remains separable across the system versus ancilla(e) split. Formally, it has been proven in \cite{streltsov2011linking,piani2011all,piani2012quantumness} that
\begin{eqnarray}\label{Equation:ClassicalStateEntanglementDefinitionA}
\hspace*{-1.5cm}\rho_{AB} \in \mathscr{C}_{A} \,\,\,\,\,\,\,\, &\Leftrightarrow& \,\,\,\,\,\,\,\, \exists \mbox{\ \ a local basis}\,\,\,\, \{\ket{a}_A\} \,\,\,\, \mbox{such that} \,\,\,\, \rho'^{\{\ket{a}_A\}}_{ABA'} \in \mathscr{S}_{AB:A'},
\\
\label{Equation:ClassicalStateEntanglementDefinitionAB}
\hspace*{-1.5cm}\rho_{AB} \in \mathscr{C}_{AB} \,\,\,\,\,\,\,\, &\Leftrightarrow& \,\,\,\,\,\,\,\, \exists \mbox{\ \ local bases}\,\,\,\, \{\ket{a}_A, \ket{b}_B\} \,\,\,\, \mbox{such that} \,\,\,\, \rho'^{\{\ket{a}_A, \ket{b}_B\}}_{ABA'B'} \in \mathscr{S}_{AB:A'B'},
\end{eqnarray}
where we denote by $\mathscr{S}_{X:Y}$ the set of separable states, Eq.~(\ref{Equation:SSet}), with the explicit indication of the  considered bipartition $X:Y$.
These two statements formalise the defining Property~\ref{Property 3:} of classical states, as illustrated in Fig.~\ref{Fig:Defs}(b). Equivalently, one may say that classical correlations are not always {\it activated} into entanglement by the act of local measurements \cite{piani2011all,adesso2014experimental}.

\subsubsection{Invariance under a local non-degenerate unitary}\label{Section:Facts-Unitary}

The possibility of invariance under a non-trivial local unitary evolution is another distinctive characteristic of classical states. Consider a local unitary $U_{A}^{\Gamma}$  with a spectrum $\Gamma$ that acts on subsystem $A$ of a bipartite system in the state $\rho_{AB}$. With a slight abuse of notation, we can define the transformed state after the action of such local unitary as $U_{A}^{\Gamma} [\rho_{AB}]$, with
\begin{equation}\label{Equation:UA}
U_{A}^{\Gamma} [\rho_{AB}] := \left( U_{A}^{\Gamma} \otimes \mathbb{I}_{B}\right) \rho_{AB} \left(U_{A}^{\Gamma} \otimes \mathbb{I}_{B}\right)^{\dagger}.
\end{equation}

In general, if we fix a non-degenerate spectrum $\Gamma$, the transformed state $U_{A}^{\Gamma} [\rho_{AB}]$ will be different from the initial $\rho_{AB}$. However, if, and only if, $\rho_{AB}$ is classical-quantum, then one can find at least one local non-degenerate unitary which leaves this state invariant. Clearly, such a unitary $U_{A}^{\Gamma}$ is characterised by having its eigenbasis given precisely by the orthonormal basis $\{\ket{i}_A\}$ entering the definition of the classical-quantum state $\rho_{AB}$ as in Eq.~(\ref{Equation:CQ}). Formally, it thus holds that \cite{gharibian2012quantifying,giampaolo2013quantifying,roga2014discord}
\begin{equation}\label{Equation:ClassicalStateUnitaryDefinitionA}
\rho_{AB} \in \mathscr{C}_{A} \quad \Leftrightarrow \quad \exists \mbox{\ \ a local unitary $U_A^\Gamma$ with non-degenerate $\Gamma$ such that $U_{A}^{\Gamma}[\rho_{AB}] = \rho_{AB}$},
\end{equation}
which formalises the defining Property~\ref{Property 4:} of classical states, as illustrated in Fig.~\ref{Fig:Defs}(c).

The spectrum $\Gamma$ must be non-degenerate to exclude such trivialities as the choice of $\mathbb{I}_{A}$ as a local unitary (which would leave any state, not only those classical-quantum, invariant). Notice that, in contrast with the case of a local measurement discussed in Section~\ref{Section:Facts-Measuring}, a local unitary does not alter the correlations between $A$ and $B$ at all, yet can assess their nature based on the global change (or lack thereof) of the state $\rho_{AB}$ in {\it response} to the action of any such local unitary \cite{roga2014discord}.

Curiously, unlike all the other definitions discussed in the current section, a necessary and sufficient condition analogous to Eq.~(\ref{Equation:ClassicalStateUnitaryDefinitionA}) does not hold for classical-classical states in this case. Namely, considering also a local unitary $U_{B}^{\Gamma}$ acting on $B$ with the same spectrum $\Gamma$, and defining the joint action of $U_A^\Gamma$ and $U_B^\Gamma$ on $\rho_{AB}$ as
\begin{equation}\label{Equation:UAB}
U_{AB}^{\Gamma} [\rho_{AB}] := \left( U_{A}^{\Gamma} \otimes U_{B}^{\Gamma}\right) \rho_{AB} \left(U_{A}^{\Gamma} \otimes U_{B}^{\Gamma}\right)^{\dagger},
\end{equation}
then if $\Gamma$ is non-degenerate it still holds that
\begin{equation}\label{Equation:ClassicalStateUnitaryDefinitionAB}
\rho_{AB} \in \mathscr{C}_{AB} \quad \Rightarrow \quad \exists \mbox{\ \ local unitaries $U_A^\Gamma, U_B^\Gamma$ with non-degenerate $\Gamma$ such that $U_{AB}^{\Gamma}[\rho_{AB}] = \rho_{AB}$},
\end{equation}
but the reverse implication is no longer true. In fact, there are even entangled states $\rho_{AB}$ which may be left unchanged by a joint non-trivial local unitary evolution; an example is the maximally entangled two-qubit Bell state $\ket{\Phi}_{AB} = \frac1{\sqrt2}(\ket{0}_A\ket{0}_B + \ket{1}_A \ket{1}_B)$, which is invariant under the action of any tensor product of two matching Pauli matrices,
\begin{equation}\label{Equation:NoABU}
({\sigma_i}_A \otimes {\sigma_i}_B) \ket{\Phi}\bra{\Phi}_{AB} ({\sigma_i}_A \otimes {\sigma_i}_B) = \ket{\Phi}\bra{\Phi}_{AB},
 \end{equation}
 for $i=1,2,3$ (notice that the Pauli matrices are unitary and Hermitian and with non-degenerate spectrum $\Gamma=\{\pm1\})$.
It is an open question how to give a suitable local unitary based condition characterising the set of classical-classical states, with one possibility being to fix a different spectrum between $A$ and $B$, or to involve multiple local unitaries.

\subsubsection{Incoherence in a local basis}\label{Section:Facts-Coherence}
The previous analysis reveals that classical-quantum states commute with at least a local unitary $U^\Gamma_A$ on subsystem $A$. An alternative way of describing this is by saying that classical-quantum states are locally {\it incoherent} with respect to the eigenbasis of $U^\Gamma_A$. To be more precise, let us recall some basic terminology used in the characterisation of quantum coherence, see e.g.~\cite{aaberg2006quantifying,baumgratz2014quantifying,winter2016operational,napoli2016robustness}.

Quantum coherence represents superposition with respect to a fixed reference orthonormal basis. In the case of a single system, fixing a reference basis $\{\ket{i}\}$, the corresponding set $\mathscr{I}^{\{\ket{i}\}}$ of incoherent states (i.e., states with vanishing coherence) is defined as the set of those states diagonal in the reference basis,
\begin{equation}\label{Equation:ISet}
\mathscr{I}^{\{\ket{i}\}} := \Big\lbrace\ \rho  \quad | \quad \rho = \sum_i p_i \ket{i}\bra{i}\Big\rbrace\,,
\end{equation}
with $\{p_i\}$ a probability distribution. Any other state $\rho \not\in \mathscr{I}^{\{\ket{i}\}}$ exhibits quantum coherence with respect to the fixed reference basis $\{\ket{i}\}$.

Consider now a bipartite system $AB$. We can fix local reference bases $\{\ket{a}_A\}$ for subsystem $A$ and $\{\ket{b}_B\}$ for subsystem $B$, and characterise coherence with respect to these. In particular, we can define two different sets of locally incoherent states. Namely, the set $\mathscr{I}^{\{\ket{a}_A\}}_A$ of incoherent-quantum (or incoherent on $A$) states  is defined as \cite{chitambar2016assisted,ma2016converting}
\begin{equation}\label{Equation:IASet}
\mathscr{I}^{\{\ket{a}_A\}}_A := \Big\lbrace\ \rho_{AB} \quad | \quad \rho_{AB} = \sum_{a} p_{a} \ket{a}\bra{a}_{A} \otimes \rho_{B}^{(a)} \Big\rbrace\,,
\end{equation}
for any quantum states $\{\rho_B^{(a)}\}$ of subsystem $B$, with $\{p_a\}$ a probability distribution. Similarly, the set $\mathscr{I}^{\{\ket{a}_A,\ket{b}_B\}}_{AB}$ of incoherent-incoherent (or incoherent on $A$ and $B$) states  is defined as \cite{bromley2015frozen,streltsov2015measuring,winter2016operational}
\begin{equation}\label{Equation:IABSet}
\mathscr{I}^{\{\ket{a}_A,\ket{b}_B\}}_{AB} := \Big\lbrace\ \rho_{AB} \quad | \quad \rho_{AB}=\sum_{ab} p_{ab} \ket{a}\bra{a}_{A} \otimes  \ket{b}\bra{b}_{B} \Big\rbrace\,,
\end{equation}
with  $\{p_{ab}\}$ a joint probability distribution.
Notice that in the above definitions the reference bases are fixed, which makes Eqs.~(\ref{Equation:IASet}) and (\ref{Equation:IABSet}) different from the definitions of the sets of classical-quantum and classical-classical states, Eqs.~(\ref{Equation:CQSet}) and (\ref{Equation:CCSet}), respectively. However, it is straightforward to realise that a state $\rho_{AB}$ is classical-quantum (classical-classical) if, and only if, there exist a local basis for $A$ (and for $B$) with respect to which $\rho_{AB}$ is incoherent on $A$ (and $B$). In formulae,
\begin{eqnarray}
\hspace*{-1cm}
\label{Equation:ClassicalStateCoherenceDefinitionA}
\rho_{AB} \in \mathscr{C}_{A} \quad &\Leftrightarrow& \quad \exists \mbox{\ \ a local basis}\,\,\,\, \{\ket{a}_A\} \,\,\,\, \mbox{such that} \,\,\,\, \rho_{AB} \in \mathscr{I}^{\{\ket{a}_A\}}_A,
\\
\hspace*{-1cm}
\label{Equation:ClassicalStateCoherenceDefinitionAB}
\rho_{AB} \in \mathscr{C}_{AB} \quad &\Leftrightarrow& \quad\exists \mbox{\ \ local bases}\,\,\,\, \{\ket{a}_A, \ket{b}_B\} \,\,\,\, \mbox{such that} \,\,\,\, \rho_{AB} \in \mathscr{I}^{\{\ket{a}_A,\ket{b}_B\}}_{AB}.
\end{eqnarray}
These characterisations amount to the defining Property~\ref{Property 5:} of classical states, as illustrated in Fig.~\ref{Fig:Defs}(d). In the present paradigm, classicality is thus pinned down as a very intuitive notion, intimately related to the lack of quantum superposition in a specific local reference frame.

\subsubsection{Invariance under an entanglement-breaking channel}\label{Section:Facts-Breaking}
Finally, a classical state can be identified as a fixed point of an entanglement-breaking channel \cite{seshadreesan2015fidelity}.
Entanglement-breaking channels \cite{horodecki2003entanglement} acting on subsystem $A$ of a bipartite system are defined as those CPTP operations $\Lambda^{\text{EB}}_A$ which map any state $\rho_{AB}$ to a separable state, i.e.,
\begin{equation}
\Lambda^{\text{EB}}_A[\rho_{AB}] :=(\Lambda^{\text{EB}}_A \otimes \mathbb{I}_B) [\rho_{AB}] \in \mathscr{S}_{AB}\,\ \ \ \forall \rho_{AB}\in\mathscr{D}_{AB}.
 \end{equation}
 Equivalently, the entanglement-breaking channels are all and only the maps that can be written as the concatenation of a local measurement followed by a preparation \cite{horodecki2003entanglement}, i.e.,
\begin{equation}\label{Equation:EntanglementBreaking}
\Lambda_A^{\text{EB}}[\rho_{A}]:=\sum_i \mbox{Tr}({M_i} \rho_{A}) \rho_{A}^{(i)},
\end{equation}
where the $\{{M_i}\}$ constitute a POVM while the $\{\rho_{A}^{(i)}\}$ are arbitrary quantum states of subsystem $A$. In other words, an entanglement-breaking channel applied to the subsystem state $\rho_{A}$ can be simulated classically as follows: a sender performs an LGM with POVM elements $\{{M_i}\}$ on subsystem $A$ and sends the outcome $i$,  which occurs with probability $p_i=\mbox{Tr}({M_i}\rho_{A})$, via a classical channel to a receiver, who then prepares the subsystem $A$ in a corresponding pre-arranged state $\rho_{A}^{(i)}$ \cite{horodecki2003entanglement}.

It then turns out that a state $\rho_{AB}$ is classical-quantum if, and only if, it is left invariant by at least one entanglement-breaking channel acting on subsystem $A$ \cite{seshadreesan2015fidelity}, i.e.,
\begin{equation}\label{Equation:ClassicalStateEBDefinitionA}
\rho_{AB} \in \mathscr{C}_{A} \quad \Leftrightarrow \quad  \exists \mbox{\ \ an entanglement-breaking channel\ \  $\Lambda_{A}^{\text{EB}}$ \ \ such that \ \  $\Lambda_{A}^{\text{EB}}[\rho_{AB}] = \rho_{AB}$}.
\end{equation}
To see this, let us first assume that $\rho_{AB} \in \mathscr{C}_{A}$, i.e., $\rho_{AB}$ is defined as in Eq.~(\ref{Equation:CQ});
%) = \sum_{i} p_{i} \ket{i}\bra{i}_{A} \otimes \rho_{B}^{(i)}$ for some orthonormal basis $\ket{i}_A$, arbitrary states $\rho_{B}^{(i)}$ and probability distribution $p_{i}$, then
we have then from Section~\ref{Section:Facts-Measuring} that such a state is invariant under the complete LPM with rank one projectors $\{\ket{i}\bra{i}_{A}\}$, which is a particular case of an entanglement-breaking channel. On the other hand, if there exists at least one entanglement-breaking channel $\Lambda_{A}^{\text{EB}}$ such that $\Lambda_{A}^{\text{EB}}[\rho_{AB}] = \rho_{AB}$, then also the channel $\overline{\Lambda}_{A}^{\text{EB}}$, defined as
$\overline{\Lambda}_{A}^{\text{EB}} = \lim_{N\rightarrow \infty} \frac{1}{N} \sum_{n=1}^N \left(\Lambda_{A}^{\text{EB}}\right)^n$,
leaves $\rho_{AB}$ invariant. In \cite{fukuda2015quantum} it has been shown that this map is an entanglement-breaking channel as well, whose action can be written as follows,
$\overline{\Lambda}_{A}^{\text{EB}}[\rho_{A}] = \sum_i \mbox{Tr}(M_i \rho_{A}) \sigma_A^{(i)}$,
where the $\{M_i\}$ form a POVM while the $\{\sigma_A^{(i)}\}$ are states of subsystem $A$ with orthogonal support. Therefore the channel $\overline{\Lambda}_{A}^{\text{EB}}$ transforms $\rho_{AB}$ into a classical-quantum state, and since $\rho_{AB}$ is invariant with respect to the action of
this channel, then $\rho_{AB}$ must be classical-quantum itself.

Analogously, one can see that a state $\rho_{AB}$ is classical-classical if, and only if, it is left invariant by at least one entanglement-breaking channel $\Lambda_{AB}^{\text{EB}} := \Lambda_{A}^{\text{EB}} \otimes \Lambda_B^{\text{EB}}$ acting on both subsystems $A$ and $B$ \cite{seshadreesan2015fidelity}, that is,
\begin{equation}\label{Equation:ClassicalStateEBDefinitionAB}
\rho_{AB} \in \mathscr{C}_{AB} \quad \Leftrightarrow \quad \exists \mbox{\ \ an entanglement-breaking channel\ \  $\Lambda_{AB}^{\text{EB}}$\ \  such that \ \  $\Lambda_{AB}^{\text{EB}}[\rho_{AB}] = \rho_{AB}$}.
\end{equation}

Notice that, apart from the special case of local measurements, entanglement-breaking channels do not generally destroy the QCs content in a state $\rho_{AB}$, as they only vanquish its entanglement. Nevertheless, similarly to the case of local unitaries in Section~\ref{Section:Facts-Unitary}, it is the global change (or lack thereof) of the state $\rho_{AB}$ under the action of any entanglement-breaking channel that discerns the nature of the correlations in $\rho_{AB}$. Unlike the case of local unitaries, though, the invariance under suitable entanglement-breaking channels allows us to characterise both sets of classical-quantum and classical-classical states. The two statements in Eqs.~(\ref{Equation:ClassicalStateEBDefinitionA}) and (\ref{Equation:ClassicalStateEBDefinitionAB}) thus formalise the defining Property~\ref{Property 6:} of classical states, as illustrated in Fig.~\ref{Fig:Defs}(e).

\subsection{Defining and characterising quantumly correlated states}\label{Section:Facts-Q}

Having familiarised ourselves with classically correlated states and their physical characterisations, we can now formally define {\it quantumly correlated} states in a composite system --- the true object of our interest --- as simply those which are {\it not} classically correlated. Any state which is not a classical state will be said to possess some form of QCs.
Specifically, we shall adopt the following terminology, in line with the majority of the current literature:
 \begin{itemize}
 \item A state $\rho_{AB} \not\in \mathscr{C}_{A}$ is said to have {\it one-sided} QCs, or equivalently is {\it quantumly correlated} with respect to subsystem $A$, or in short quantum on $A$;
      \item A state $\rho_{AB} \not\in \mathscr{C}_{AB}$ is said to have {\it two-sided} QCs, or equivalently is {\it quantumly correlated} with respect to either subsystem $A$ or $B$, or in short is quantum on $A$ or $B$.
          %;
       %\item A state $\rho_{AB} \not\in \mathscr{C}_{A} \cup \mathscr{C}_B$ has {\it two-way} QCs, or equivalently is {\it quantumly correlated} with respect to both subsystems $A$ and $B$, or in short is quantum on $A$ and $B$.
\end{itemize}
A brief remark on the semantics is in order. Here the attributes `one-sided' and `two-sided' are motivated by the set of  states on which the corresponding  QCs vanish, namely one-sided QCs are intended as those which vanish on one-sided classical states (alias classical-quantum states) belonging to the set $\mathscr{C}_{A}$, while two-sided QCs are intended as those which vanish on two-sided classical states (alias classical-classical states) belonging to the set $\mathscr{C}_{AB}$. In this sense, note that the set of states with two-sided QCs as defined above is strictly larger than the set of states with one-sided QCs, which in turn is strictly larger than the set of states  $\rho_{AB} \not\in \mathscr{C}_{A} \cup \mathscr{C}_B$, which are quantumly correlated with respect to {\it both} subsystems $A$ and $B$, or in short quantum on $A$ {\it and} $B$.

In general, to be precise, QCs capture the genuinely quantum character of the marginal subsystems of a composite quantum system, a feature which can yet manifest only in case the subsystems are correlated. In this paradigm, in fact, any state of a single system is necessarily regarded as classical, since it can always be diagonalised in its orthonormal eigenbasis (hence being invariant under a complete projective measurement in such eigenbasis).

Referring to the analysis of Section~\ref{Section:Facts-C}, by negation of the various defining Properties \ref{Property 1:}--\ref{Property 6:} of classical states, we can hence provide the following equivalent definitions of bipartite states possessing QCs, as illustrated schematically in the right column of Fig.~\ref{Fig:Defs}.
\begin{description}
\setcounter{desccount}{0}
\descitem{Definition 1:}{A state $\rho_{AB}$ has QCs if it is not a classical state, i.e., if $\rho_{AB} \notin \mathscr{C}_{A}$ then $\rho_{AB}$ has one-sided QCs [negation of Eq.~(\ref{Equation:CQ})], whilst if $\rho_{AB} \notin \mathscr{C}_{AB}$ then $\rho_{AB}$ has two-sided QCs [negation of Eq.~(\ref{Equation:CC})].}
\descitem{Definition 2:}{A state $\rho_{AB}$ has QCs if it is always perturbed by any local complete rank one projective measurement, i.e., if for every complete rank one LPM on $A$ it holds $\pi_{A}[\rho_{AB}] \neq \rho_{AB}$ then $\rho_{AB}$ has one-sided QCs [negation of Eq.~(\ref{Equation:ClassicalStateProjectiveDefinitionA})], whilst if for every complete rank one LPM %$\pi_{AB} = \pi_A \otimes \pi_B$
on both $A$ and $B$ it holds $\pi_{AB}[\rho_{AB}] \neq \rho_{AB}$ then $\rho_{AB}$ has two-sided QCs [negation of Eq.~(\ref{Equation:ClassicalStateProjectiveDefinitionAB})]; see Fig.~\ref{Fig:Defs}(a).}
\descitem{Definition 3:}{A state $\rho_{AB}$ has QCs if it always becomes entangled with an apparatus during any local complete rank one projective measurement, i.e., if for every local basis $\{\ket{a}_A\}$ the pre-measurement state $\rho'^{\{\ket{a}_A\}}_{ABA'} \not\in \mathscr{S}_{AB:A'}$ then $\rho_{AB}$ has one-sided QCs [negation of Eq.~(\ref{Equation:ClassicalStateEntanglementDefinitionA})], whilst if for every local bases $\{\ket{a}_A, \ket{b}_B\}$ the pre-measurement state $\rho'^{\{\ket{a}_A,\ket{b}_B\}}_{ABA'B'} \not\in \mathscr{S}_{AB:A'B'}$ then $\rho_{AB}$ has two-sided QCs  [negation of Eq.~(\ref{Equation:ClassicalStateEntanglementDefinitionAB})]; see Fig.~\ref{Fig:Defs}(b).}
\descitem{Definition 4:}{A state $\rho_{AB}$ has QCs if it is always altered by any local non-degenerate unitary, i.e.~if for every local unitary $U_{A}^{\Gamma}$ with non-degenerate spectrum $\Gamma$  it holds $U_{A}^{\Gamma}[\rho_{AB}]\neq\rho_{AB}$ then $\rho_{AB}$ has one-sided QCs  [negation of Eq.~(\ref{Equation:ClassicalStateUnitaryDefinitionA})]; see Fig.~\ref{Fig:Defs}(c).}
\descitem{Definition 5:}{A state $\rho_{AB}$ has QCs if it exhibits quantum coherence in all local bases,  i.e., if for every local basis $\{\ket{a}_A\}$ it holds $\rho_{AB} \not \in \mathscr{I}^{\{\ket{a}_A\}}_A$ then $\rho_{AB}$ has one-sided QCs [negation of Eq.~(\ref{Equation:ClassicalStateCoherenceDefinitionA}).], whilst if for every local bases $\{\ket{a}_A, \ket{b}_B\}$ it holds $\rho_{AB} \not\in \mathscr{I}^{\{\ket{a}_A,\ket{b}_B\}}_{AB}$ then $\rho_{AB}$  has two-sided QCs  [negation of Eq.~(\ref{Equation:ClassicalStateCoherenceDefinitionAB})]; see Fig.~\ref{Fig:Defs}(d).}
\descitem{Definition 6:}{A state $\rho_{AB}$ has QCs if it is always altered by any local entanglement-breaking channel, i.e.~if for every entanglement-breaking channel $\Lambda_{A}^{\text{EB}}$ on $A$ it holds $\Lambda_{A}^{\text{EB}}[\rho_{AB}] \neq \rho_{AB}$ then $\rho_{AB}$ has one-sided QCs [negation of Eq.~(\ref{Equation:ClassicalStateEBDefinitionA})], whilst if for every entanglement-breaking channel $\Lambda_{AB}^{\text{EB}}=\Lambda_{A}^{\text{EB}}\otimes \Lambda_{B}^{\text{EB}}$ on both $A$ and $B$ it holds $\Lambda_{AB}^{\text{EB}}[\rho_{AB}] \neq \rho_{AB}$ then $\rho_{AB}$ has two-sided QCs [negation of Eq.~(\ref{Equation:ClassicalStateEBDefinitionAB})]; see Fig.~\ref{Fig:Defs}(e).}
\end{description}

In the next Section, we promote each of the above definitions from the qualitative to the quantitative domain, and report on some of the most up-to-date progress available on the quest for the quantification of QCs in bipartite quantum systems.

\section{Measures of quantum correlations}\label{Section:Measures}

In the previous Section we have discussed the defining characteristics of quantumly correlated states. It is a natural question now to ask whether one can meaningfully compare the QCs present in a pair or more of such states. This is achieved by imposing a quantitative ordering on the set of quantum states according to a suitable  {\it measure} of QCs. However, as it is often the case in quantum information theory \cite{plenio2007an}, it turns out that there is not a unique, universal measure, but rather different measures resulting from different physical motivations and imposing different orderings on the set of quantum states.

In this Section we will review a variety of such measures. Our approach will be thematic more than historic, and guided by the general intuition that a measure of QCs in a state $\rho_{AB}$ should capture essentially how much $\rho_{AB}$ deviates from being a classical state. Having introduced different ways to characterise classical states in the previous Section, in the following we will accordingly classify existing (and potentially future) measures of QCs into suitably different categories, each reflecting a particular signature of QCs, in accordance with the guiding scheme of Fig.~\ref{Fig:Defs}. We will accompany our discussion with several Tables, useful to summarise our notation as well as the main properties of the measures and comparisons between them, and we will also link in with the applications of some representatives of these measures to quantum information processing and beyond, given later in Section~\ref{Section:Applications}.

Before delving into the zoo of QCs measures, however, we must consider what the requirements are for a given quantifier to be a true {\it bona fide} measure of QCs. Namely, while a measure must naturally incarnate the spirit of QCs, there are as well some important supplementary conditions, or potentially welcome properties, that any given quantifier should satisfy for mathematical and physical consistency. Henceforth we proceed by first discussing such a set of conditions.

\subsection{Requirements for a {\it bona fide} measure of quantum correlations}\label{Section:Requirements}

Despite the intense research drive into characterising QCs over the past few years, the completion of the rule book of criteria for a given quantifier to be a valid measure of QCs is still an open question, with no clear consensus in the literature. Various proposals for mandatory, optional and desirable requirements for a measure of QCs have been discussed e.g.~in \cite{streltsov2011linking,brodutch2012criteria,modi2012classical,girolami2013characterizing,aaronson2013comparative,spehner2014quantum,cianciaruso2015universal,roga2015geometric}, to which the reader is referred for additional details.

Here we focus on an essential set of five Requirements that reflect (in our opinion) some very natural {\it desiderata} that any measure of QCs should obey. While the first four such Requirements can be regarded as reasonably well established,  the fifth Requirement is only a proposed one, whose validation or reconsideration will necessitate further research.

A one-sided (asymmetric) measure of QCs $Q_{A}(\rho_{AB})$ on bipartite quantum states $\rho_{AB}$ must be a real, non-negative function satisfying the following Requirements:

%\begin{enumerate}
\renewcommand\labelenumi{(\roman{enumi})}
\renewcommand\theenumi\labelenumi
\begin{description}
\reqitem{Requirement (i):}{$Q_{A}(\rho_{AB})=0$ if  $\rho_{AB} \in \mathscr{C}_{A}$ is a classical-quantum state as defined in Eq.~(\ref{Equation:CQ});}\label{Requirements:Faithfulness}
\reqitem{Requirement (ii):}{$Q_{A}(\rho_{AB})$ is invariant under local unitaries, i.e.~$Q_{A}\big((U_{A} \otimes U_{B})\rho_{AB}(U_{A}^{\dagger} \otimes U_{B}^{\dagger})\big)=Q_{A}(\rho_{AB})$, for any state $\rho_{AB}$ and any local unitary operation $U_{A}$ and $U_{B}$ acting on subsystem $A$ and $B$, respectively;}\label{Requirements:UnitaryInvariance}
\reqitem{Requirement (iii):}{$Q_{A}(\rho_{AB})$ reduces to a measure of entanglement $E(\rho_{AB})$ on pure states, i.e.~$Q_{A}(\ket{\psi}_{AB})=E(\ket{\psi}_{AB})$ for any pure bipartite state ${\ket{\psi}_{AB}}$.}\label{Requirements:EntanglementPure}
\end{description}
Similar constraints hold for a two-sided (symmetric) measure of QCs $Q_{AB}(\rho_{AB})$, where one simply substitutes the classical-quantum set $\mathscr{C}_{A}$ in Requirement~(\ref{Requirements:Faithfulness}) with the classical-classical set $\mathscr{C}_{AB}$ defined in Eq.~(\ref{Equation:CCSet}).

Requirement~(\ref{Requirements:Faithfulness}) formalizes the basic requirement that a valid measure of QCs must vanish on classical (i.e., free) states.  We anticipate that all the measures reviewed in this article are in fact {\em faithful}, meaning that they vanish on and {\it only} on classical states.
Requirement~(\ref{Requirements:UnitaryInvariance}) ensures that any measure of QCs is locally basis independent, as expected for any quantifier of correlations \cite{henderson2001classical}. Moreover, as already discussed, all forms of non-classical correlations reduce to entanglement for pure states, hence  one expects any valid measure of QCs to reduce to a corresponding valid measure of entanglement in such a special case. This is  imposed by Requirement~(\ref{Requirements:EntanglementPure}), which has appeared frequently in recent literature on QCs \cite{aaronson2013comparative,girolami2013characterizing,girolami2014quantum,farace2014discriminating,ciccarello2014toward,roga2014discord,cianciaruso2015universal,roga2015geometric}

Another important consideration is how a measure of QCs should behave under the action of local quantum channels. For example, any valid entanglement measure must be non-increasing on average under the action of LOCC \cite{horodecki2009quantum}, so what are the analogous constraints for a measure of general QCs? A strict Requirement is the following, first formalised in \cite{streltsov2011linking},
\begin{description}
\reqitem{Requirement (iv):}{$Q_{A}(\rho_{AB})$ is monotonically non-increasing under any local operation on the party whose quantumness is not being measured (in our convention subsystem $B$), that is, ${Q_{A}\big((\mathbb{I}_{A}\otimes \Lambda_{B})[\rho_{AB}]\big) \leq Q_{A}(\rho_{AB})}$ for any state $\rho_{AB}$ and any CPTP map $\Lambda_B$ on $B$.}\label{Requirements:UnmeasuredParty}
\end{description}
This Requirement guarantees that it must be physically impossible to  increase the quantumness degree of party $A$ by operating on party $B$ alone. Obviously, an analogous constraint cannot be stated for two-sided measures $Q_{AB}(\rho_{AB})$ of QCs. However, it remains to analyse which operations on $A$ (or also on $B$ in the case of two-sided measures) may or should not lead to a creation or an increase of QCs in a bipartite state $\rho_{AB}$.
  To help with this task, one can view QCs as a resource within the framework of {\em resource theories}~\cite{horodecki2013quantumness,coecke2014mathematical,brandao2015second}.

In a resource theory, the \emph{free states} and the \emph{free operations} are identified as, respectively, the states without any resource content and the operations that are freely implementable, i.e., that map free states into free states. The resource theory framework then imposes that any valid measure of a resource should vanish on the free states and should be monotonically non-increasing under the action of the corresponding free operations, or in other words that one can only produce resource by implementing operations that themselves cost resource. For example, turning again to entanglement, the free states are the separable states of Eq.~(\ref{Equation:Separable}), the free operations are the LOCC, and any entanglement measure must be monotonically non-increasing under LOCC \cite{vidal2000entanglement,plenio2007an,horodecki2009quantum}. It is thus clear that the freely implementable operations associated to QCs must be identified in order to impose a similar monotonicity constraint.

Unfortunately, there is no consensus yet on the free operations for an eventual resource theory of QCs \cite{horodecki2013quantumness}. What is fully characterised, though, is the maximal set of local operations that cannot create a state with QCs from a classical (free) state. This set is constituted by so-called local commutativity preserving operations (LCPO), defined as follows. A CPTP map $\Lambda$ is commutativity preserving if
\begin{equation}\label{Equation:CommutativityPreserving}
\left[\Lambda[\rho],\Lambda[\sigma]\right] = 0 \quad \forall \ \ \rho, \sigma \mbox{ such that } \left[\rho,\sigma\right]=0\,,
\end{equation}
with $[ \cdot, \cdot]$ denoting the commutator.
In a bipartite system, a CPTP map $\Lambda_A^{\text{LCPO}} := \Lambda_A \otimes \mathbb{I}_B$ is LCPO on subsystem $A$ if $\Lambda_A$ is commutativity preserving. Similarly, a CPTP map $\Lambda_{AB}^{\text{LCPO}} := \Lambda_A \otimes \Lambda_B$ is LCPO on subsystems $A$ and $B$ if both $\Lambda_A$ and $\Lambda_B$ are commutativity preserving.
It then holds that \cite{hu2012necessary}
\begin{eqnarray}
\label{Equation:LCPOA}
\hspace*{-2.5cm} (\Lambda_A \otimes \mathbb{I}_B)[\rho_{AB}] \in \mathscr{C}_A \quad \forall  \ \rho_{AB} \in \mathscr{C}_A \quad &\Leftrightarrow& \quad \mbox{$\Lambda_A$ is commutativity preserving}, \\
\label{Equation:LCPOAB}
\hspace*{-2.5cm}(\Lambda_A \otimes \Lambda_B)[\rho_{AB}] \in \mathscr{C}_{AB} \quad \forall  \ \rho_{AB} \in \mathscr{C}_{AB} \quad &\Leftrightarrow& \quad \mbox{$\Lambda_A$ and $\Lambda_B$ are commutativity preserving}.
\end{eqnarray}

In other words, LCPO  are all and only the admissible local channels, acting on the subsystem(s) whose quantumness is being measured, that preserve the set of classical states. Any other operation, even if locally performed on party $A$ only, can create QCs out of a classically correlated state. This may seem counterintuitive, as local operations alone cannot create entanglement or even classical correlations instead \cite{henderson2001classical}. However, if we recall the meaning of QCs as discussed in Section~\ref{Section:Facts-Q}, we can convince ourselves that implementing local operations which involve, say, the creation of coherence on $A$, may consequently map a classical state out of its set, breaking the conditions of Section~\ref{Section:Facts-C}; the possibility of local creation of QCs was first studied in \cite{ciccarello2012creating,streltsov2011behavior}. It becomes important, then, to provide a more detailed characterisation of the LCPO set, for which such a possibility is denied.
It was shown in~\cite{streltsov2011behavior} that, if subsystem $A$ is a qubit, LCPO on $A$ amount either to unital channels, $\Lambda_{A}[\mathbb{I}_{A}] = \mathbb{I}_{A}$, i.e.~preserving the maximally mixed state, or to completely decohering maps, $\Lambda_{A}[\rho_{A}] = \sum_{i}p_{i}^{(\rho_A)} \ket{i}\bra{i}_A$, i.e.~mapping any $\rho_{A}$ to a state diagonal in some basis  $\{\ket{i}_A\}$ with probabilities $\{p_{i}^{(\rho_{A})}\}$. It was further shown in \cite{hu2012necessary,guo2013necessary} that, if instead subsystem $A$ is a qudit with dimension $d_A>2$,  LCPO on $A$ can be either isotropic channels, $\Lambda_A[\rho_A] = t \Upsilon[\rho_A] + (1-t) \mbox{Tr}(\rho_A)\frac{\mathbb{I}}{d_A}$, with $\Upsilon$ being either a unitary operation or unitarily equivalent to transposition  (see~\cite{bromley2016there} for further details on these channels, including their Kraus operator sum representation), or completely decohering maps.

Any physically consistent set of free local operations for QCs should therefore be contained into the set of LCPO, or coincide with it. In entanglement theory, for example, LOCC are  singled out by the physical paradigm of distant laboratories, and form a strict subset of the maximal set of operations preserving separable states \cite{brandao2008reversible} (as the latter also contain certain nonlocal operations such as the \textsc{swap} gate), while in the resource theories of quantum coherence there are currently several alternative proposals for free operations, with different characterisations, and a still ongoing debate on their physical consistency \cite{marvian2013the,baumgratz2014quantifying,marvian2015quantum,winter2016operational,yadin2015quantum,napoli2016robustness,chitambar2016are,marvian2016how}.

Here, following \cite{cianciaruso2015universal}, we postulate a {\it strong monotonicity} desideratum for measures of QCs under the maximal set of local operations preserving classical states, given by
\begin{description}
\reqitem{Requirement (v):}{$Q_{A}(\rho_{AB})$ is monotonically non-increasing under any LCPO, that is,  $Q_{A}\big(\Lambda^{\text{LCPO}}_A[\rho_{AB}]\big) \leq Q_{A}(\rho_{AB})$, for any state $\rho_{AB}$ and any LCPO $\Lambda^{\text{LCPO}}_A$ on subsystem $A$.}\label{Requirements:LCPO}
\end{description}
A similar desideratum is postulated for a two-sided measure of QCs $Q_{AB}(\rho_{AB})$, where one then considers LCPO $\Lambda^{\text{LCPO}}_{AB}$ acting on both subsystems $A$ and $B$. Notice that Requirement~(\ref{Requirements:LCPO}) is only a proposed one, and should not be regarded yet as mandatory, unlike the previous four Requirements. Nonetheless, it makes sense to consider it for the following reasons.

If a measure of QCs is proven to obey Requirement~(\ref{Requirements:LCPO}), then it will be a valid monotone in any resource theory of QCs with free operations restricted to local channels. On the other hand, in case an otherwise valid quantifier is found to fail Requirement~(\ref{Requirements:LCPO}), it may still happen to be monotonic under a smaller (as yet uncharacterised) set of local operations, which might form the basis for a physically consistent resource theory of QCs, with an accordingly modified Requirement~(\ref{Requirements:LCPO}). In particular, since local completely decohering operations on one or both subsystems always map any state $\rho_{AB}$ to a classical state belonging respectively to $\mathscr{C}_A$ or $\mathscr{C}_{AB}$, every measure of QCs which obeys Requirement~(\ref{Requirements:Faithfulness}) is automatically a monotone under local completely decohering operations. For any valid measure, therefore, testing Requirement~(\ref{Requirements:LCPO}) amounts to verifying monotonicity under local unital channels when the dimension of the measured party (or parties) is $2$, and under local isotropic channels when the dimension of the measured party (or parties) is higher.

Let us now briefly mention a couple of additional potential desiderata for measures of QCs. First, {\it continuity} has been analysed in \cite{brodutch2012criteria}: this captures the reasonable requirement that QCs should not change too much if a state is slightly perturbed. Moreover, we recall that our Requirements~(\ref{Requirements:UnitaryInvariance}) and (\ref{Requirements:UnmeasuredParty}) together imply that measures of one-sided QCs $Q_A$ are maximal on pure maximally entangled states if $\text{dim}(\mathscr{H}_A) \leq \text{dim}(\mathscr{H}_B)$ \cite{streltsov2012are,spehner2014quantum}. The authors of \cite{roga2015geometric} additionally postulate that measures of one-sided QCs should be maximal {\it only} on such states. Notice however that these properties are no longer valid for measures of two-sided QCs $Q_{AB}$, which (quite surprisingly) can reach higher values on mixed partially entangled states than on pure maximally entangled states \cite{piani2011all}.

Let us further remark that, while {\it convexity} is an optional property that one may wish to have for entanglement measures \cite{plenio2007an} (even though there exist valid entanglement monotones that are not convex \cite{plenio2005logarithmic}), general measures of QCs are (and should be) neither convex nor concave. In fact, while mixing e.g.~a quantum state $\rho_{AB}$ with the maximally mixed one $\propto \mathbb{I}_{AB}$ typically reduces the QCs content of $\rho_{AB}$, one can also create nonzero QCs by mixing two classical states, provided they are locally diagonal in two different bases.

In what follows, we shall adopt the minimal set of (four plus a potential one) Requirements~(\ref{Requirements:Faithfulness})--(\ref{Requirements:LCPO}) introduced above, and aim to validate QCs quantifier against them.

\subsection{Types of measures of quantum correlations}\label{Section:Types}

We begin our exploration of the zoology of established measures of QCs by introducing a classification which groups them into different types, based upon how the QCs in an arbitrary bipartite state $\rho_{AB}$ are captured conceptually by the characterisations of Section~\ref{Section:Facts-Q} (see Fig.~\ref{Fig:Defs}), as summarised in Table~\ref{Table:Types}.

\renewcommand{\baselinestretch}{1.2}
\begin{table}[t!]
\centering
\begin{tabular}{cccc}
\hline \hline
 \hspace{1 cm} Type \hspace{1 cm} & \hspace{0.2 cm} One-sided \hspace{0.2 cm} & \hspace{0.2 cm} Two-sided \hspace{0.2 cm} &  Section \\ \hline
Geometric & $Q_{A}^{G_{\delta}}$ & $Q_{AB}^{G_{\delta}}$ & \ref{Section:Measures-Types-Geometric}, \ref{Section:Measures-Geometric}  \\
Measurement induced geometric & $Q_{A}^{M_{\delta}}$ & $Q_{AB}^{M_{\delta}}$ & \ref{Section:Measures-Types-MIG}, \ref{Section:Measures-MIG}  \\
Measurement induced informational & $Q_{A}^{I_{\epsilon}}$ & $Q_{AB}^{I_{\epsilon}}$ & \ref{Section:Measures-Types-Informational}, \ref{Section:Measures-Informational}  \\
Entanglement activation & $Q_{A}^{E^{\zeta}}$ & $Q_{AB}^{E^{\zeta}}$ & \ref{Section:Measures-Types-Activation}, \ref{Section:Measures-Entanglement} \\
Unitary response & $Q_{A}^{U_{\delta}}$ & $-$ & \ref{Section:Measures-Types-Unitary}, \ref{Section:Measures-Response}  \\
Coherence based & $Q_{A}^{C^{\eta}}$ & $Q_{AB}^{C^{\eta}}$ & \ref{Section:Measures-Types-Coherence}, \ref{Section:Measures-Coherence}  \\
Recoverability & $Q_{A}^{R_{\delta}}$ & $Q_{AB}^{R_{\delta}}$ & \ref{Section:Measures-Types-Recoverability}, \ref{Section:Measures-Recoverability}  \\ \hline \hline
\end{tabular}\renewcommand{\baselinestretch}{1}
\caption{\label{Table:Types} The various types of measures of QCs, their notation and location within this review.}
\end{table}
\renewcommand{\baselinestretch}{1}

\begin{itemize}
\item
{\bf Geometric measures} (Section~{\ref{Section:Measures-Types-Geometric}}) quantify how distant $\rho_{AB}$ is from the set of classical states, cf.~Fig.~\ref{Fig:Sets}(a).

\item
{\bf Measurement induced geometric measures} (Section~\ref{Section:Measures-Types-MIG}) quantify how distinguishable $\rho_{AB}$ is from the output of a least perturbing local measurement, as in~Fig.~\ref{Fig:Defs}(a).

\item
{\bf Measurement induced informational measures} (Section~\ref{Section:Measures-Types-Informational}) quantify how much information about $\rho_{AB}$ is lost due to a least perturbing local measurement, another take on~Fig.~\ref{Fig:Defs}(a).

\item
{\bf Entanglement activation measures} (Section~\ref{Section:Measures-Types-Activation}) quantify the minimum entanglement created with an apparatus in a local pre-measurement, as in~Fig.~\ref{Fig:Defs}(b).

\item
    {\bf Unitary response measures} (Section~\ref{Section:Measures-Types-Unitary}) quantify how distinguishable $\rho_{AB}$ is from its image after a least perturbing local non-degenerate unitary, as in~Fig.~\ref{Fig:Defs}(c).

    \item
{\bf Coherence based measures} (Section~\ref{Section:Measures-Types-Coherence}) quantify the minimum quantum coherence of $\rho_{AB}$ in any local basis, as in~Fig.~\ref{Fig:Defs}(d).

\item
{\bf Recoverability measures} (Section~\ref{Section:Measures-Types-Recoverability}) quantify how distant $\rho_{AB}$ is from being a fixed point of an entanglement-breaking channel, as in~Fig.~\ref{Fig:Defs}(e).
\end{itemize}

In the following subsections, we first discuss general properties of each type of QCs measures, such as their compliance with the aforementioned Requirements, and later specialise to particular instances of measures within each category and how they interlink across the categories. For each reviewed measure, we will discuss some of the history, experimental investigations, and progress on their analytical or numerical evaluation, where relevant. We will also highlight any operational interpretations linking in with Section~\ref{Section:Applications}.
Finally, a schematic summary of most of the reviewed measures will be presented in Table~\ref{Table:SummaCumLaude}.

Note that we will focus mainly on finite dimensional bipartite systems in this review. However, most of the presented measures have also been extended to continuous variable systems of two or more quantum harmonic oscillators (or modes of the radiation field), and studied in particular in the case of two-mode Gaussian states. Deferring the reader to \cite{adesso2014open} (and references therein) for a recent review on continuous variable Gaussian states including some investigations of their QCs, we will briefly mention a few relevant results in the following, where deemed necessary.

\subsubsection{Geometric measures}\label{Section:Measures-Types-Geometric}

Definition~\ref{Definition 1:} gives a negative characterisation of quantumly correlated states, i.e.~quantumly correlated states are all the states that are not classical. The most natural quantitative extension of this is to consider the distance to the set of classical states as a measure of QCs. Given a distance $D_{\delta}$, we can then define {\em geometric} measures of one-sided and two-sided QCs of a bipartite state $\rho_{AB}$ as follows,
\begin{equation}\label{Equation:GeometricMeasures}
\begin{aligned}
Q_{A}^{G_{\delta}}(\rho_{AB}) &:= \inf_{\chi_{AB} \in \mathscr{C}_{A}} D_{\delta}(\rho_{AB}, \chi_{AB}), \\
Q^{G_{\delta}}_{AB}(\rho_{AB}) &:=\inf_{\chi_{AB} \in \mathscr{C}_{AB}} D_{\delta}(\rho_{AB}, \chi_{AB}),
\end{aligned}\end{equation}
where $\delta$ labels the particular distance functional $D_{\delta}$ on the set of states $\mathscr{D}_{AB}$, and the superscript $G_{\delta}$ indicates that the measures are of the geometric type with respect to this chosen distance.
%There are many possible distances $D_{\delta}$ to choose from, as summarised in Table~\ref{Table:Distances}.

By construction, the geometric measures are zero only for classical states, hence Requirement~(\ref{Requirements:Faithfulness}) is satisfied. Requirements~(\ref{Requirements:UnitaryInvariance}), (\ref{Requirements:UnmeasuredParty}), and (\ref{Requirements:LCPO}) are also satisfied for any choice of $D_{\delta}$ that is {\it contractive} under the action of CPTP operations, i.e.~such that
\begin{equation}\label{Equation:Contractivity}
D_{\delta}\big(\Lambda[\rho],\Lambda[\sigma]\big)\leq D_{\delta}(\rho,\sigma),
\end{equation}
for any two states $\rho$ and $\sigma$ and any CPTP operation $\Lambda$. Although contractivity may not be necessary to satisfy these properties, there is good reason to restrict to contractive distances. Distances are used to quantify the distinguishability between quantum states, and non-invertible CPTP operations are the representation of noise~\cite{bengtsson2007geometry}. Intuitively, one expects noise to never increase the distinguishability of quantum states, hence the distances that contract through any CPTP operation are the mathematical realisation of this. We also note that, in a resource theoretic context, any contractive distance will induce a geometric measure of QCs that is monotonically non-increasing under the action of the free operations, regardless of their exact specification, purely by virtue of the fact that the free operations cannot create QCs.
Table~\ref{Table:Distances} summarises a selection of widely adopted contractive distances in quantum information theory, whose corresponding geometric QCs measures will be discussed in Section~\ref{Section:Measures-Geometric}.

\renewcommand{\baselinestretch}{1.1}
\begin{table}[t!]
\centering
\begin{tabular}{cccc}
\hline \hline
 \hspace{1 cm} Distance \hspace{1 cm} &  $D_{\delta}$  & \hspace{1 cm} Formula \hspace{1 cm} & Contractive  \\ \hline
Relative entropy & $D_{\text{RE}}$ & ${\cal S}(\rho\|\sigma)=\mbox{Tr}\left(\rho \log \rho - \rho \log \sigma\right)$ & Yes \\
($p^{\rm th}$ power) Schatten $p$-norm & $D_{p}$ & $\|\rho-\sigma\|_p^p=\mbox{Tr}\left(|\rho - \sigma|^{p}\right)$ %\left| \left| \rho - \sigma \right| \right|^{p}_{p}$
& Only for $p=1$ \\
Infidelity & $D_{{F}}$ & $1-\left[\mbox{Tr}\left(\sqrt{\sqrt{\sigma} \rho \sqrt{\sigma}}\right)\right]^2$ & Yes \\
Squared Bures & $D_{\text{Bu}}$ & $2\left[1-\mbox{Tr}\left(\sqrt{\sqrt{\sigma} \rho \sqrt{\sigma}}\right)\right]$ & Yes \\
Squared Hellinger & $D_{\text{He}}$ & $2\left[1-\mbox{Tr}\left(\sqrt{\rho} \sqrt{\sigma}\right)\right]$ & Yes \\ \hline \hline
\end{tabular}\renewcommand{\baselinestretch}{1}
\caption{\label{Table:Distances} Some distances $D_{\delta}$ between two quantum states $\rho$ and $\sigma$. Details on the definitions and discussions of contractivity can be found in~\cite{bengtsson2007geometry,paula2013geometric,luo2004informational}.}
\end{table}
\renewcommand{\baselinestretch}{1}

Finally, there is a subtlety regarding Requirement~(\ref{Requirements:EntanglementPure}) for the geometric measures of QCs. To each geometric measure of QCs one can associate a corresponding measure of entanglement, defined in terms of the distance to the separable states. Specifically, for a bipartite state $\rho_{AB}$, the geometric measure of entanglement given a particular choice of distance $D_{\delta}$ is \cite{plenio2007an}
\begin{equation}\label{Equation:EntanglementGeometricMeasures}
E^{G_{\delta}}(\rho_{AB}) := \inf_{\sigma_{AB} \in \mathscr{S}_{AB}} D_{\delta}(\rho_{AB}, \sigma_{AB}),
\end{equation}
where $\mathscr{S}_{AB}$ is the set of separable states defined in Eq.~(\ref{Equation:SSet}). Because of the hierarchy of Eq.~(\ref{Equation:Subsets}), as depicted in Fig.~\ref{Fig:Sets}(a), it generally holds that
\begin{equation}
Q_{AB}^{G_{\delta}}(\rho_{AB}) \geq Q_{A}^{G_{\delta}}(\rho_{AB}) \geq E^{G_{\delta}}(\rho_{AB}).
\end{equation}
For pure states ${\ket{\psi}_{AB}}$, one would naturally expect that
\begin{equation}\label{Equation:GeometricPureEquality}
Q_{AB}^{G_{\delta}}({\ket{\psi}_{AB}}) = Q_{A}^{G_{\delta}}({\ket{\psi}_{AB}}) = E^{G_{\delta}}({\ket{\psi}_{AB}}),
\end{equation}
because the set of pure classical and pure separable states all reduce to the set of pure product states. However, since the infima in Eqs.~(\ref{Equation:GeometricMeasures}) and (\ref{Equation:EntanglementGeometricMeasures}) are, respectively, over all (not necessarily pure) classical and separable states, Eq.~(\ref{Equation:GeometricPureEquality}) does not hold trivially. Instead, one must show Requirement~(\ref{Requirements:EntanglementPure}) on a case-by-case basis for each choice of distance $D_{\delta}$.
%Furthermore, it is not necessary for $Q_{A}^{G_{\delta}}({\ket{\psi}_{AB}})$ or $Q_{AB}^{G_{\delta}}({\ket{\psi}_{AB}})$ to correspond with a \emph{geometric} measure of entanglement, or for them to equate, although we will highlight the cases in which they are known to do so in the following.

\subsubsection{Measurement induced geometric measures}\label{Section:Measures-Types-MIG}

Another intuitive aspect of QCs, outlined by Definition~\ref{Definition 2:}, is that a quantumly correlated state is perturbed by any possible complete rank one LPM. If we consider a complete rank one LPM on subsystem $A$, %i.e.~consisting of projection operators $\{(\Pi_{A})_{i}\}$ summing to the identity $\sum_{i}(\Pi_{A})_{i} = \mathbb{I}$ and transforming $\rho_{AB}$ as
%\begin{equation}
%\Pi_{A} (\rho_{AB}) = \sum_{i}(\Pi_{A})_{i} \otimes \mathbb{I}_{B}\rho_{AB} (\Pi_{A})_{i} \otimes \mathbb{I}_{B},
%\end{equation}
we know that $\pi_{A} [\rho_{AB}]$ as defined in Eq.~(\ref{Equation:PostLPMA}) is necessarily different to $\rho_{AB}$ if QCs are present. It is then meaningful to think that, provided one identifies the least perturbing LPM $\pi_A$, the more distinguished $\pi_{A} [\rho_{AB}]$ is from $\rho_{AB}$, the more QCs were present originally in $\rho_{AB}$. By quantifying this distinguishability as the distance between $\rho_{AB}$ and the closest (i.e., least perturbed) output state $\pi_{A} [\rho_{AB}]$, we can define the {\it measurement induced geometric} measures of QCs of a state $\rho_{AB}$. Given a distance $D_{\delta}$, these are defined as
\begin{equation}\label{Equation:MeasurementInducedDisturbances}
\begin{aligned}
Q_{A}^{M_{\delta}}(\rho_{AB}) &:= \inf_{\pi_{A}} D_{\delta}\big(\rho_{AB}, \pi_{A} [\rho_{AB}]\big),  \\
Q^{M_{\delta}}_{AB}(\rho_{AB}) &:=\inf_{\pi_{AB}} D_{\delta}\big(\rho_{AB}, \pi_{AB} [\rho_{AB}]\big),
\end{aligned}\end{equation}
where $M_{\delta}$ denotes the measurement induced geometric type for a corresponding distance $D_{\delta}$, and we have considered LPMs on both subsystems $A$ and $B$ as defined in Eq.~(\ref{Equation:PostLPMAB}) in the case of two-sided measures.

Once more, we can choose any distance on the set of quantum states to define a measure within this family, with some common examples listed in Table~\ref{Table:Distances}. The minimisation must be performed only over complete rank one LPMs, which represent the most fine grained projective measurements since each projector corresponds to a particular outcome and no outcomes are merged into a higher rank projector (more trivially, without restricting to rank one projectors one would always satisfy the infima in Eq.~(\ref{Equation:MeasurementInducedDisturbances}) by choosing as projectors, respectively, $\mathbb{I}_{A}$ and $\mathbb{I}_{A}\otimes \mathbb{I}_{B}$).
%It is also possible to consider generalised local measurements in the form of positive operator valued measures (POVMs), which consist of positive operators $\{M_{i}\}$ satisfying $\sum_{i} M_{i} = \mathbb{I}$~\cite{nielsen2010quantum}. Each POVM element $M_{i}$ can be decomposed as $M_{i} = m_{i}^{\dagger}m_{i}$ into measurement operators $m_{i}$, and hence a local general measurement $M_{A}$ on subsystem $A$ consisting of measurement operators $\{(m_{A})_{i}\}$ transforms $\rho_{AB}$ as
%\begin{equation}
%M_{A}(\rho_{AB}) = \sum_{i}(m_{A})_{i} \otimes \mathbb{I}_{B}\rho_{AB} (m_{A})_{i}^{\dagger} \otimes \mathbb{I}_{B}.
%\end{equation}
%By optimising instead over local POVMs in Eq.~(\ref{Equation:MeasurementInducedDisturbances}) we arrive at the similar measures
We can use the fact that $\pi_{A} [\rho_{AB}] \in \mathscr{C}_{A}$ and $\pi_{AB} [\rho_{AB}] \in \mathscr{C}_{AB}$, as we have already discussed in Eqs.~(\ref{Equation:ClassicalStateProjectiveDefinitionA}) and (\ref{Equation:ClassicalStateProjectiveDefinitionAB}), to see that
\begin{equation}\label{Equation:MIPGeometricInequality}
Q_{A}^{G_{\delta}}(\rho_{AB}) \leq Q_{A}^{M_{\delta}}(\rho_{AB}), \quad \mbox{and}\quad
Q_{AB}^{G_{\delta}}(\rho_{AB}) \leq Q_{AB}^{M_{\delta}}(\rho_{AB}).
\end{equation}
For certain choices of $D_{\delta}$, these inequalities are in fact saturated, and we will highlight when this occurs in the following (Sections~\ref{Section:Measures-Geometric} and \ref{Section:Measures-MIG}). It is immediate to see that Requirement~(\ref{Requirements:Faithfulness}) for measurement induced geometric measures is satisfied because a state can be left invariant under a complete rank one LPM if, and only if, it is classical. Requirements~(\ref{Requirements:UnitaryInvariance}) and (\ref{Requirements:UnmeasuredParty}) also hold for any contractive distance, however it is unknown whether this is true for Requirement~(\ref{Requirements:LCPO}) as well. Again, Requirement~(\ref{Requirements:EntanglementPure}) must be checked for each particular distance $D_{\delta}$.

\subsubsection{Measurement induced informational measures}\label{Section:Measures-Types-Informational}

An alternative viewpoint from which we can look at QCs is based on how much information about the state of a composite system can be extracted by accessing it locally. When only classical correlations are present, we can in principle extract all such information after a local measurement, while this is no longer the case when also QCs come into play.
%An alternative way of capturing quantitatively the unavoidable disturbance due to local measurements on quantum states, still inspired by Definition~\ref{Definition 2:}, is by analysing the change in some appropriate informational quantifier following the action of a minimally perturbing local measurement.
%One can adopt an information theoretic perspective and compare the information present before and after an LGM.
As discussed in Section~\ref{Section:Facts-Measuring}, LGMs $\tilde{\Pi}_{A}$ and $\tilde{\Pi}_{AB}$ represent recording the measurement information about $\rho_{AB}$ onto classical registers. It is intuitive, therefore, to adopt an information theoretic perspective and regard the minimal difference between the original information in the state $\rho_{AB}$ and the post-measurement information stored in the classical register as a measure of quantumness. Given an informational quantifier $I_{\epsilon}$, we can then define {\it measurement induced informational} (or simply {\em informational}) measures of QCs of a state $\rho_{AB}$ as
\begin{equation}\label{Equation:InformationalMeasuresPOVM}
\begin{aligned}
Q_{A}^{\tilde{I}_{\epsilon}}(\rho_{AB}) &:= \inf_{\tilde{\Pi}_{A}} \left[I_{\epsilon}\left(\rho_{AB}\right) - I_{\epsilon}\left(\tilde{\Pi}_{A} [\rho_{AB}]\right) \right], \\
Q_{AB}^{\tilde{I}_{\epsilon}}(\rho_{AB}) &:= \inf_{\tilde{\Pi}_{AB}} \left[ I_{\epsilon}\left(\rho_{AB}\right) -I_{\epsilon}\left(\tilde{\Pi}_{AB} [\rho_{AB}]\right) \right].\end{aligned}
\end{equation}
A list of the most prominent choices of $I_{\epsilon}$ is presented in Table~\ref{Table:InformationalQuantifiers}. We elaborate more on these choices in Section~\ref{Section:Measures-Informational}, also highlighting the relationships between informational and other types of measures of QCs.

%Rather than quantifying the difference between $\rho_{AB}$ and $\Pi_{A} (\rho_{AB})$ with a distance, we can instead take an information theoretic perspective and think about how the information changes due to the LPM, which was indeed the original inspiration for the QCs. Consider some informational quantifier $I_{\epsilon}(\rho_{AB})$ of a bipartite quantum state $\rho_{AB}$, we can consider the minimum amount of change in this information before a projective measurement and after as a measure of QCs, i.e.~%\begin{eqnarray}\label{Equation:InformationalMeasures}
%Q_{A}^{I_{\epsilon}}(\rho_{AB}) \equiv \inf_{\Pi_{A}} \left[I_{\epsilon}\left(\rho_{AB}\right) - I_{\epsilon}\left(\Pi_{A} (\rho_{AB})\right) \right] \nonumber \\
%Q_{AB}^{I_{\epsilon}}(\rho_{AB}) \equiv \inf_{\Pi_{AB}} \left[ I_{\epsilon}\left(\rho_{AB}\right) -I_{\epsilon}\left(\Pi_{AB} (\rho_{AB})\right) \right].
%\end{eqnarray}
%A list of possible choices of $I_{\epsilon}$ is presented in Table~\ref{Table:InformationalQuantifiers}. We elaborate more on these choices in the following, also discussing the links between $Q_{A}^{I_{\mathcal{I}}}(\rho_{AB})$ and the other types of measures of QCs.

\renewcommand{\baselinestretch}{1.1}
\begin{table}[t!]
\centering
\begin{tabular}{ccc}
\hline \hline
 \hspace{1 cm} Informational quantity \hspace{1 cm} & \hspace{0.5 cm} $I_{\epsilon}$ \hspace{0.5 cm} & \hspace{0.5 cm} Formula \hspace{0.7 cm} \\ \hline
(negative) von Neumann entropy & $I_{\mathcal{S}}$ & $-\mathcal{S}(\rho_{AB})=\mbox{Tr}\left(\rho_{AB} \log \rho_{AB}\right)$ \\ Mutual information & $I_{\mathcal{I}}$ & $\mathcal{I}(\rho_{AB})=\mathcal{S}(\rho_{A}) + \mathcal{S}(\rho_{B}) - \mathcal{S}(\rho_{AB})$
\\ \hline \hline
\end{tabular}\renewcommand{\baselinestretch}{1}
\caption{\label{Table:InformationalQuantifiers}Some informational quantifiers $I_{\epsilon}$ of a bipartite state $\rho_{AB}$ with reduced states $\rho_{A} = \mbox{Tr}_{B}(\rho_{AB})$ and $\rho_{B} = \mbox{Tr}_{A}(\rho_{AB})$.}
\end{table}
\renewcommand{\baselinestretch}{1}

The measures in Eq.~(\ref{Equation:InformationalMeasuresPOVM}) obey Requirement~(\ref{Requirements:Faithfulness}) since for all and only the classical states there is always an LPM (which is within the class of all LGMs) that leaves them unchanged. However, one should be careful when choosing $I_{\epsilon}$ to ensure that $Q_{A}^{\tilde{I}_{\epsilon}}(\rho_{AB})$ and $Q_{AB}^{\tilde{I}_{\epsilon}}(\rho_{AB})$ are non-negative (hence the choice of the negative von Neumann entropy in Table~\ref{Table:InformationalQuantifiers}). Requirement~(\ref{Requirements:UnitaryInvariance}) holds for any $I_{\epsilon}$ invariant under local unitaries, which in turn should always hold as a measure of information or correlations is not expected to be dependent upon the local bases chosen. However, Requirements~(\ref{Requirements:EntanglementPure}), (\ref{Requirements:UnmeasuredParty}) and (\ref{Requirements:LCPO}) must be checked on a case-by-case basis for each selection of $I_{\epsilon}$. Notice also that for some choices of $I_\epsilon$ one may need to impose further restrictions to the allowed LGMs in the optimisations appearing in Eqs.~(\ref{Equation:InformationalMeasuresPOVM}) in order to avoid trivial results, as we discuss in Section~\ref{Section:Measures-Informational-Entropy}.

%Similarly to the measurement induced geometric measures of QCs,
We can explicitly consider the special case of informational measures with optimisation restricted over complete rank one LPMs, hence defining
\begin{equation}\label{Equation:InformationalMeasures}\begin{aligned}
Q_{A}^{I_{\epsilon}}(\rho_{AB}) &:= \inf_{\Pi_{A}} \left[I_{\epsilon}\left(\rho_{AB}\right) - I_{\epsilon}\left(\Pi_{A} [\rho_{AB}]\right) \right],\\
Q_{AB}^{I_{\epsilon}}(\rho_{AB}) &:=\inf_{\Pi_{AB}} \left[ I_{\epsilon}\left(\rho_{AB}\right) -I_{\epsilon}\left(\Pi_{AB} [\rho_{AB}]\right) \right].
\end{aligned}\end{equation}
It then holds by construction that
\begin{equation}
Q_{A}^{\tilde{I}_{\epsilon}}(\rho_{AB}) \leq Q_{A}^{I_{\epsilon}}(\rho_{AB}), \quad \mbox{and}\quad
Q_{AB}^{\tilde{I}_{\epsilon}}(\rho_{AB}) \leq Q_{AB}^{I_{\epsilon}}(\rho_{AB}).
\end{equation}
To be more precise, invoking Naimark's theorem as in Section~\ref{Section:Facts-Measuring}, we recall that an LGM on $\rho_{AB}$ can equivalently be performed by embedding $\rho_{AB}$ into a larger space and carrying out an LPM.  This means that
\begin{equation}
Q_{A}^{\tilde{I}_{\epsilon}}(\rho_{AB}) = \inf_{A'} Q_{A'}^{I_{\epsilon}}(\rho_{A'B}), \quad \mbox{and} \quad
Q_{AB}^{\tilde{I}_{\epsilon}}(\rho_{AB}) = \inf_{A',B'} Q_{A'B'}^{I_{\epsilon}}(\rho_{A'B'}),
\end{equation}
or in words that the one-sided LGM based informational measures are obtained as a minimisation of the corresponding one-sided LPM based informational measures over all embeddings of $A$ into a larger system $A'$, and similarly for the two-sided measures considering embeddings of both $A$ and $B$ in larger systems $A'$ and $B'$, respectively.

Taking care again to ensure non-negativity, the LPM subclass of informational measures obey Requirement~(\ref{Requirements:Faithfulness}) and Requirement~(\ref{Requirements:UnitaryInvariance}) if $I_{\epsilon}$ is invariant under local unitaries, but all the other requirements must be checked for each $I_{\epsilon}$ on a case-by-case basis. The meaning of the complete rank one LPM based informational measures in Eqs.~(\ref{Equation:InformationalMeasures}) is even more intuitive. Indeed, by recalling that the output of the quantum-to-classical maps $\Pi_A$ (resp., $\Pi_{AB}$) can be considered equivalent to the post-measurement states $\pi_A$ (resp., $\pi_{AB}$), we have that such informational measures capture directly the minimum amount of change in the information content of the state $\rho_{AB}$ itself  (without the need for invoking a classical register) before and after a local von Neumann measurement. This reproduces the original spirit of the quantum discord \cite{ollivier2001quantum}, which can be classified indeed as an informational measure, as we show in Section~\ref{Section:Measures-Informational}.

\subsubsection{Entanglement activation measures}\label{Section:Measures-Types-Activation}
Definition~\ref{Definition 3:} suggests that measures of QCs can be readily defined from measures of entanglement.  In particular, the QCs content of a state $\rho_{AB}$ can be  quantified in terms of the minimum entanglement created between the system $AB$ and an ancillary system, given by $A'$ for the one-sided case, or  by the composite $A'B'$ for the two-sided case, in the pre-measurement state corresponding to a least perturbing complete rank one LPM on $A$ or on both $A$ and $B$, respectively. Given a specific bipartite entanglement monotone $E^\zeta$, this conceptual framework initially put forward in \cite{piani2011all,streltsov2011linking} defines the {\it entanglement activation} measures of QCs of a state $\rho_{AB}$,
\begin{equation}\label{Equation:EntanglementActivationMeasures}\begin{aligned}
Q_{A}^{E^{\zeta}}(\rho_{AB}) &:= \inf_{\{\ket{a}_A\}} E^{\zeta}_{AB:A'}\left(\rho'^{\{\ket{a}_A\}}_{ABA'}\right),\\
Q_{AB}^{E^{\zeta}}(\rho_{AB}) &:= \inf_{\{\ket{a}_A, \ket{b}_B\}} E^{\zeta}_{AB:A'B'}\left(\rho'^{\{\ket{a}_A, \ket{b}_B\}}_{ABA'B'}\right),
\end{aligned}\end{equation}
where the optimisation is over the local bases of the pre-measurement interaction as detailed in Section~\ref{Section:Facts-Entanglement}, and the pre-measurement states $\rho'^{\{\ket{a}_A\}}_{ABA'}$ and $\rho'^{\{\ket{a}_A, \ket{b}_B\}}_{ABA'B'}$ are specifically defined in Eqs.~(\ref{Equation:Premeasurement}) and (\ref{Equation:PremeasurementAB}); notice that we indicate the explicit bipartition $X:Y$ with respect to which entanglement is being calculated with the notation $E^\zeta_{X:Y}(\rho_{XY})$.

\renewcommand{\baselinestretch}{1.1}
\begin{table}[t!]
\centering
\begin{tabular}{ccc}
\hline \hline
 \hspace{0.4cm} Entanglement measure \hspace{0.4cm} & \hspace{0.2cm} $E^\zeta$ \hspace{0.2cm} &  Formula  \\ \hline
 Geometric measure(s)   &  $E^{G_{\delta}}$ & $E^{G_{\delta}}(\rho_{AB}) = \underset{{\sigma_{AB} \in \mathscr{S}_{AB}}}{\inf} D_{\delta}(\rho_{AB}, \sigma_{AB})$ \\
 Distillable entanglement & $E^d$ & See Ref.~\cite{plenio2007an,horodecki2009quantum} \\
 Entanglement negativity & $E^N$ & $E^N(\rho_{AB}) = \|\rho_{AB}^{T_A}\|_1 - 1$ \\
Entanglement of formation & $E^f$ & $E^f(\rho_{AB})=   \underset{\mbox{${}^{\!\!\!\!{{\{p_i,\ \ket{\psi_i}_{AB}\}\,,}}\!\!\!\! }$}\atop  \mbox{${}^{ \!\!\!\!\!\!\!\!\!\!\!\!{{\rho_{AB}=\sum_i p_i \ket{\psi_i}\bra{\psi_i}_{AB}}}\!\!\!\!\!\!\!\!\!\!\!\!}$}}{{}_{\quad}\inf_{\quad}} \sum_i p_i \ {\cal S}\big(\mbox{Tr}_B(\ket{\psi_i}\!\bra{\psi_i}_{AB})\big)$
\\ \hline \hline
\end{tabular}\renewcommand{\baselinestretch}{1}
\caption{\label{Table:Entanglement}Some entanglement monotones $E^{\zeta}$ for a bipartite state $\rho_{AB}$.}
\end{table}
\renewcommand{\baselinestretch}{1}

One of the key strengths of this approach is that there is a huge variety of entanglement measures $E^\zeta$ available in the quantum information literature \cite{plenio2007an} (we list a small selection of choices in Table~\ref{Table:Entanglement}): for each of those, Eqs.~(\ref{Equation:EntanglementActivationMeasures}) define corresponding measures of more general one-sided and two-sided QCs. Crucially, it turns out that this correspondence respects the hierarchy of non-classical correlations illustrated in Figs.~\ref{Fig:CorrHier} and \ref{Fig:Sets}(a). Indeed, it was proven in \cite{piani2012quantumness} that
\begin{equation}\label{Equation:HierardyMustBeRespected}
Q_{AB}^{E^\zeta}(\rho_{AB}) \geq Q_{A}^{E^\zeta}(\rho_{AB}) \geq E^\zeta_{A:B}(\rho_{AB})\,,
\end{equation}
for all bipartite states $\rho_{AB}$ and any $E^\zeta$ monotonically non-increasing under LOCC. This follows because the original state $\rho_{AB}$ can be obtained from the pre-measurement states in Eqs.~(\ref{Equation:EntanglementActivationMeasures}) by means of an LOCC with respect to the system versus ancilla(e) split \cite{piani2012quantumness}. The inequalities in Eq.~(\ref{Equation:HierardyMustBeRespected}) all turn into equalities for pure states $\ket{\psi}_{AB}$, which means that Requirement~(\ref{Requirements:EntanglementPure}) is satisfied for all entanglement activation measures of QCs \cite{piani2012quantumness}. It is rightful to interpret Eq.~(\ref{Equation:HierardyMustBeRespected}) as a quantitative indication that QCs truly go {\it beyond } entanglement, in a very precise sense, and reduce to it in the case of pure states, as it should be.  Requirement~(\ref{Requirements:Faithfulness}) holds by construction due to the characterisation of classical states according to Property~\ref{Property 3:}, and Requirements~(\ref{Requirements:UnitaryInvariance}) and (\ref{Requirements:UnmeasuredParty}) are satisfied as well for any entanglement activation measure of QCs defined from a valid entanglement monotone $E^\zeta$. Requirement~(\ref{Requirements:LCPO}) has not been investigated in general and remains to be checked on a case-by-case basis.

We discuss some specific examples of entanglement activation type measures of QCs in Section~\ref{Section:Measures-Entanglement}.

\subsubsection{Unitary response measures}\label{Section:Measures-Types-Unitary}

As outlined by Definition~\ref{Definition 4:}, QCs give also rise to the unavoidable disturbance of a bipartite quantum state in response to the action of non-degenerate unitary operations on one of the subsystems. This qualitative aspect of QCs can be turned into a QCs quantifier by considering how far a given bipartite state $\rho_{AB}$ is from being invariant under a non-degenerate local unitary operation, i.e., by considering the distinguishability between $\rho_{AB}$ and its image after a least disturbing non-degenerate local unitary. For a given distance $D_\delta$, the (one-sided) {\em unitary response} measure of QCs of a state $\rho_{AB}$ is then defined as \cite{gharibian2012quantifying,giampaolo2013quantifying,nakano2013negativity,roga2014discord}
\begin{equation}\label{Equation:UnitaryResponseQuantifiersGamma}
Q_A^{{U}^\Gamma_\delta} (\rho_{AB}):= \inf_{U^\Gamma_A} D_\delta\big(\rho_{AB},{{U}^\Gamma_A}[\rho_{AB}]\big),
\end{equation}
where the optimisation is over all local unitaries ${{U}^\Gamma_A}$ with non-degenerate spectrum $\Gamma$, as defined in Eq.~(\ref{Equation:UA}).

Among all possible choices of the spectrum $\Gamma$, a special role is played by the harmonic spectrum $\Gamma^{\star}$ given by the $d_A$-th complex roots of unity, with $d_A$ being the dimension of subsystem $A$. The corresponding local unitary operations, also known as root-of-unity operations, will be denoted simply by  $U_A \equiv {U}^{{\Gamma}^{\star}}_A$ (dropping the spectrum superscript). They may be expected to be the most informative ones that we can restrict to in order to quantify QCs, in the sense that the more the unitaries in the aforementioned optimisation are fully non-degenerate (with no spacing between any two eigenvalues becoming arbitrarily small) the better the corresponding response measures should be able to discern between different states~\cite{monras2011entanglement,gharibian2012quantifying,giampaolo2013quantifying,roga2014discord}. For this reason, we will primarily focus on the special case of unitary response measures with optimisation restricted to harmonic spectra,
\begin{equation}\label{Equation:UnitaryResponseQuantifiers}
Q_A^{U_\delta} (\rho_{AB}) := \inf_{{U}_A} D_\delta\big(\rho_{AB},{U}_A[\rho_{AB}]\big).
\end{equation}

The unitary response measures of QCs clearly satisfy Requirement~(\ref{Requirements:Faithfulness}) since any classical-quantum state is invariant under a non-degenerate local unitary~\cite{gharibian2012quantifying,giampaolo2013quantifying,roga2014discord}. Moreover, they satisfy Requirement~(\ref{Requirements:UnitaryInvariance}) and, provided $D_\delta$ is contractive, Requirement~(\ref{Requirements:UnmeasuredParty})~\cite{roga2014discord}, whereas it is still an open question if they comply with Requirement~(\ref{Requirements:LCPO}). For pure states $\ket{\psi}_{AB}$, any measure $Q_A^{U_\delta}$ reduces to the corresponding ``entanglement of response'' \cite{monras2011entanglement},
\begin{equation}\label{Equation:EntanglementofResponse}
E^{U_\delta} (\ket{\psi}_{AB}) := \inf_{{U}_A} D_\delta\big(\ket{\psi}\bra{\psi}_{AB},{U}_A[\ket{\psi}\bra{\psi}_{AB}]\big),
\end{equation}
which entails that Requirement~(\ref{Requirements:EntanglementPure}) is satisfied as well for any unitary response measure of QCs.

Examples of measures $Q_A^{U_\delta}$ for typical distances $D_\delta$ as given in Table~\ref{Table:Distances}, and their connections to geometric and measurement induced geometric measures will be explored in Section~\ref{Section:Measures-Response}. We finally remark that, by virtue of the discussion in Section~\ref{Section:Facts-Unitary}, the unitary response approach does not appear to be straightforwardly extendible to define two-sided measures of QCs, as the latter would fail Requirement~(\ref{Requirements:Faithfulness}), unlike the case of all the other types of measures covered in this review.

\subsubsection{Coherence based measures}\label{Section:Measures-Types-Coherence}

Following Definition~\ref{Definition 5:}, QCs can be quantified in terms of measures of quantum coherence, minimised over all local bases.
Measures of quantum coherence have been studied quite extensively in the last few years, see e.g.~\cite{aaberg2006quantifying,marvian2013the,baumgratz2014quantifying,bromley2015frozen,streltsov2015measuring,yadin2015quantum,winter2016operational,napoli2016robustness,chitambar2016assisted,girolami2014observable,marvian2014extending,marvian2015quantum,ma2016converting} and the very recent review~\cite{streltsov2016review}.
While several connections between coherence, entanglement, and QCs have been pointed out quite recently, including quantitative interconversion schemes \cite{xi2015quantum,streltsov2015measuring,killoran2016converting,chitambar2016assisted,ma2016converting} somehow analogous to the entanglement activation framework \cite{piani2011all,streltsov2011linking}, a consistent definition of QCs quantifiers in terms of coherence has not been reported in full generality (to our knowledge), and will be introduced here.

For a single system, consider a measure of coherence $C^\eta$ with respect to a fixed reference basis $\{\ket{i}\}$, that is, a real, non-negative function $C^\eta(\rho)$ which vanishes on, and only on, the corresponding set $\mathscr{I}^{\{\ket{i}\}}$ of incoherent states  (see Section~\ref{Section:Facts-Coherence}), and is monotonically non-increasing under a suitable set of  incoherent operations. The incoherent operations may be, for instance, those defined in \cite{baumgratz2014quantifying}, which admit an operator sum representation where all Kraus operators map the set of incoherent states into itself, or a more restricted set such as the so-called covariant operations, which have been defined in the context of the resource theory of asymmetry \cite{marvian2013the}; we defer the reader to recent studies where merits and setbacks of various proposals for incoherent operations are discussed \cite{marvian2015quantum,winter2016operational,yadin2015quantum,napoli2016robustness,chitambar2016are,marvian2016how}. We list a sample of coherence measures $C^\eta$ of relevance in various resource theoretic contexts in Table~\ref{Table:Coherence}.

\renewcommand{\baselinestretch}{1.2}
\begin{table}[t!]
\centering
\begin{tabular}{ccc}
\hline \hline
 Coherence measure  &  $C^\eta$   &  Formula  \\ \hline
 Relative entropy  & $C^{\text{RE}}$ & $C^{\text{RE}}(\rho) = \underset{\chi \in \mathscr{I}^{\{\ket{i}\}}}{\inf} D_{\text{RE}}(\rho, \chi) = {\cal S}\big(\pi[\rho]\big) - {\cal S}(\rho)$ \\
 Infidelity   & $C^F$ & $C^F(\rho) = 1-\underset{\chi \in \mathscr{I}^{\{\ket{i}\}}}{\sup} F(\rho, \chi)$ \\
$\ell_1$ norm & $C^{\ell_1}$ & $C^{\ell_1}(\rho) = \|\rho\|^{\{\ket{i}\}}_{\ell_1}-1 = \sum_{i\neq j} |\rho_{ij}|$ \\
Robustness & $C^{\text{Ro}}$ & $C^{\text{Ro}}(\rho) = \underset{\tau}{\min} \left\{s \geq 0\ \ \big| \ \ \displaystyle\frac{\rho + s \tau}{1+s} \in \mathscr{I}^{\{\ket{i}\}}\right\}$  \\
 Wigner-Yanase skew information & $C^{\text{WY}}$ & $C^{\text{WY}}(\rho, K) = -\frac12 \mbox{Tr}\left([\!\sqrt{\rho},K]^2\right)$ \\
 Quantum Fisher information & $C^{\text{QF}}$ & $C^{\text{QF}}(\rho, K) = 4 \sum_{m<n,\, q_m+q_n \neq 0} \frac{(q_m-q_n)^2}{q_m+q_n}\left|\bra{\psi_m}K\ket{\psi_n}\right|^2$\vspace*{.1cm}\\ \hline \hline
\end{tabular}\renewcommand{\baselinestretch}{1}
\caption{\label{Table:Coherence} Some coherence measures for a state $\rho$ with respect to a reference basis $\{\ket{i}\}$. The first four \cite{aaberg2006quantifying,baumgratz2014quantifying,streltsov2015measuring,napoli2016robustness} are monotones under the more general incoherent operations of \cite{baumgratz2014quantifying}, while the last two \cite{girolami2014observable,marvian2014extending,marvian2015quantum}, in which the reference basis is identified as the eigenbasis of the observable $K = \sum_i k_i \ket{i}\!\bra{i}$ with non-degenerate spectrum $\Gamma = \{k_i\}$,  are monotones under the more restricted set of phase covariant operations defined in asymmetry theory \cite{marvian2013the}. Furthermore, in the first row we denote by $\pi[\rho]$ the full dephasing in the reference basis, $\pi[\rho] = \sum_i \ket{i}\!\bra{i} \rho \ket{i}\!\bra{i}$, in the third row we evaluate the $\ell_1$ norm for $\rho$ written in the reference basis, in the fourth row we indicate by $\tau$ an arbitrary state, while in the final row we write $\rho$ in its spectral decomposition, $\rho= \sum_n q_n \ket{\psi_n}\!\bra{\psi_n}$.}
\end{table}
\renewcommand{\baselinestretch}{1}

Given now a bipartite system $AB$ in the state $\rho_{AB}$, and fixing local reference bases $\{\ket{a}_A\}$ for subsystem $A$ and $\{\ket{b}_B\}$ for subsystem $B$ as in Section~\ref{Section:Facts-Coherence}, assume that to a chosen coherence measure $C^\eta$ we can formally associate: a one-sided local coherence quantifier
${C^{\eta\;{\{\ket{a}_A\}}}_{A}}(\rho_{AB})$ with the requirements of vanishing on (and only on) the set $\mathscr{I}^{\{\ket{a}_A\}}_A$ of incoherent-quantum states defined in Eq.~(\ref{Equation:IASet}), and of being monotonic under (the chosen set of) local incoherent operations on $A$ and arbitrary local operations on $B$; and a two-sided local coherence quantifier ${C^{\eta\;{\{\ket{a}_A, \ket{b}_B\}}}_{AB}}(\rho_{AB})$ with the requirements of vanishing on (and only on) the set $\mathscr{I}^{\{\ket{a}_A,\ket{b}_B\}}_{AB}$ of incoherent-incoherent states defined in Eq.~(\ref{Equation:IABSet}), and of being monotonic under (the chosen set of) local incoherent operations on $A$ and $B$. In particular, the two-sided quantifier can be defined as the standard extension of $C^\eta$ to bipartite systems, choosing the product basis $\{\ket{a}_A \otimes \ket{b}_B\}$ as reference basis for the composite $AB$ \cite{bromley2015frozen,streltsov2015measuring,winter2016operational}.
 Notice however that, depending on the definition of $C^\eta$, it may not be straightforwardly possible to construct both one-sided and two-sided corresponding quantifiers of local coherence; we will discuss this in more detail in Section~\ref{Section:Measures-Coherence}.

For a given coherence quantifier $C^\eta$, we then define the {\it coherence based} measures of QCs of a state $\rho_{AB}$ as
\begin{equation}\label{Equation:CoherenceMeasures}\begin{aligned}
Q_{A}^{C^{\eta}}(\rho_{AB}) &:= \inf_{\{\ket{a}_A\}} {C^{\eta\;{\{\ket{a}_A\}}}_{A}}(\rho_{AB}),\\
Q_{AB}^{C^{\eta}}(\rho_{AB}) &:= \inf_{\{\ket{a}_A, \ket{b}_B\}} {C^{\eta\;{\{\ket{a}_A, \ket{b}_B\}}}_{AB}}(\rho_{AB}),
\end{aligned}\end{equation}
where the minimisation is over the local reference bases $\{\ket{a}_A\}$ of $A$ and $\{\ket{b}_B\}$ of $B$.

 By construction, due to Property~\ref{Property 5:} and the constraints imposed on the local coherence quantifiers above, coherence based measures of QCs satisfy Requirements~(\ref{Requirements:Faithfulness}) and (\ref{Requirements:UnmeasuredParty}), and also Requirement~(\ref{Requirements:UnitaryInvariance}), which holds manifestly in the two-sided case due to the involved minimisation over local bases on both parties and is implied by (\ref{Requirements:UnmeasuredParty}) in the one-sided case. However, there are no general results yet concerning Requirements~(\ref{Requirements:EntanglementPure}) and (\ref{Requirements:LCPO}), which have to be checked on a case-by-case basis.

In Section~\ref{Section:Measures-Coherence} we show how the coherence based approach encompasses some relevant quantifiers of QCs, and discuss their interplay with other types of measures.

\subsubsection{Recoverability measures}\label{Section:Measures-Types-Recoverability}
The last facet of QCs that we take into account is the following: when the QCs between two subsystems are small, almost all the information about the overall bipartite system is locally recoverable after performing a measurement on one of the subsystems. Since every entanglement-breaking channel can be written as a sequence of a measurement map followed by a preparation (recovery) map, as discussed in Section~\ref{Section:Facts-Breaking}, such a facet of QCs can be quantitatively captured by how far the state $\rho_{AB}$ of the overall bipartite system is from being a fixed point of an entanglement-breaking channel \cite{seshadreesan2015fidelity}, along the lines of Definition~\ref{Definition 6:}. Given a distance $D_\delta$, we can then define the {\em recoverability} type measures of QCs as
\begin{equation}\label{Equation:GeometricRecoverability}
\begin{aligned}
Q_{A}^{R_\delta}(\rho_{AB}) := \inf_{\Lambda^{\text{EB}}_A} D_{\delta}\big(\rho_{AB},\Lambda^{\text{EB}}_A[\rho_{AB}]\big), \\
Q_{AB}^{R_\delta}(\rho_{AB}) := \inf_{\Lambda^{\text{EB}}_{AB}} D_{\delta}\big(\rho_{AB},\Lambda^{\text{EB}}_{AB}[\rho_{AB}]\big),
\end{aligned}\end{equation}
%\begin{equation}\label{Equation:SurprisalMeasurementRecoverability}
%Q_{A}^{R_F} \equiv - \log \sup_{\Lambda^{\text{EB}}_A} F(\rho_{AB},\Lambda^{\text{EB}}_A(\rho_{AB})),
%\end{equation}
where the optimisation is over the convex set of entanglement-breaking channels $\Lambda^{\text{EB}}_A$ acting locally on $A$ and $\Lambda^{\text{EB}}_{AB}$ acting locally on $A$ and on $B$, respectively for one-sided and two-sided measures.

%More generally, one can consider the following geometric measures of recoverability~\cite{seshadreesan2015fidelity}:
%\begin{equation}\label{Equation:GeometricRecoverability}
%Q_{A}^{R_\delta} \equiv \inf_{\Lambda^{\text{EB}}_A} D_{\delta}(\rho_{AB},\Lambda^{\text{EB}}_A(\rho_{AB})),
%\end{equation}
%where $D_{\delta}$ is any contractive distance.

The recoverability based measures comply with Requirement~(\ref{Requirements:Faithfulness}) as all and only the classical state are fixed points of some entanglement-breaking channel. They further satisfy Requirements~(\ref{Requirements:UnitaryInvariance})  and (\ref{Requirements:UnmeasuredParty}) provided $D_\delta$ is a contractive distance, while Requirements~(\ref{Requirements:EntanglementPure}) and  Requirement~(\ref{Requirements:LCPO}) have to be tested for each $D_\delta$. Observe that, given an arbitrary state $\rho_{AB}$, the recoverability based measures of Eqs.~(\ref{Equation:GeometricRecoverability}) are in general smaller or equal than the corresponding measurement induced geometric measures of Eqs.~(\ref{Equation:MeasurementInducedDisturbances}) defined via the same $D_{\delta}$, since
every LPM is in particular a local entanglement-breaking channel.

Instances of recoverability measures of QCs as defined in recent literature \cite{seshadreesan2015fidelity} are presented in  Section~\ref{Section:Measures-Recoverability}.

\subsection{Geometric measures}\label{Section:Measures-Geometric}

Here we cover in detail geometric measures of QCs, as defined in Eqs.~(\ref{Equation:GeometricMeasures}), for some representative choices of the distance $D_\delta$, including those summarised in Table~\ref{Table:Distances}.

\subsubsection{Relative entropy}\label{Section:Measures-Geometric-RE}

The first definition of a geometric measure of QCs was given in~\cite{horodecki2005local,groisman2007quantumness}, where the relative entropy $D_{\text{RE}}$ (see Table~\ref{Table:Distances}) was considered. Although not strictly a distance~\cite{bengtsson2007geometry}, the relative entropy (or Kullback-Leibler divergence) is a widely used tool in classical information theory to measure the distinguishability between two probability distributions, and it is naturally extended to quantum states~\cite{vedral2002role}. The relative entropy is contractive \cite{lindblad1975completely,uhlmann1977relative}, therefore we need only to check Requirement~(\ref{Requirements:EntanglementPure}) for the corresponding QCs measures $Q_{A}^{G_{\text{RE}}}$ and $Q_{AB}^{G_{\text{RE}}}$. The latter is verified, as for pure states ${\ket{\psi}_{AB}}$ it holds
\begin{equation}
Q_{A}^{G_{\text{RE}}}({\ket{\psi}_{AB}}) = Q_{AB}^{G_{\text{RE}}}({\ket{\psi}_{AB}}) = E^{G_{\text{RE}}} ({\ket{\psi}_{AB}}) = \mathcal{S}(\rho_{A}) = \mathcal{S}(\rho_{B}),
\end{equation}
where $\mathcal{S}(\rho_{A}) = \mathcal{S}(\rho_{B})$ amounts to the canonical  measure of entanglement for the pure state $\ket{\psi}_{AB}$, known as the entanglement entropy~\cite{vedral1998entanglement,bengtsson2007geometry}, with the marginal state $\rho_{A}$ ($\rho_{B}$) obtained from $\ket{\psi}\bra{\psi}_{AB}$ by partial trace over $B$ ($A$).

In~\cite{modi2010unified} it was shown that the closest classical state $\chi_{AB}$ to any state $\rho_{AB}$, which solves the optimisation in Eqs.~(\ref{Equation:GeometricMeasures}) for the relative entropy distance, is always the one obtained as the output of least perturbing complete rank one LPMs $\pi_{A} [\rho_{AB}]$ and $\pi_{AB} [\rho_{AB}]$ for the one-sided and two-sided measures, respectively. This implies that the geometric and measurement induced geometric measures based upon the relative entropy distance coincide, i.e.~\begin{equation}
Q_{A}^{G_{\text{RE}}}(\rho_{AB}) = Q_{A}^{M_{\text{RE}}}(\rho_{AB}), \quad \mbox{and} \quad Q_{AB}^{G_{\text{RE}}}(\rho_{AB}) = Q_{AB}^{M_{\text{RE}}}(\rho_{AB}).
\end{equation}
Furthermore, there is an agreement with the LPM informational measures based on the negative von Neumann entropy~\cite{zurek2003quantum,horodecki2005local} (see Table~\ref{Table:InformationalQuantifiers})
\begin{eqnarray}\label{Equation:RelativeEntropyMeasuresGICoincidence}
Q_{A}^{G_{\text{RE}}}(\rho_{AB}) = \inf_{\Pi_{A}} \left[ \mathcal{S}(\Pi_{A} [\rho_{AB}]) - \mathcal{S}(\rho_{AB})\right] = Q_{A}^{I_{\mathcal{S}}}(\rho_{AB}), \nonumber \\
Q_{AB}^{G_{\text{RE}}}(\rho_{AB}) = \inf_{\Pi_{AB}} \left[ \mathcal{S}(\Pi_{AB} [\rho_{AB}]) - \mathcal{S}(\rho_{AB})\right] = Q_{AB}^{I_{\mathcal{S}}}(\rho_{AB}),
\end{eqnarray}% These are the work informational ones, i.e.~the defecit
as well as with the entanglement activation measures based upon relative entropy and distillable entanglement~\cite{streltsov2011linking,piani2011all}
\begin{equation}\label{Equation:RelativeEntropyMeasuresGECoincidence}
Q_{A}^{G_{\text{RE}}}(\rho_{AB}) = Q_{A}^{E^{G_{\text{RE}}}}(\rho_{AB})=Q_{A}^{E^d}(\rho_{AB}), \quad \mbox{and} \quad Q_{AB}^{G_{\text{RE}}}(\rho_{AB}) = Q_{AB}^{E^{G_{\text{RE}}}}(\rho_{AB}) =Q_{AB}^{E^d}(\rho_{AB}),
\end{equation}
and also with the coherence based measures in terms of relative entropy $Q_{A}^{C^{\text{RE}}}(\rho_{AB})$ and $Q_{AB}^{C^{\text{RE}}}(\rho_{AB})$, defined later in Eq.~(\ref{Equation:CoherenceMeasuresRE})~\cite{modi2010unified,baumgratz2014quantifying,yao2015quantum,ma2016converting}.

There is clearly a quite remarkable convergence on
the one-sided and two-sided measures $Q_{A}^{G_{\text{RE}}}$ and $Q_{AB}^{G_{\text{RE}}}$, which are also known in the literature as  ``relative entropy of discord'' and ``relative entropy of quantumness'' respectively \cite{horodecki2005local,groisman2007quantumness,modi2010unified,piani2011all},
across a range of different (generally inequivalent) categories of QCs with accordingly different physical interpretations.

As it happens customarily in quantum information theory, however, the best behaved quantities are often the most difficult to compute \cite{tufarelli2013geometric}. In the case of the geometric measures of QCs based on relative entropy, $Q^{G_{\text{RE}}}_{A}$ and $Q^{G_{\text{RE}}}_{AB}$, analytical formulae are available only for  special classes of states, such as Bell diagonal states of two qubits \cite{luo2008quantum} and highly symmetric (i.e., Werner and isotropic) states of two qudits \cite{chitambar2012quantum}. For both of these classes of states, one-sided and two-sided measures of QCs coincide.

%mazzola2011activating experiment

%response... experiment

\subsubsection{Hilbert-Schmidt distance}

By using the $2$-norm induced distance $D_{2}$ (see Table~\ref{Table:Distances}), equal to the squared Hilbert-Schmidt distance, another geometric quantifier of one-sided QCs, corresponding to $Q_{A}^{G_{2}}(\rho_{AB})$ as defined in  Eq.~(\ref{Equation:GeometricMeasures}), was proposed in~\cite{dakic2010necessary}, where a simple closed form of this quantity for arbitrary states of two qubits was also found. Extensions beyond two qubits were subsequently derived, including in particular analytical formulae for all states where subsystem $A$ is a qubit while the dimension of subsystem $B$ is arbitrary~\cite{luo2010geometric,luo2012evaluating,vinjanampathy2012quantum,tufarelli2012quantum}, and an operational interpretation was provided by linking $Q_{A}^{G_{2}}(\rho_{AB})$ to the task of remote state preparation~\cite{dakic2012quantum} (see Section~\ref{Section:RSP} for a critical discussion). It was later shown that~\cite{luo2010geometric}
\begin{equation}\label{Equation:HSGeometricMID}
Q_{A}^{G_{2}}(\rho_{AB}) = Q_{A}^{M_{2}}(\rho_{AB}),%= Q^{R_{2}}_A(\rho_{AB}).
\end{equation}
while, if $A$ is a qubit, it also holds that \cite{gharibian2012quantifying,giampaolo2013quantifying}
\begin{equation}\label{Equation:HSGeometricU2}
Q_{A}^{G_{2}}(\rho_{AB}) = \frac14 Q^{U_{2}}_A(\rho_{AB}).
\end{equation}

Experimental quantifications of $Q_{A}^{G_{2}}(\rho_{AB})$ have been carried out~\cite{dakic2012quantum,silva2013measuring,passante2012measuring,jin2012direct} and lower bounds provided~\cite{girolami2012observable,hassan2012tight,rana2012tight}. The  two-sided quantifier $Q_{AB}^{G_2}$ was defined in \cite{jiang2013symmetric}.

The Hilbert-Schmidt based geometric measure of QCs enjoyed considerable popularity due to its analytical simplicity and became initially known just as the ``geometric discord'', but an important problem was subsequently identified, due to the fact that the Hilbert-Schmidt distance $D_{2}$ is not contractive in general, as mentioned in Table~\ref{Table:Distances}. Although $D_{2}$ is still unitarily invariant, which implies Requirement~(\ref{Requirements:UnitaryInvariance}), this means that Requirements~(\ref{Requirements:UnmeasuredParty}) and (\ref{Requirements:LCPO}) are not automatically satisfied for $Q_{A}^{G_{2}}$. Indeed,~\cite{piani2012problem} provided a counter argument to show that $Q_{A}^{G_{2}}(\rho_{AB})$ does not obey Requirement~(\ref{Requirements:UnmeasuredParty}), which goes as follows. Consider appending an uncorrelated ancilla state $\rho_{C}$ to the bipartite state $\rho_{AB}$, it then holds that
\begin{equation}\label{Equation:HSAncillaProblem}
Q_{A}^{G_{2}}(\rho_{AB}\otimes \rho_{C}) = Q_{A}^{G_{2}}(\rho_{AB}) \mbox{Tr}(\rho_{C}^{2}),
\end{equation}
due to the multiplicativity of the squared Hilbert-Schmidt norm on tensor products. Since $\mbox{Tr}(\rho_{C}^{2})$ is just the purity of $\rho_{C}$, we know that adding/removing an impure ancilla will decrease/increase $Q_{A}^{G_{2}}$. However, addition and removal of an ancilla is a reversible quantum operation on the unmeasured subsystem $B$, for which we require the one-sided QCs never to increase. Thus, $Q_{A}^{G_{2}}$ does not obey this crucial requirement. One proposed resolution to this particular problem was to rescale $Q_{A}^{G_{2}}(\rho_{AB})$ by the purity of $\rho_{AB}$~\cite{tufarelli2013geometric}, but in this case it is still possible to find other counterexamples showing that Requirement~(\ref{Requirements:UnmeasuredParty}) is violated~\cite{hu2013quantum}. Another suggestion given in~\cite{chang2013remedying} was to exploit the link between the Hilbert-Schmidt geometric and measurement induced geometric measures in Eq.~(\ref{Equation:HSGeometricMID}) and instead consider $Q_{A}^{M_{2}}(\sqrt{\rho_{AB}})$. %, which for a qubit is equal to $Q_{A}^{U_{\text{He}}}(\sqrt{\rho_{AB}})/2$.
While maintaining the analytical simplicity of $Q_{A}^{G_{2}}(\rho_{AB})$, this remedied modification no longer experiences the problem in Eq.~(\ref{Equation:HSAncillaProblem}), and for pure states ${\ket{\psi}_{AB}}$ it reduces to $Q_{A}^{M_{2}}(\rho_{AB})$ because $\sqrt{\ket{\psi}\bra{\psi}_{AB}} = \ket{\psi}\bra{\psi}_{AB}$. Finally, Requirement~(\ref{Requirements:EntanglementPure}) is satisfied for $Q_{A}^{G_{2}}(\rho_{AB})$, since for pure states one finds~\cite{luo2012evaluating}
\begin{equation}\label{Equation:ITangle}
Q_{A}^{G_{2}}({\ket{\psi}_{AB}}) = 1 - \mbox{Tr}(\rho_{A}^{2})= 1 - \mbox{Tr}(\rho_{B}^{2}),
\end{equation}
which is half of the entanglement monotone known as the I-tangle~\cite{rungta2003concurrence}. The latter reduces to the squared entanglement negativity \cite{zyczkowski1998volume,vidal2002computable} if $A$ is a qubit, where we define the entanglement negativity $E^{N}(\rho_{AB})$ as in Table~\ref{Table:Entanglement}, with the suffix $T_A$ denoting partial transposition with respect to subsystem $A$. More generally, $2 Q_{A}^{G_{2}}(\rho_{AB}) \geq [E^{N}(\rho_{AB})]^2$ for all two-qubit states $\rho_{AB}$ \cite{girolami2011interplay}, but this is not true anymore for higher dimensional states \cite{rana2012entanglement}. Geometric measures of QCs based on the Hilbert-Schmidt distance have been studied for two-mode Gaussian states in \cite{adesso2011gaussian,tufarelli2013geometric} (specifically, in the measurement induced geometric approach); however, the only classical Gaussian states are product states \cite{adesso2010quantum,mista2014no}, which makes any geometric approach to the quantification of their QCs ambiguous (i.e., geometric quantum and total correlations collapse to the same quantity when restricting distances to the subsets $\mathscr{C}_{A}$ or $\mathscr{C}_{AB}$ and $\mathscr{P}_{A}$ of Gaussian states, respectively).

\subsubsection{Trace distance}\label{Section:Measures-Geometric-Trace}

The trace distance, arising (modulo a $\frac12$ normalisation factor) from the $1$-norm induced distance $D_{1}$ (see Table~\ref{Table:Distances}), has also been adopted to define geometric measures of QCs as in Eqs.~(\ref{Equation:GeometricMeasures}) \cite{debarba2012witnessed,rana2013comment,nakano2013negativity,paula2013geometric,aaronson2013hierarchy}. Because $D_{p}$ is only contractive for $p=1$, the trace distance based geometric measure, alias ``trace distance discord'', is the only one of its class that directly obeys Requirements~(\ref{Requirements:UnitaryInvariance}), (\ref{Requirements:UnmeasuredParty}), and (\ref{Requirements:LCPO}), hence emerging as the most suitable choice out of all the Schatten $D_{p}$ norm induced geometric measures. For pure states ${\ket{\psi}_{AB}}$, only when subsystem $A$ is a qubit it has been shown in~\cite{nakano2013negativity} that this  measure of QCs reduces to the entanglement negativity~\cite{zyczkowski1998volume,vidal2002computable}, i.e.~\begin{equation}
Q_{A}^{G_{1}}({\ket{\psi}_{AB}}) = E^{N}({\ket{\psi}_{AB}}),
\end{equation}
while the behaviour of $Q_{A}^{G_{1}}({\ket{\psi}_{AB}})$ and $Q_{AB}^{G_{1}}({\ket{\psi}_{AB}})$ for arbitrary pure states is not known, leaving the fulfillment of Requirement~(\ref{Requirements:EntanglementPure}) as an open question in general. A closed formula for $Q_{A}^{G_{1}}(\rho_{AB})$ on two-qubit states with X-shaped density matrices (X states) and two-qubit quantum-classical states was provided in~\cite{ciccarello2014toward}, where also the optimisation for general two-qubit states was greatly simplified. It was shown in~\cite{nakano2013negativity} that, if subsystem $A$ is a qubit, then
\begin{equation}\label{Equation:NakanoLinks}
Q_{A}^{G_{1}}(\rho_{AB}) = Q_{A}^{M_{1}}(\rho_{AB}) = Q_{A}^{E^{N}}(\rho_{AB})=\frac12 Q_{A}^{U_{1}}(\rho_{AB}).
\end{equation}
Operationally, the trace distance is related to the minimum probability of error in the discrimination between states~\cite{helstrom1976quantum}. Experimental investigations of $Q_{A}^{G_{1}}(\rho)$ have been carried out in a two-qubit nuclear magnetic resonance setup~\cite{silva2013measuring,paula2013observation}.

\subsubsection{Bures distance}\label{Section:Measures-Geometric-Bures}

The squared Bures distance $D_{\text{Bu}}$, as well as the monotonically related infidelity $D_{F}$ (see Table~\ref{Table:Distances}), can induce another pair of geometric measures of QCs defined as in Eqs.~(\ref{Equation:GeometricMeasures}), that satisfy all of the five  Requirements of Section~\ref{Section:Requirements}~\cite{streltsov2011linking,aaronson2013comparative,spehner2013geometric,spehner2013geometric2,bromley2014unifying}. Indeed, $D_{\text{Bu}}$ is contractive, and for pure states ${\ket{\psi}_{AB}}$ it holds that
\begin{equation}
Q_{A}^{G_{\text{Bu}}}({\ket{\psi}_{AB}}) = Q_{AB}^{G_{\text{Bu}}}({\ket{\psi}_{AB}}) = E^{G_{\text{Bu}}}({\ket{\psi}_{AB}}) = 2(1 - \sqrt{s_{\max}}),
\end{equation}
where $s_{\max}$ is the maximum Schmidt coefficient of ${\ket{\psi}_{AB}}$~\cite{spehner2013geometric}, which defines the geometric measure of entanglement~\cite{wei2003geometric}. An operational interpretation of $Q_{A}^{G_{\text{Bu}}}(\rho_{AB})$ can be provided in terms of the optimal success probability of an ambiguous quantum state discrimination task~\cite{spehner2013geometric}, as discussed later in Section~\ref{Section:Applications-Discrimination}. Closed formulae for $Q_{A}^{G_{\text{Bu}}}(\rho_{AB})$ have been supplied for Bell diagonal states of two qubits~\cite{aaronson2013comparative,spehner2013geometric2}.

%under CPTP operations and so induces a measure of QCs obeying Requirements (\ref{Requirements:UnitaryInvariance}), (\ref{Requirements:UnmeasuredParty}) and (\ref{Requirements:LCPO}). Furthermore, for pure states it holds that $Q_{A}^{G_{\text{Bu}}}(\rho) = Q^{G_{\text{Bu}}}(\rho) = 2(1 - \sqrt{s_{max}})$ where $s_{max}$ is the maximum Schmidt coefficient of $\ket{\psi}$~\cite{spehner2013geometric}, which is the geometric measure of entanglement~\cite{wei2003geometric} and so Requirement~(\ref{Requirements:EntanglementPure}) is also obeyed. An operational interpretation of $Q_{A}^{G_{\text{Bu}}}(\rho)$ is the optimal success probability of an ambiguous quantum state discrimination task~\cite{spehner2013geometric}. Closed formulae for $Q_{A}^{G_{\text{Bu}}}(\rho)$ have been supplied for qubit-qudit systems where the qubit is the measured subsystem $A$~\cite{spehner2013geometric2}.

\subsubsection{Hellinger distance}

Similarly, using the squared Hellinger distance $D_{\text{He}}$ (see Table~\ref{Table:Distances}) one obtains as well valid measures of geometric QCs obeying all the Requirements. An in-depth study of $Q_{A}^{G_{\text{He}}}(\rho_{AB})$ was carried out in~\cite{roga2015geometric}, where they simplified the problem of evaluating this measure for arbitrary bipartite states, and
showed in particular that for pure states ${\ket{\psi}_{AB}}$ it holds
\begin{equation}
Q_{A}^{G_{\text{He}}}({\ket{\psi}_{AB}}) = 2 \left( 1 - \sqrt{\mbox{Tr}(\rho_{A}^{2})}  \right) =  2 \left( 1 - \sqrt{\mbox{Tr}(\rho_{B}^{2})}\right),
\end{equation}
which is a measure of entanglement~\cite{vidal2000entanglement}. An interesting link to the Hilbert-Schmidt based geometric measure was also established \cite{roga2015geometric},
\begin{equation}\label{Equation:HellingerGeometric}
Q_{A}^{G_{\text{He}}}(\rho_{AB}) = 2-2\sqrt{1-Q_{A}^{G_{2}}\left(\sqrt{\rho_{AB}}\right)},
\end{equation}
which shows that, without compromising its reliability, the squared Hellinger distance based geometric measure maintains the same computational simplicity of the Hilbert-Schmidt distance based geometric measure corrected as in \cite{chang2013remedying}, and is thus amenable to a closed analytical formula for all bipartite states $\rho_{AB}$ when subsystem $A$ is a qubit. Geometric measures of QCs based on the Hellinger distance have been discussed for Gaussian states in \cite{marian2015hellinger}. Operationally, the Hellinger distance admits an interpretation in terms of asymptotic state discrimination~\cite{audenaert2007discriminating,roga2015geometric}, as detailed in Section~\ref{Section:Applications-Discrimination}.

\subsubsection{Hierarchy of geometric measures}

It is also interesting to compare the various geometric measures of QCs arising from different choices of distance $D_\delta$. We can use general inequalities between the distances in Table~\ref{Table:Distances} to infer inequalities between the corresponding geometric measures, i.e.~if for two choices of distance $\delta_{1}$ and $\delta_{2}$ it holds that $f(D_{\delta_{1}} (\rho,\sigma)) \leq g(D_{\delta_{2}} (\rho,\sigma))$ for any $\rho$ and $\sigma$ and for two increasing functions $f$ and $g$, then we know that
\begin{equation}
f\left(Q_{A}^{G_{\delta_{1}}}(\rho_{AB})\right) \leq g\left(Q_{A}^{G_{\delta_{2}}}(\rho_{AB})\right), \quad \mbox{and} \quad f\left(Q_{AB}^{G_{\delta_{1}}}(\rho_{AB})\right) \leq g\left(Q_{AB}^{G_{\delta_{2}}}(\rho_{AB})\right).
\end{equation}
The following inequalities are therefore useful~\cite{bengtsson2007geometry,luo2004informational,nielsen2010quantum,spehner2013geometric,fuchs1999cryptographic,roga2015geometric}
\begin{eqnarray}
&&\frac{1}{r}D_{1}(\rho,\sigma)^{2} \leq  D_{2}(\rho,\sigma) \leq D_{1}(\rho,\sigma)^{2},  \nonumber \\
&&\frac{1}{2} D_{1}(\rho ,\sigma)^{2} \leq  D_{\text{RE}}(\rho ,\sigma),  \\
&&D_{\text{Bu}} (\rho,\sigma) \leq D_{\text{He}} (\rho,\sigma) \leq  D_{1}(\rho,\sigma) \leq \sqrt{1-\left(1- \frac12 D_{\text{Bu}}(\rho,\sigma)\right)^{2}},\nonumber
\end{eqnarray}
where $r$ is the rank of $\rho - \sigma$. These bounds are also applicable to the other types of measures of QCs involving a distance. One can further derive bounds between geometric measures of QCs that do not stem from fundamental inequalities between the corresponding distances. More generally, bounds amongst the different types of measures of QCs can be found, but these are sometimes tailored to the dimension of the relevant subsystems. A wealth of such inequalities are provided in~\cite{roga2015geometric}.

%\subsubsection{Summary of geometric measures}
%
%The key literature on geometric measures is now summarised in Table~\ref{Table:GeometricMeasures} below.
%
%\begin{table}[!h]
%\begin{tabular}{c|cc|cc|cc}%%!! THESE QUESTION MARKS SHOULD BE CHECKED AND REPLACED WITH A HYPHEN
%\hline \hline
% $D_{\delta}$  &  $Q_{A}^{G_{\delta}}(\rho_{AB})$ &  Formulae & $Q_{AB}^{G_{\delta}}(\rho_{AB})$ & Formulae & Operational & Experimental\\ \hline
% $D_{\text{RE}}$ &~\cite{horodecki2005local} &~\cite{modi2010unified} &~\cite{horodecki2005local} & ? &~\cite{modi2010unified} & ? \\
% $D_{2}$ &~\cite{dakic2010necessary} &~\cite{luo2012evaluating,vinjanampathy2012quantum} & \cite{jiang2013symmetric} & ? &~\cite{dakic2012quantum} %&~\cite{passante2012measuring,jin2012direct} \\
% $D_{1}$ &~\cite{paula2013geometric} &~\cite{ciccarello2014toward} & ? & ? & ? &~\cite{silva2013measuring} \\
% $D_{\text{Bu}}$ &~\cite{spehner2013geometric} &~\cite{spehner2013geometric2} &~\cite{spehner2013geometric} & ? &~\cite{spehner2013geometric} & ? \\
%  $D_{\text{He}}$ &~\cite{roga2015geometric} &~\cite{roga2015geometric} & ? & ? &~\cite{roga2015geometric} & ? \\ \hline \hline
%\end{tabular}
%\caption{\label{Table:GeometricMeasures} A summary of the literature on geometric measures of QCs for some common distances $D_{\delta}$, including %where the one-sided and two-sided measures were introduced, the best available closed formulae for general classes of states, known operational %interpretations and any relevant experimental results.}
%\end{table}

\subsubsection{Other distances}\label{Section-Measures-Geometric-Other}

There are other possible choices of $D_{\delta}$ in Eqs.~(\ref{Equation:GeometricMeasures}) that can be used to define geometric measures of QCs. First of all, note that both the Bures distance and Hellinger distance are closely related, in fact they become identical when considering the distance between pairs of commuting states. Indeed, both are examples of geodesic distances belonging to a family of monotone and Riemannian metrics defined on the set of quantum states, characterised by the Morozova-{\v{C}}entsov-Petz theorem~\cite{morozova1991markov,petz1996monotone,bengtsson2007geometry}. These infinitely many monotone and Riemannian metrics can each give a unique contractive geodesic distance that could in principle be used to define a geometric measure of QCs, but finding the geodesic distance in each case is a formidable problem and the information theoretic relevance of each distance would need to be considered.

Generalisations of the relative entropy could also be used to define geometric measures of QCs. Two recently suggested generalisations are the  sandwiched relative R{\'{e}}nyi entropies~\cite{wilde2014strong,muller2013quantum} (see also \cite{frank2013monotonicity})
\begin{equation}\label{Equation:SandRenyi}
D_{{\mathcal{R}_{\alpha}}} (\rho,\sigma) = \frac{\log  \mbox{Tr}\left[\left(\sigma^{\frac{1-\alpha}{2\alpha}}\rho \sigma^{\frac{1-\alpha}{2\alpha}} \right)^{\alpha}\right]}{\alpha - 1}
\end{equation}
and the sandwiched relative Tsallis entropies~\cite{rajagopal2014quantum}
\begin{equation}\label{Equation:SandTsallis}
D_{{\mathcal{T}_{\alpha}}} (\rho,\sigma) = \frac{\mbox{Tr}\left[\left(\sigma^{\frac{1-\alpha}{2\alpha}}\rho \sigma^{\frac{1-\alpha}{2\alpha}}\right)^{\alpha}\right] -1}{\alpha - 1}.
\end{equation}
These recent variations of the traditional relative R{\'{e}}nyi and Tsallis entropies, designed to capture the noncommutative nature of the density matrix, are contractive quasi-distances for $\alpha \in (1/2,\infty]$  and thus any geometric measure of QCs based upon either choice would naturally obey Requirements~(\ref{Requirements:Faithfulness}), (\ref{Requirements:UnitaryInvariance}), (\ref{Requirements:UnmeasuredParty}) and (\ref{Requirements:LCPO}), with Requirement~(\ref{Requirements:EntanglementPure}) needing to be tested for each $\alpha$. Certain values of $\alpha$ recover well known relative entropies, in particular when $\alpha \rightarrow 1$ we have
\begin{equation}
D_{{\mathcal{R}_{1}}} (\rho,\sigma) = D_{{\mathcal{T}_{1}}} (\rho,\sigma) = D_{\text{RE}} (\rho,\sigma).
\end{equation}

%There are other possible distances that could be used to define a geometric measure of QCs. A prime example is the Hellinger distance~\cite{luo2004informational}, which is closely related to the Bures distance. Indeed, both the Hellinger and Bures distances are geodesic distances of a family of monotone and Riemannian metrics defined on the set of quantum states characterised by the Morozova-{\v{C}}entsov-Petz theorem~\cite{morozova1991markov,petz1996monotone,bengtsson2007geometry}. These infinitely many monotone and Riemannian metrics can each give a unique contractive geodesic distance that could be in principle used to define a geometric measure of QCs, but finding the geodesic distance in each case is a formidable problem and the informational relevance of each distance would need to be considered. Another possibility could be to harness the relative versions of the aforementioned generalised entropies. In~\cite{misra2015quantum,frank2013monotonicity} it was suggested that the sandwiched relative R{\'{e}}nyi entropy posseses most of the properties required to make a valid geometric measure of QCs.

\subsection{Measurement induced geometric measures}\label{Section:Measures-MIG}

As we have discussed in Section \ref{Section:Measures-Geometric}, the measurement induced geometric measures of QCs with minimisation over LPMs, defined by Eqs.~(\ref{Equation:MeasurementInducedDisturbances}), often have an intuitive crossover with the conventional geometric measures of QCs.

\subsubsection{Hilbert-Schmidt and trace distance}
The measurement induced geometric type of measure was first introduced in~\cite{luo2008using} for the squared Hilbert-Schmidt distance and has subsequently been primarily studied in parallel with the geometric measures. In particular, due to the link in Eq.~(\ref{Equation:HSGeometricMID}), $Q_{A}^{M_{2}}(\rho_{AB})$ is not a valid measure of QCs as it fails Requirement~(\ref{Requirements:UnmeasuredParty}), and similar inconsistencies can be expected for $Q_{AB}^{M_{2}}(\rho_{AB})$ due to the lack of contractivity of the Hilbert-Schmidt distance $D_{2}$.

While we have coincidence between the one-sided geometric measures and the one-sided measurement induced geometric measures of QCs based on the trace distance for a qubit \cite{nakano2013negativity}, as expressed in Eq.~(\ref{Equation:NakanoLinks}), it is still a relevant question to ask whether $Q_{A}^{M_{1}}(\rho_{AB})$ and also $Q_{AB}^{M_{1}}(\rho_{AB})$ are valid measures of QCs in general. As we have discussed in Section \ref{Section:Measures-Types-MIG}, Requirement~(\ref{Requirements:Faithfulness}) is naturally obeyed, and the contractivity of $D_{1}$ implies Requirements~(\ref{Requirements:UnitaryInvariance}) and (\ref{Requirements:UnmeasuredParty}), however it is not known whether Requirement~(\ref{Requirements:LCPO}) holds. Regarding Requirement~(\ref{Requirements:EntanglementPure}), it was proven in~\cite{piani2014quantumness} by showing that $Q_{A}^{M_{1}}({\ket{\psi}_{AB}}) = Q_{AB}^{M_{1}}({\ket{\psi}_{AB}})$ and both amount to a measure of entanglement.

\subsubsection{Bures and Hellinger distance}
Two measurement induced geometric measures of QCs that do not have such a close link with the corresponding geometric measures are the ones induced by the squared Bures distance and the squared Hellinger distance (see Table~\ref{Table:Distances}). The one-sided versions, considered in~\cite{roga2015geometric}, are valid measures of QCs obeying all the Requirements~(\ref{Requirements:Faithfulness})--(\ref{Requirements:UnmeasuredParty}), while Requirement~(\ref{Requirements:LCPO}) has not been tested yet.  In particular, for pure states $\ket{\psi}_{AB}$ it holds
\begin{eqnarray}
Q_{A}^{M_{\text{Bu}}}({\ket{\psi}_{AB}}) = Q_{A}^{G_{\text{He}}}({\ket{\psi}_{AB}}) = 2 \left( 1 - \sqrt{\mbox{Tr}(\rho_{A}^{2})}  \right) =  2 \left( 1 - \sqrt{\mbox{Tr}(\rho_{B}^{2})}\right), \\
Q_{A}^{M_{\text{He}}}({\ket{\psi}_{AB}}) = 2 \left( 1 - \mbox{Tr}(\rho_{A}^{3/2})\right)  = 2 \left( 1 - \mbox{Tr}(\rho_{B}^{3/2})\right),
\end{eqnarray}
which are measures of entanglement~\cite{vidal2000entanglement}. We point again to~\cite{roga2015geometric} for a compendium of informative bounds.

Finally, the extensions to other distances suggested in Section~\ref{Section-Measures-Geometric-Other} can likewise be considered for the measurement induced geometric  measures.

\subsection{Measurement induced informational measures}\label{Section:Measures-Informational}
This category encompasses some of the seminal and most studied measures of QCs.

\subsubsection{von Neumann entropy based measures: Quantum deficit}\label{Section:Measures-Informational-Entropy}

A simple choice to gauge the change of information due to local measurements is the (negative) von Neumann entropy in Table~\ref{Table:InformationalQuantifiers}. By using $I_{\mathcal{S}} = -\mathcal{S}$ in Eqs.~(\ref{Equation:InformationalMeasures}) one can see that $Q_{A}^{I_{\mathcal{S}}}(\rho_{AB})$ and $Q_{AB}^{I_{\mathcal{S}}}(\rho_{AB})$ quantify the global information lost by (rank one) LPMs. % Indeed, it is well known that projective measurements always increase the von Neumann entropy (hence decrease the information) whenever the final result of the measurement is not known~\cite{nielsen2010quantum}, which justifies choosing the negative of the von Neumann entropy to guarantee the non-negativity of the corresponding measures of QCs.

The measures $Q_{A}^{I_{\mathcal{S}}}(\rho_{AB})$ and $Q_{AB}^{I_{\mathcal{S}}}(\rho_{AB})$ were introduced in~\cite{horodecki2005local}, being called the ``one-way quantum deficit'' and ``zero-way quantum deficit'', respectively. They have fundamental operational interpretations in thermodynamics, as we discuss in Section~\ref{Section:Thermodynamics}. The coincidence between these measures and the relative entropy based geometric measures, Eq.~(\ref{Equation:RelativeEntropyMeasuresGICoincidence}), was highlighted in~\cite{horodecki2005local}, and the coincidence with the distillable entanglement based entanglement activation measures, Eq.~(\ref{Equation:RelativeEntropyMeasuresGECoincidence}), was highlighted in~\cite{piani2011all,streltsov2011linking}. The one-sided measure $Q_{A}^{I_{\mathcal{S}}}(\rho_{AB})$ was also independently studied~\cite{zurek2003quantum} and referred to as ``thermal discord'' in~\cite{modi2012classical}. All the Requirements are satisfied for these measures as discussed in Section~\ref{Section:Measures-Geometric-RE}.

The corresponding LGM based informational measures in terms of von Neumann entropy, $Q_{A}^{\tilde{I}_{\mathcal{S}}}(\rho_{AB})$ and $Q_{AB}^{\tilde{I}_{\mathcal{S}}}(\rho_{AB})$, were considered in~\cite{lang2011entropic,horodecki2013quantumness}, with optimisation
restricted to local rank one POVMs.   These measures are less used in the literature and Requirements~(\ref{Requirements:UnmeasuredParty}) and (\ref{Requirements:LCPO}) were not tested, while Requirement~(\ref{Requirements:EntanglementPure}) was proven in \cite{lang2011entropic}.

A final word of clarification is needed regarding the restriction to rank one measurements for von Neumann entropy based informational measures of QCs. Notice that, if one allowed instead the optimisation over all LGMs (or even all LPMs), then  the optimal measurement could be the trivial `measurement' of not doing anything (i.e., the identity), resulting in ill-defined quantifiers. The restriction to rank one measurements is therefore necessary to arrive at informative (and non-negative) QCs measures, and is anyway an intuitive restriction since rank one measurements represent  the most fine grained type of measurements~\cite{wu2009correlations}.

\subsubsection{Mutual information based measures: Quantum discord}\label{Section:Measures-Informational-MutualInformation}

Since we are interested in quantum \emph{correlations}, it seems appropriate to consider as an informational gauge the mutual information $I_{\cal I} = {\cal I}$, which is itself a measure of correlations. Specifically, resulting from a generalisation of the classical mutual information to the quantum setting, the mutual information in Table~\ref{Table:InformationalQuantifiers} can be seen as the canonical quantifier of the total correlations between subsystem $A$ and $B$ in a state $\rho_{AB}$. We can think of the total correlations remaining in the
%post-measurement
states   $\tilde{\Pi}_{A}[\rho_{AB}]$ or $\tilde{\Pi}_{AB}[\rho_{AB}]$ after a (general) LGM on subsystem $A$ or on both subsystems $A$ and $B$ as effectively classical, because these are the correlations that are extracted from $\rho_{AB}$ by a local measurement and can be stored in a classical register. Hence, to find the corresponding QCs one should subtract these classical correlations from the original total correlations of $\rho_{AB}$. Since different local measurements may lead to the extraction of different amounts of classical correlations, and given that we aim to quantify only the genuinely quantum share of the total correlations, we need to minimise this difference (or equivalently maximise the extractable classical correlations) over all LGMs, hence justifying the form of the mutual information based measures of QCs in Eqs.~(\ref{Equation:InformationalMeasuresPOVM}) and Eqs.~(\ref{Equation:InformationalMeasures}). These measures will therefore be understood as quantifying, in general, the minimum amount of correlations lost in the state $\rho_{AB}$ by the act of a
%least perturbing
maximally informative local measurement on one or both subsystems.

%Quantum correlations were originally conceived in~\cite{ollivier2001quantum,henderson2001classical} by extending two classically identical definitions of the mutual information to the quantum setting. One mutual information $\mathscr{I}$, the mutual information we consider, is simply extended, while the other mutual information $\mathcal{J}$ is conditional upon $A$. In the classical setting Bayes' rule tells us that $\mathscr{I} = \mathcal{J}$, but in the quantum setting we must construct a local measurement on subsytem $A$ to arrive at $\mathcal{J}$. We can then think of $\mathscr{I}$ as the total correlations and $\mathcal{J}$ as the classical correlations, and consider $\mathscr{I} - \mathcal{J}$ minimised over all LPMs as a measure of quantum correlations. It turns out that this quantum discord is exactly $Q_{A}^{I_{\mathcal{I}}}(\rho_{AB})$.

It turns out that the one-sided mutual information based measure of QCs coincides with the original definition of ``quantum discord'', which was first conceived in~\cite{ollivier2001quantum,henderson2001classical} by interpreting the disagreement (or discord, indeed) between two possible extensions of classically identical definitions of the mutual information to the quantum setting. In~\cite{ollivier2001quantum} Ollivier and Zurek considered $Q_{A}^{I_{\mathcal{I}}}(\rho_{AB})$, i.e., minimisation over LPMs, while in~\cite{henderson2001classical} Henderson and Vedral considered $Q_{A}^{\tilde{I}_{\mathcal{I}}}(\rho_{AB})$, i.e., minimisation over LGMs. Having clarified the fundamental meaning of quantum discord within the broader scope of the informational approach to QCs, for completeness we will now provide a summary of the original analysis behind it by focusing on the more general case of LGMs, although the same holds analogously for LPMs.

Consider two classical random variables $A$ and $B$ with a joint probability distribution $p_{ab}$ and marginal probability distributions $p_{a} = \sum_{b \in B} p_{ab}$ and $p_{b} = \sum_{a \in A} p_{ab}$. The correlations between $A$ and $B$ are given by the mutual information%, which has two equivalent definitions
\begin{equation}
\label{Equation:MutualInformationClassical}
%\begin{aligned}
\mathcal{I}(A:B) := \mathcal{H}(A)+\mathcal{H}(B)-\mathcal{H}(AB),\\
%\mathcal{J}(A:B) &= \mathcal{H}(B) - \mathcal{H}(B|A)
%\end{aligned}
\end{equation}
with $\mathcal{H}(A) = - \sum_{a \in A} p_{a} \log p_{a}$ and $\mathcal{H}(B) = - \sum_{b \in B} p_{b} \log p_{b}$ the Shannon entropies of the individual random variables $A$ and $B$, and $\mathcal{H}(AB) = - \sum_{a \in A}\sum_{b \in B} p_{ab} \log p_{ab}$ the joint Shannon entropy of $A$ and $B$~\cite{shannon1948mathematical,cover2012elements}. An equivalent expression for the mutual information is given by
\begin{equation}\label{Equation:ClassicalJ}
\mathcal{J}(A:B) := \mathcal{H}(B) - \mathcal{H}(B|A),
\end{equation}
where
\begin{equation}\label{Equation:ConditionalShannonEntropy}
\mathcal{H}(B|A)  :=  \sum_{a \in A} p_{a} \mathcal{H}(B|a)%= - \sum_{a \in A} p_{a} \sum_{b \in B} p_{b|a} \log p_{b|a}
\end{equation}
is the conditional Shannon entropy, i.e.~the average of $\mathcal{H}(B|a)=-\sum_{b \in B} p_{b|a} \log p_{b|a}$ over $a \in A$, with $p_{b|a}:=p_{ab}/p_{a}$ the conditional probability distribution of $B$ given result $a$. %and $\mathcal{H}(B|a)=-\sum_{b \in B} p_{b|a} \log p_{b|a}$.
By substituting the definition of $p_{b|a}$ into the above expression of $\mathcal{H}(B|A)$, we can easily see that
\begin{equation}\label{Equation:ConditionalShannonEntropies}
\mathcal{H}(B|A) = \mathcal{H}(AB) - \mathcal{H}(A),
\end{equation}
establishing the equivalence of $\mathcal{I}(A:B)$ and $\mathcal{J}(A:B)$.

By extending the above to the quantum setting, the joint probability distribution $p_{ab}$ generalises to a state $\rho_{AB}$ of the composite system and the marginal probability distributions $p_{a}$ and $p_{b}$ become states of the respective subsystems, i.e.~$\rho_{A} = \mbox{Tr}_{B}(\rho)$ and $\rho_{B} = \mbox{Tr}_{A}(\rho)$. The first definition $\mathcal{I}(A:B)$  of the mutual information then naturally extends to
\begin{equation}\label{Equation:MutualInformation}
\mathcal{I}(\rho_{AB}) := \mathcal{S}(\rho_{A}) + \mathcal{S}(\rho_{B}) - \mathcal{S}(\rho_{AB})\,,
\end{equation}
with the von Neumann entropy $\mathcal{S}$ replacing the Shannon entropy. Extending the second definition $\mathcal{J}(A:B)$ of the mutual information is less simple because it is not unambiguously clear how to deal with the conditional entropy $\mathcal{H}(B|A)$. The most direct way would be to emulate Eq.~(\ref{Equation:ConditionalShannonEntropies}) with the conditional entropy
\begin{equation}\label{Equation:ConditionalVNEntropies}
\mathcal{S}_{B|A}(\rho_{AB}) := \mathcal{S}(\rho_{AB}) - \mathcal{S}(\rho_{A}).
\end{equation}
%which is of interest operationally in quantum state merging, see Section~\ref{Section:Merging}.
However, to truly capture the conditioning of ``$B$ given $A$'' in the quantum setting, one must first extract information on $A$ by performing local measurements.
% $\mathcal{S}_{B|A}(\rho_{AB})$ can be negative for some $\rho_{AB}$ (consider, for example, any pure entangled state).
Hence, following Eq.~(\ref{Equation:ConditionalShannonEntropy}), the conditional entropy can be interpreted as the average entropy of the post-measurement ensemble of states of subsystem $B$ following an LGM
%$\tilde{\Pi}_A$
on $A$ with subnormalised maps $\{(\tilde{\pi}_{A})_{a}\}$. Invoking the definitions in Section~\ref{Section:Facts-Measuring}, one can then define the LGM based conditional von Neumann entropy as
\begin{equation}\label{Equation:ConditionalEntropyLGM}
{\mathcal{S}}^{\tilde{\pi}_A}_{B|A}(\rho_{AB}) := \sum_{a}p_{a} \mathcal{S}(\rho_{B|a})\,,\end{equation}
which gives an LGM based quantum version of the classical mutual information as
\begin{equation}\label{Equation:ClassicalCorrelationsLGM}
\tilde{\mathcal{J}}^{\tilde{\pi}_A}_{A}(\rho_{AB}) := \mathcal{S}(\rho_{B}) - {\mathcal{S}}^{\tilde{\pi}_A}_{B|A}(\rho_{AB})\,.
\end{equation}
As argued before,  $\mathcal{I}(\rho_{AB})$ in Eq.~(\ref{Equation:MutualInformation}) represents the total correlations, while the maximisation of the quantity in Eq.~(\ref{Equation:ClassicalCorrelationsLGM}) over all  LGMs,
\begin{equation}\label{Equation:ClassicalCorrelations}
\tilde{\mathcal{J}}_A(\rho_{AB}) := \sup_{\tilde{\pi}_A}\left[\tilde{\mathcal{J}}^{\tilde{\pi}_A}_{A}(\rho_{AB})\right]\,,
\end{equation}
can be regarded as a measure of one-sided classical correlations (with respect to subsystem $A$). Their difference yields the (LGM based) quantum discord with respect to subsystem $A$,
\begin{equation}\label{Equation:QuantumDiscord}
\tilde{\mathcal{D}}_{A}(\rho_{AB}) := \mathcal{I}(\rho_{AB}) - \tilde{\mathcal{J}}_A(\rho_{AB}) = \inf_{\tilde{\pi}_{A}} \left[ {\mathcal{I}}(\rho_{AB}) - \tilde{\mathcal{J}}^{\tilde{\pi}_A}_{A}(\rho_{AB})\right] \nonumber \\,
\end{equation}
which may be equivalently seen as the difference between two quantum versions of the classical conditional entropy
\begin{equation}
\tilde{\mathcal{D}}_{A}(\rho_{AB}) =\inf_{\tilde{\pi}_{A}} \left[ {\mathcal{S}}^{\tilde{\pi}_A}_{B|A}(\rho_{AB}) - \mathcal{S}_{B|A}(\rho_{AB})\right] .
\end{equation}
To see that this quantity coincides with the informational QCs measure that we label $Q_{A}^{\tilde{I}_{\mathcal{I}}}(\rho_{AB})$, as defined in Eqs.~\ref{Equation:InformationalMeasuresPOVM}, we can write explicitly~(see e.g.~\cite{spehner2014quantum})
%The coincidence with $Q_{A}^{\tilde{I}_{\mathcal{I}}}(\rho_{AB})$ in Eq.~(\ref{Equation:InformationalMeasuresPOVM}) arises ??because??
%\begin{equation}
%\mathcal{J}_{A}(\rho_{AB}) = I_{\mathcal{I}}\left( \tilde{\Pi}_{A}\left( \rho_{AB}\right)\right).
%\end{equation}%% how the hell do we prove this?
\begin{eqnarray}
\tilde{\mathcal{J}}^{\tilde{\pi}_A}_{A}(\rho_{AB}) &=& \mathcal{S}(\rho_{B}) - {\mathcal{S}}^{\tilde{\pi}_A}_{B|A}(\rho_{AB}) \nonumber
= \mathcal{S}(\rho_{B}) - \sum_{a}p_{a} \mathcal{S}(\rho_{B|a}) \nonumber \\
&=& \mathcal{S}(\rho_{B}) -\sum_{a}p_{a}\log p_{a} - \mathcal{S}\bigg( \sum_{a}p_{a} \ket{a}\bra{a}_{A'} \otimes \rho_{B|a} \bigg) \nonumber \\
&=& {\mathcal{I}}\bigg( \sum_{a}p_{a} \ket{a}\bra{a}_{A'} \otimes \rho_{B|a}\bigg)  = {\mathcal{I}}\left( \tilde{\Pi}_{A}\left[ \rho_{AB}\right]\right),
\end{eqnarray}
where in the second line we use the joint entropy theorem~\cite{nielsen2010quantum}, and in the last  we use Eq.~(\ref{Equation:PostLGMA}).% and in the fourth equality we use the fact that $\mbox{Tr}_{A}\left(\sum_{i}p_{i} \ket{i}\bra{i}_{A} \otimes \rho_{B|i}\right) = \rho_{B}$ and $\mbox{Tr}_{B}\left(\sum_{i}p_{i} \ket{i}\bra{i}_{A} \otimes \rho_{B|i}\right) = \sum_{i}p_{i} \ket{i}\bra{i}_{A}$.??

The two-sided informational measure based upon mutual information has also been extensively studied in the literature. In~\cite{luo2008quantum}, a modification of $Q_{AB}^{I_{\mathcal{I}}}(\rho_{AB})$ was considered by omitting the optimisation over local measurements and called the ``measurement induced disturbance''. Local projections fixed in the respective eigenbases of the reduced states were used instead in its definition, i.e.~if $\rho_{A} = \sum_{a}p_{a}\ket{a}\bra{a}_{A}$ and $\rho_{B} = \sum_{b} p_{b} \ket{b}\bra{b}_{B}$ then the chosen LPM would consist of projectors $\{(\pi_{A})_{a} \otimes (\pi_{A})_{b}\}$ with $(\pi_{A})_{a}[\rho] = \ket{a}\bra{a}_{A} \rho \ket{a}\bra{a}_{A}$ and $(\pi_{B})_{b}[\rho] = \ket{b}\bra{b}_{B}\rho \ket{b}\bra{b}_{B}$. The motivation behind this choice is so that the local projection leaves the reduced states of $\rho_{AB}$ invariant, simplifying the measurement induced disturbance to the difference in global von Neumann entropies and thus leading to an upper bound to $Q_{AB}^{I_{\mathcal{S}}}(\rho_{AB})$. However, ignoring the optimisation over local measurements generally causes an overestimation of the QCs and also gives rise to pathologies such as the measurement induced disturbance being undefined on states whose marginals have a degenerate spectrum \cite{wu2009correlations}, or even approaching a maximum on some classical-classical states where QCs  vanish instead~\cite{girolami2011faithful}. To overcome this problem, the  measure $Q_{AB}^{I_{\mathcal{I}}}(\rho_{AB})$ including an optimisation over LPMs on $A$ and $B$ was studied in~\cite{girolami2011faithful} as the ``ameliorated measurement induced disturbance'', while the more general LGM based measure  $Q_{AB}^{\tilde{I}_{\mathcal{I}}}(\rho_{AB})$, sometimes referred to as ``symmetric quantum discord'' had been considered already  in~\cite{piani2008no,wu2009correlations}. In particular, %in the operational context of locking of classical correlations (as discussed in Section~\ref{Section:Applications-Locking}),
the authors of \cite{terhal2002entanglement,divincenzo2004locking,hall2006maximum} defined the (two-sided) ``classical mutual information''
\begin{equation}\label{Equation:ClassicalCorrelationsAB}
\tilde{\mathcal{J}}_{AB}(\rho_{AB}) := \sup_{\tilde{\Pi}_{AB}}{\cal I}\big(\tilde{\Pi}_{AB}[\rho_{AB}]\big),
\end{equation}
as the maximum classical correlations obtainable from a quantum state $\rho_{AB}$ by purely local processing. The quantity in Eq.~(\ref{Equation:ClassicalCorrelationsAB}) amounts to the generalisation of  Eq.~(\ref{Equation:ClassicalCorrelations}) when LGMs $\tilde{\Pi}_{AB} = \tilde{\Pi}_A\otimes \tilde{\Pi}_B$ on both $A$ and $B$ are considered as in Eq.~(\ref{Equation:PostLGMAB}). In this way, the two-sided mutual information based measure of QCs can then be written simply as
\begin{equation}\label{Equation:MIAB}
Q_{AB}^{\tilde{I}_{\mathcal{I}}}(\rho_{AB})= {\cal I}(\rho_{AB}) - \tilde{{\cal J}}_{AB}(\rho_{AB})\,,
\end{equation}
that is, as the difference between total and classical mutual information, in analogy to Eq.~(\ref{Equation:QuantumDiscord}).

Non-negativity of $Q_{A}^{\tilde{I}_{\mathcal{I}}}(\rho_{AB})$ follows directly from the strong subadditivity of the von Neumann entropy \cite{wehrl1978general}  (see also \cite{ollivier2001quantum,datta2008studies,datta2010a}). This implies non-negativity of all the other mutual information based informational measures of QCs, see Eq.~(\ref{Equation:InequalityWheel}) in the following for clarification.

%??Non-negativity of the mutual information based informational measures is ensured by the contractivity of the mutual information under any quantum operation??,
%??Non-negativity of the mutual information based informational measures was shown in~\cite{datta2008studies,wu2009correlations}??, and
As all informational measures, Requirements~(\ref{Requirements:Faithfulness}) and (\ref{Requirements:UnitaryInvariance}) are satisfied. Furthermore, Requirement~(\ref{Requirements:UnmeasuredParty}) was proven to hold in~\cite{streltsov2011linking,piani2012problem}, and Requirement~(\ref{Requirements:EntanglementPure}) holds as well, since for pure states ${\ket{\psi}_{AB}}$ we have~~\cite{bengtsson2007geometry,spehner2014quantum,lang2011entropic}
\begin{equation}
Q_{A}^{I_{\mathcal{I}}}({\ket{\psi}_{AB}})=Q_{AB}^{I_{\mathcal{I}}}({\ket{\psi}_{AB}})=Q_{A}^{\tilde{I}_{\mathcal{I}}}({\ket{\psi}_{AB}})=Q_{AB}^{\tilde{I}_{\mathcal{I}}}({\ket{\psi}_{AB}})=\mathcal{S}(\rho_{A})=\mathcal{S}(\rho_{B}),
\end{equation}
which is the entanglement entropy. However, it is still an open question whether Requirement~(\ref{Requirements:LCPO}) is obeyed for this class of measures.

Regarding computability, the main obstacle in calculation of the mutual information based informational measures of QCs is the non-trivial optimisation over LGMs or LPMs. Indeed, the computational complexity of this optimisation is NP-complete~\cite{huang2014computing}, falling into the hardest class of problems within NP and with no efficient way currently known to provide a solution, so that the running time of any present algorithm grows exponentially with the dimension of the Hilbert space. Nevertheless, in the LGM case it is sufficient to optimise only over local POVMs that are extremal and of rank one (see Section~\ref{Section:Facts-Measuring}), as proven in~\cite{datta2008studies,hamieh2004positive,d2005classical}. Notice that, unlike the case of the von Neumann $I_{\mathcal{S}}$ based informational measures in Section~\ref{Section:Measures-Informational-Entropy}, the mutual information $I_{\mathcal{I}}$ based measures defined in this section can be optimised in principle over \emph{all} LGMs or LPMs, without the need to restrict their definitions a priori, yet the optimal measurements always happen to consist of rank one operators~\cite{datta2008studies,lang2011entropic,wu2009correlations}.

Comparisons between minimisation over LGMs and LPMs have been carried out, in particular for the quantum discord, and even for two-qubit states there are many cases where $Q_{A}^{\tilde{I}_{\mathcal{I}}}(\rho_{AB}) < Q_{A}^{I_{\mathcal{I}}}(\rho_{AB})$, although the two quantities are typically quite close~\cite{galve2011orthogonal,lang2011entropic,chen2011quantum,shi2012optimal}. Nevertheless, there has been some analytical progress in the calculation of these quantities, including: the evaluation of $Q_{A}^{\tilde{I}_{\mathcal{I}}}(\rho_{AB})$ in~\cite{shi2011quantum} for all two-qubit states of rank $2$, and in~\cite{ali2010quantum,chen2011quantum,lu2011optimal,shi2011geometric,shi2012optimal,chitambar2012quantum,sabapathy2013quantum} for certain two-qubit X states (note that the original result of~\cite{ali2010quantum} was claimed to apply to all two-qubit X states, but this claim was later shown to be incorrect in~\cite{chen2011quantum,lu2011optimal,huang2013quantum}), and in~\cite{chitambar2012quantum} for highly symmetric two-qudit states; and the evaluation of $Q_{A}^{I_{\mathcal{I}}}(\rho_{AB})$ in~\cite{luo2008quantum} for two-qubit Bell diagonal states (in which case $Q_{A}^{I_{\mathcal{I}}}(\rho_{AB})=Q_{AB}^{I_{\mathcal{I}}}(\rho_{AB})$), in~\cite{fanchini2010non,li2011quantum,yurischev2015quantum,jing2016analytical} for two-qubit X states, in~\cite{vinjanampathy2012quantum} for qubit-qudit extended X states, in~\cite{rossignoli2012measurements} for certain two-qutrit states, in~\cite{ciliberti2010quantum,ciliberti2013discord} for two-qubit reduced states  of ground states of spin chains, and in~\cite{girolami2011quantum,wu2015quantum} by providing an efficient numerical algorithm for general two-qubit states. The quantum discord $Q_{A}^{\tilde{I}_{\mathcal{I}}}(\rho_{AB})$ has been generalised to continuous variable Gaussian states in \cite{giorda2010gaussian,adesso2010quantum}, and a closed computable formula obtained for all two-mode Gaussian states with optimisation restricted to Gaussian LGMs \cite{adesso2010quantum}; the latter measurements turn out to be in fact optimal (even among non-Gaussian measurements) for a large class of two-mode Gaussian states \cite{pirandola2014optimality}.
%An operational interpretation for discord in a class of Gaussian states is mentioned in Section~\ref{Section:Applications-Cryptography} in the context of quantum cryptography.
Two-sided measures of QCs based on the mutual information $Q_{AB}^{\tilde{I}_{\mathcal{I}}}(\rho_{AB})$ have also been studied for Gaussian states \cite{mista2011measurement}, showing that in this case Gaussian LGMs are generally not optimal.

We will detail some of the most concrete and direct operational interpretations of the quantum discord and other mutual information based informational measures of QCs in Section~\ref{Section:Applications}. We will also discuss the curious connection between $Q_{A}^{I_{\mathcal{I}}}(\rho_{AB})$ and the entanglement activation type measures in Section \ref{Section:Measures-Entanglement}. Finally, let us mention an interesting and useful inequality involving the quantum discord $Q^{\tilde{I}_{\mathcal{I}}}_A$ and the entanglement of formation $E^f$ (see Table~\ref{Table:Entanglement}), in arbitrary tripartite states $\rho_{ABC}$ (with equality on pure states), given by \cite{koashi2004monogamy,fanchini2011conservation}
\begin{equation}\label{Equation:Koashi}
Q^{\tilde{I}_{\mathcal{I}}}_A(\rho_{AB}) \geq {\cal S}(\rho_A) - {\cal S}(\rho_{AB}) + E^f_{B:C}(\rho_{BC})\,.
\end{equation}

\subsubsection{Hierarchy of informational measures and their regularisation}\label{Section:Measures-Informational-Regularisation}
The following wheel of inequalities summarises the hierarchical structure of the informational measures of QCs introduced so far, when all evaluated on the same arbitrary bipartite state $\rho_{AB}$:
\begin{equation}\label{Equation:InequalityWheel}
\begin{array}{ccccccc}
Q_{A}^{\tilde{I}_{\mathcal{S}}} & & & \leq & & & Q_{AB}^{\tilde{I}_{\mathcal{S}}} \\
 & \rotatebox[origin=c]{135}{$\leq$} & & & & \rotatebox[origin=c]{45}{$\leq$} & \\
 & & Q_{A}^{\tilde{I}_{\mathcal{I}}} & \leq & Q_{AB}^{\tilde{I}_{\mathcal{I}}} & & \\
 &&&&&& \\
 \rotatebox[origin=c]{270}{$\leq$} & & \rotatebox[origin=c]{270}{$\leq$} & & \rotatebox[origin=c]{270}{$\leq$} & & \rotatebox[origin=c]{270}{$\leq$} \\
 &&&&&& \\
 & & Q_{A}^{I_{\mathcal{I}}} & \leq & Q_{AB}^{I_{\mathcal{I}}} & & \\
  & \rotatebox[origin=c]{225}{$\leq$} & & & & \rotatebox[origin=c]{-45}{$\leq$} & \\
Q_{A}^{I_{\mathcal{S}}} & & & \leq & & & Q_{AB}^{I_{\mathcal{S}}} \\
\end{array}
\end{equation}
The vertical inequalities arise because an optimisation over LGMs necessarily includes an optimisation over LPMs. The horizontal inequalities are shown in \cite{wu2009correlations,horodecki2005local}. % [need to check top one, is it obvious].
Finally, the radial inequalities are explored in~\cite{lang2011entropic}.%????
%The two circular sets of inequalities arise from the definitions of Eq.~(\ref{Equation:InformationalMeasuresPOVM}) and (\ref{Equation:InformationalMeasures}) because LGM optimisation includes optimisation over LPMs, and optimisation over measurements on both subsystems $A$ and $B$ includes optimisation over just subsystem $A$. The radial inequalities are explored in~\cite{lang2011entropic}.????

%\subsubsection{Regularisation}\label{Section:Measures-Informational-Regularisation}

From an information theoretic perspective, it is often  relevant to consider also an asymptotic scenario in which  many independent and identically distributed (i.i.d.) copies of a quantum state are available. Given a real, non-negative function $f(\rho)$ on the set of quantum states, the corresponding \emph{regularisation} (if the limit exists) is defined as
\begin{equation}
f(\rho)_{\infty} := \lim_{N \rightarrow \infty} \frac{1}{N} f(\rho^{\otimes N}).
\end{equation}
Regularisation has been used extensively in entanglement theory \cite{horodecki2009quantum}, in particular to define additive entanglement measures (like the entanglement cost) from non-additive ones (like the entanglement of formation), since for any $f(\rho)$ the regularised $f(\rho)_{\infty}$ is by construction additive on tensor product states.

One can then define accordingly regularised versions of all of the measures of QCs mentioned in this review.
In particular, for the informational type of measures, a remarkable convergence occurs upon regularisation, collapsing the whole left half of the hierarchy in Eq.~(\ref{Equation:InequalityWheel}). Specifically, it was shown in~\cite{devetak2005distillation} that the regularisations of LGM and LPM versions of both the mutual information and the von Neuman entropy based informational QCs measures all reduce to the same quantity, i.e.~\begin{equation}\label{Equation:RegularisedDiscord}
Q_{A}^{\tilde{I}_{\mathcal{I}}}(\rho_{AB})_{\infty} = Q_{A}^{I_{\mathcal{I}}}(\rho_{AB})_{\infty} = Q_{A}^{\tilde{I}_{\mathcal{S}}}(\rho_{AB})_{\infty} = Q_{A}^{I_{\mathcal{S}}}(\rho_{AB})_{\infty}\,.
\end{equation}

This means in particular that the regularised one-way information deficit and the regularised one-sided quantum discord coincide in general. The quantity in Eq.~(\ref{Equation:RegularisedDiscord}) can equivalently be written as
\begin{equation}\label{Equation:RegularisedDiscordJ}
Q_{A}^{\tilde{I}_{\mathcal{I}}}(\rho_{AB})_{\infty} = \mathcal{I}(\rho_{AB})-\tilde{\mathcal{J}}_A(\rho_{AB})_{\infty}\,,
\end{equation}
which is obtained from Eq.~(\ref{Equation:QuantumDiscord}) upon regularisation, taking into account that the mutual information is already additive, $\mathcal{I}(\rho_{AB})_\infty = \mathcal{I}(\rho_{AB})$.
The complementary quantity $\tilde{\mathcal{J}}_A(\rho_{AB})_{\infty}$ in Eq.~(\ref{Equation:RegularisedDiscord}), which is the regularised version of the one-sided measure of classical correlations  \cite{henderson2001classical} defined in Eq.~(\ref{Equation:ClassicalCorrelations}), is known as ``distillable common randomness'' \cite{devetak2004distilling,horodecki2013quantumness}.

\subsubsection{Other informational measures}

There are natural extensions of the two informational measures suggested in Table~\ref{Table:InformationalQuantifiers}. Generalisations of the von Neumann entropy are considered in~\cite{rossignoli2010generalized}, in particular the R{\'{e}}nyi entropies
\begin{equation}
\mathcal{R}_\alpha(\rho):=-\frac{\log \mbox{Tr}\left(\rho^{\alpha}\right)}{\alpha-1}
\end{equation}
and the Tsallis entropies
\begin{equation}
\mathcal{T}_\alpha(\rho):=\frac{1-\mbox{Tr}\left(\rho^{\alpha}\right)}{\alpha-1},
\end{equation}
with, as $\alpha \rightarrow 1$,
\begin{equation}
\mathcal{R}_1(\rho) = \mathcal{T}_1(\rho) = \mathcal{S}(\rho).
\end{equation}
These generalised entropies can give extensions of the von Neumann entropy based informational measures of QCs by choosing in Eqs.~(\ref{Equation:InformationalMeasures}) the informational quantities $I_{\mathcal{R}_\alpha}  = -\mathcal{R}_\alpha$ and $I_{\mathcal{T}_\alpha} = -\mathcal{T}_\alpha$. Indeed, in~\cite{rossignoli2010generalized,rossignoli2011quantum} the authors consider $Q_{A}^{I_{\mathcal{R}_\alpha}}(\rho_{AB})$, $Q_{AB}^{I_{\mathcal{R}_\alpha}}(\rho_{AB})$, $Q_{A}^{I_{\mathcal{T}_\alpha}}(\rho_{AB})$ and $Q_{AB}^{I_{\mathcal{T}_\alpha}}(\rho_{AB})$, and even more general forms of entropy. These quantities are all non-negative because a complete rank one LPM always increases the R{\'{e}}nyi and Tsallis entropies when the result is not known. Furthermore, $I_{\mathcal{R}_\alpha}(\rho_{AB})$ and $I_{\mathcal{T}_\alpha}(\rho_{AB})$ are invariant under local unitaries and hence Requirements~(\ref{Requirements:Faithfulness}) and (\ref{Requirements:UnitaryInvariance}) hold. For pure states ${\ket{\psi}_{AB}}$, one has
\begin{eqnarray}
Q_{A}^{I_{\mathcal{R}_\alpha}}({\ket{\psi}_{AB}}) &=& Q_{AB}^{I_{\mathcal{R}_\alpha}}({\ket{\psi}_{AB}}) =  \mathcal{R}_\alpha(\rho_{A}) = \mathcal{R}_\alpha(\rho_{B})\,, \\
Q_{A}^{I_{\mathcal{T}_\alpha}}({\ket{\psi}_{AB}}) &=& Q_{AB}^{I_{\mathcal{T}_\alpha}}({\ket{\psi}_{AB}}) =  \mathcal{T}_\alpha(\rho_{A}) = \mathcal{T}_\alpha(\rho_{B})\,,
\end{eqnarray}
which are both valid generalisations of the entanglement entropy~\cite{bengtsson2007geometry}, thus validating Requirement~(\ref{Requirements:EntanglementPure}). However, Requirement~(\ref{Requirements:UnmeasuredParty}) does not hold for the family of informational measures corresponding to the Tsallis entropies for any choice of $\alpha \neq 1$ \cite{boysk2016unified}, because these measures can arbitrarily change with the addition and removal of an impure ancilla, as in Eq.~(\ref{Equation:HSAncillaProblem}). Indeed, when $\alpha = 2$ the Tsallis entropies reduces to the linear entropy $\mathcal{T}_2(\rho)=1 - \mbox{Tr}(\rho^{2})$ and it holds that~\cite{bellomo2012dynamics}
\begin{equation}
Q_{A}^{I_{\mathcal{T}_2}}(\rho_{AB}) = Q_{A}^{M_{2}}(\rho_{AB}) = Q_{A}^{G_{2}}(\rho_{AB}),
\end{equation}
which is the Hilbert-Schmidt based geometric measure of QCs. It was then suggested in~\cite{boysk2016unified} to rescale these measures by a type of generalised purity, analogous to the suggestion of~\cite{tufarelli2013geometric}, but it is still not known whether Requirement~(\ref{Requirements:UnmeasuredParty}) may be recovered in this way. The family of informational measures corresponding to the R{\'{e}}nyi entropies are left unchanged by the addition and removal of an impure ancilla~\cite{canosa2015quantum}, but it is still unknown whether the more general Requirement~(\ref{Requirements:UnmeasuredParty}) holds. Finally, it is not known whether Requirement~(\ref{Requirements:LCPO}) is obeyed either.

One can also look at generalisations of the mutual information adopted as informational quantifier $I_{\mathcal{I}}$. Recall that the mutual information ${\mathcal{I}}(\rho_{AB})$, defined in Eq.~(\ref{Equation:MutualInformation}), can be equivalently rewritten as the distance from $\rho_{AB}$ to the set of product states according to the relative entropy \cite{vedral2002role,modi2010unified}, i.e.~\begin{equation}
{\mathcal{I}} (\rho_{AB}) = \inf_{\sigma_{AB} \in \mathscr{P}_{AB}} D_{\text{RE}}(\rho_{AB},\sigma_{AB}),
\end{equation}
where $\mathscr{P}_{AB}$ is the set of product states, and the minimisation is achieved in general by $\sigma_{AB}=\rho_A \otimes \rho_B$, the product of the marginals of $\rho_{AB}$.
One could instead consider in Eq.~(\ref{Equation:InformationalMeasures}) the informational quantities
\begin{equation}
\begin{aligned}
I_{\mathcal{I}^\mathcal{R}_\alpha} (\rho_{AB}) &= \inf_{\chi_{AB} \in \mathcal{P}} D_{\mathcal{R}_\alpha}(\rho_{AB},\chi_{AB}), \\
I_{\mathcal{I}^\mathcal{T}_\alpha} (\rho_{AB}) &= \inf_{\chi_{AB} \in \mathcal{P}} D_{\mathcal{T}_\alpha}(\rho_{AB},\chi_{AB}),
\end{aligned}
\end{equation}
arising from the sandwiched relative R{\'{e}}nyi and Tsallis entropies defined in Eqs.~(\ref{Equation:SandRenyi}) and (\ref{Equation:SandTsallis}), to define generalised mutual information based measures of QCs $Q_{A}^{I_{\mathcal{I}^\mathcal{R}_\alpha}}(\rho_{AB})$ and $Q_{A}^{I_{\mathcal{I}^\mathcal{T}_\alpha}}(\rho_{AB})$ with optimisation over LPMs, as was done in~\cite{misra2015quantum}. The corresponding two-sided LPM and one-sided and two-sided LGM measures may be defined as well. Being non-negative, zero on classical states and local unitarily invariant, all these quantities obey Requirements~(\ref{Requirements:Faithfulness}) and (\ref{Requirements:UnitaryInvariance}). However, it is unknown whether Requirements~(\ref{Requirements:EntanglementPure}), (\ref{Requirements:UnmeasuredParty}), and (\ref{Requirements:LCPO}) hold, and one may need to check in each case for particular values of $\alpha$ (beyond the standard $\alpha \rightarrow 1$).

We finally highlight that similar approaches to defining measures of QCs by generalising entropies have been reported~\cite{berta2015renyi,chi2013generalized,jurkowski2013quantum,majtey2012new,adesso2012measuring,costa2013bayes}.

%Another possibility is to consider the quantum conditional entropy
%\begin{equation}\label{Equation:ConditionalVNEntropies}
%\mathcal{S}_{B|A}(\rho_{AB}) = \mathcal{S}(\rho_{AB}) - \mathcal{S}(\rho_{A}).
%\end{equation}
%as a different generalisation of the conditional Shannon entropy in Eq.~(\ref{Equation:ConditionalShannonEntropies}). This quantity can be negative for %some quantum states (for example, any pure entangled state). The informational quantity $I_{\mathcal{S}_{B|A}}(\rho_{AB}) = - %\mathcal{S}_{B|A}(\rho_{AB})$ was suggested in~\cite{lang2011entropic} to define $Q_{AB}^{\tilde{I}_{\mathcal{S}_{B|A}}}(\rho_{AB})$ as a measure of QCs %when optimised over rank one LGMs. Although the LGM acts on both subsystems $A$ and $B$, this measure is asymmetric because %$\mathcal{S}_{B|A}(\rho_{AB}) \neq \mathcal{S}_{A|B}(\rho_{AB})= \mathcal{S}(\rho_{AB}) - \mathcal{S}(\rho_{B})$. It also holds that %$Q_{A}^{\tilde{I}_{\mathcal{S}_{B|A}}}(\rho_{AB})=Q_{A}^{\tilde{I}_{\mathcal{I}}}(\rho_{AB})$ and %$Q_{A}^{I_{\mathcal{S}_{B|A}}}(\rho_{AB})=Q_{A}^{I_{\mathcal{I}}}(\rho_{AB})$??.

\subsection{Entanglement activation measures}\label{Section:Measures-Entanglement}
The entanglement activation approach to QCs, leading to the measures in Eqs.~(\ref{Equation:EntanglementActivationMeasures}), was defined almost simultaneously in \cite{piani2011all} and \cite{streltsov2011linking}. In particular, \cite{piani2011all} formalised the problem as an adversarial game, named the ``activation protocol'', in which the goal is to establish entanglement between a given principal system and a register of ancillary systems against an adversary who could scramble the local bases on the principal system before the control-\textsc{not} interactions. The authors focused on fully symmetric measures of QCs (i.e., generalisations of the two-sided measures to a multipartite setting) as prerequisites for the necessary activation of bipartite entanglement between system and ancillae in the output state of the protocol. Their approach was later generalised to one-sided and generally partial quantumness measures in \cite{piani2012quantumness}.  Independently, \cite{streltsov2011linking} studied an essentially equivalent protocol, focusing primarily on the one-sided case, yet providing the fundamental interpretation in terms of the von Neumann model of local measurements, as described in Section~\ref{Section:Facts-Entanglement}.

In the following, we cover a few particular cases of entanglement activation measures, corresponding to some relevant choices of $E^\zeta$ as given in Table~\ref{Table:Entanglement}. The case of the entaglement of formation $E^f$ is not treated explicitly here, but details are available in \cite{modi2012classical}.

\subsubsection{Distillable entanglement and relative entropy}
Both \cite{piani2011all} and \cite{streltsov2011linking} reached the conclusion that the entanglement activation measures of QCs based upon relative entropy $E^{\text{RE}}$ and distillable entanglement $E^d$ (see Tables~\ref{Table:Distances} and \ref{Table:Entanglement}) coincide with each other and reduce to their geometric counterparts, see Eq.~(\ref{Equation:RelativeEntropyMeasuresGECoincidence}), and to the informational quantum deficits, as amply discussed in Section~\ref{Section:Measures-Geometric-RE}. This is a consequence of the fact that the pre-measurement states as defined in the entanglement activation approach belong to the special class of so-called maximally correlated states, for which $E^d=E^{\text{RE}}$ \cite{hiroshima2004finding}; more generally, this observation is crucial to establish Requirement~(\ref{Requirements:Faithfulness}) for all the QCs measures $Q_{A}^{E^\zeta}$ and $Q_{AB}^{E_\zeta}$ even when using entanglement monotones $E^\zeta$ that may vanish on some non-separable states, such as the distillable entanglement $E^d$.

It is interesting to note that the entanglement activation approach can also be adapted to recover the original measure of quantum discord $Q_{A}^{I_{\mathcal{I}}}(\rho_{AB})$ optimised over LPMs, as anticipated in Section~\ref{Section:Measures-Informational-MutualInformation} \cite{streltsov2011linking}. This requires a modification of the one-sided definition in Eqs.~(\ref{Equation:EntanglementActivationMeasures}), to consider the minimisation of the {\it partial} entanglement in the pre-measurement state,
\begin{equation}\label{Equation:EntanglementActivationPartial}
Q_{A}^{\Delta E^{\zeta}}(\rho_{AB}) := \inf_{\{\ket{a}_A\}} \left[E^{\zeta}_{AB:A'}\left(\rho'^{\{\ket{a}_A\}}_{ABA'}\right)-E^{\zeta}_{A:A'}\left(\mbox{Tr}_B \big(\rho'^{\{\ket{a}_A\}}_{ABA'}\big)\right)\right],
\end{equation}
that is, the difference given by the entanglement created between the whole system $AB$ and the ancilla $A'$, minus the entanglement created between the probed subsystem $A$ and the ancilla $A'$ in their reduced state, discarding the other subsystem $B$. It then turns out that \cite{streltsov2011linking}
\begin{equation}\label{Equation:DiscordPartialEntanglement}
Q_{A}^{\Delta E^{\zeta}}(\rho_{AB}) = Q_{A}^{I_{\mathcal{I}}}(\rho_{AB})\,,
\end{equation}
with $E^\zeta$ being either $E^{\text{RE}}$ or $E^{d}$, thus providing an alternative interpretation to the quantum discord \cite{ollivier2001quantum}.

Note that one could define more general measures of QCs $Q_{AB}^{\Delta E^{\zeta}}$ based on the partial entanglement activation approach of Eq.~(\ref{Equation:EntanglementActivationPartial}) by considering other choices for $E^\zeta$, however in this case one needs $E^\zeta$ to be convex and non-increasing on average under stochastic LOCC (which are stronger requirements than standard LOCC monotonicity) \cite{plenio2007an}, in order to generate valid measures of QCs, in particular to fulfill Requirement~(\ref{Requirements:UnmeasuredParty}) \cite{streltsov2011linking}.

\subsubsection{Negativity of quantumness}\label{Section:Measures-NoQ}
The ``negativity of quantumness'' was defined in \cite{piani2011all} and studied in detail in \cite{nakano2013negativity}. One-sided and two-sided versions of this measure, $Q_{A}^{E^{N}}(\rho_{AB})$ and $Q_{AB}^{E^{N}}(\rho_{AB})$, are obtained by choosing the entanglement negativity $E^N$ (see Table~\ref{Table:Entanglement}) \cite{zyczkowski1998volume,vidal2002computable} in Eqs.~(\ref{Equation:EntanglementActivationMeasures}). As already noticed in \cite{piani2011all,nakano2013negativity}, the negativity of quantumness amounts to the minimum coherence in all local bases quantified by the $\ell_1$ norm (see Table~\ref{Table:Coherence}). We can formalise this equivalence using the notation of Section~\ref{Section:Measures-Types-Coherence}. Namely, for a state $\rho_{AB}$ and fixing local orthonormal reference bases $\{\ket{a}_A\}$ and $\{\ket{b}_B\}$, let us define
\begin{equation}
\label{Equation:Coherencel1}
\begin{aligned}
C^{\ell_1\;{\{\ket{a}_A\}}}_{A}(\rho_{AB}) &:= \sum_{ij} \|\, {}_A\!\bra{a_i} \rho_{AB} \ket{a_j}\!{}_A\,\|_1-1,\\
C^{\ell_1\;{\{\ket{a}_A, \ket{b}_B\}}}_{AB}(\rho_{AB}) &:=  \|\rho_{AB}\|_{\ell_1}^{\{\ket{a}_A \otimes \ket{b}_B\}}-1\,,
\end{aligned}
\end{equation}
where we have explicitly indexed the local basis $\{\ket{a}_A\} \equiv \{\ket{a_i}_A\}_{i=1}^{d_A}$ in the one-sided case, while the notation in the two-sided case indicates that the $\ell_1$ norm is taken for the matrix representation of $\rho_{AB}$ with respect to the product basis $\{\ket{a} \otimes \ket{b}\}$ (recall that the $\ell_1$ norm, or taxicab norm, is basis dependent).
The minimisation of Eqs.~(\ref{Equation:Coherencel1}) over all local reference bases as in Eq.~(\ref{Equation:CoherenceMeasures}) defines then the one-sided and two-sided $\ell_1$ coherence based measures of QCs, which coincide with the corresponding versions of the negativity of quantumness \cite{nakano2013negativity},
\begin{equation}\label{Equation:NakanoCoincidence}
\begin{aligned}
Q_{A}^{C^{\ell_1}}(\rho_{AB})&:= \inf_{\{\ket{a}_A\}} {C^{\ell_1\;{\{\ket{a}_A\}}}_{A}}(\rho_{AB}) = Q_{A}^{E^{N}}(\rho_{AB})\,, \\
Q_{AB}^{C^{\ell_1}}(\rho_{AB}) &:= \inf_{\{\ket{a}_A, \ket{b}_B\}} {C^{\ell_1\;{\{\ket{a}_A, \ket{b}_B\}}}_{AB}}(\rho_{AB})=Q_{AB}^{E^{N}}(\rho_{AB})\,.
\end{aligned}
\end{equation}

The two-sided negativity of quantumness $Q_{AB}^{E^{N}}(\rho_{AB})$ can also be interpreted as a geometric measure as in Eq.~(\ref{Equation:GeometricMeasures}) and as a measurement induced geometric measure as in Eq.~(\ref{Equation:MeasurementInducedDisturbances}),
\begin{equation}\label{Equation:NakanoMore}
\begin{aligned}
Q_{AB}^{E^N}(\rho_{AB}) &= Q_{AB}^{G_{\ell_1}}(\rho_{AB}) = \inf_{\chi_{AB} \in \mathscr{C}_{AB}} D_{\ell_1}(\rho_{AB}, \chi_{AB})\,,  \\
Q_{AB}^{E^N}(\rho_{AB}) &= Q_{AB}^{M_{\ell_1}}(\rho_{AB}) = \inf_{\pi_{AB}} D_{\ell_1}\big(\rho_{AB}, \pi_{AB} [\rho_{AB}]\big)\,,
\end{aligned}\end{equation}
where we adopt the $\ell_1$ distance \begin{equation}
D_{\ell_1}(\rho, \sigma) := \|\rho - \sigma\|_{\ell_1}^{\{\ket{i}\}}\,,
 \end{equation}
 evaluated with respect to the basis $\{\ket{a}_A \otimes \ket{b}_B\}$, which corresponds in the first relation to the eigenbasis of the classical-classical state $\chi_{AB}$, and in the second relation to the product of local bases associated with the complete rank one LPM
 %$\pi_{AB}$
 as defined in Eq.~(\ref{Equation:PostLPMAB}).

In the special case of subsystem $A$ being a qubit, the one-sided negativity of quantumness $Q_{AB}^{E^{N}}(\rho_{AB})$ relates instead to the geometric, measurement induced, and unitary response measures based on the trace distance \cite{nakano2013negativity}, as summarised in Eq.~(\ref{Equation:NakanoLinks}).

Closed formulae for the negativity of quantumness of two-qubit Bell diagonal states and two-qudit highly symmetric states were obtained in \cite{nakano2013negativity}.
%In \cite{chaves2011noisy}, the operational role of the negativity of quantumness, thereby referred to as ``minimum entanglement potential'', was investigated  in the context of remote state preparation with noisy resources, as we mention in Section~\ref{Section:RSP}. Furthermore,
Two optical experiments recently investigated the entanglement activation framework \cite{adesso2014experimental,ciampini2015experimental}. In particular,  in \cite{adesso2014experimental} a procedure was devised and tested to verify the `if and only if' in Eq.~(\ref{Equation:ClassicalStateEntanglementDefinitionAB}) based on a finite net of data, and the quantitative equivalence $Q_{A}^{M_{1}}(\rho_{AB}) = Q_{A}^{G_{1}}(\rho_{AB})$ was demonstrated for two qubits. The experiment in \cite{ciampini2015experimental} started instead with a classically correlated state, in which QCs were then created by local noise (using dissipative non-LCPO maps \cite{ciccarello2012creating}), and finally activated into entanglement with ancillae through the protocol described in Section~\ref{Section:Facts-Entanglement}.  Both experiments measured QCs via the negativity of quantumness. Finally, the negativity of quantumness has been studied for Gaussian states in \cite{mista2014no}, where it was also shown that non-Gaussian pre-measurement interactions are necessary for the entanglement activation of QCs in Gaussian states. Alternative schemes inspired by the activation protocol have been proposed for continuous variable systems \cite{mazzola2011activating,farace2014steady}.

\subsection{Unitary response measures}\label{Section:Measures-Response}

The unitary response measures were first introduced in~\cite{giampaolo2007characterization,monras2011entanglement} to quantify entanglement. For pure states $\ket{\psi}_{AB}$, the entanglement of response was in fact defined as the distance from $\ket{\psi}_{AB}$ to its image through a least disturbing local unitary with spectrum given by the $d_A$-th complex roots of unity, as in Eq.~(\ref{Equation:EntanglementofResponse}).
 %This immediately entails that the unitary response measures of QCs defined in Eq.~(\ref{Equation:UnitaryResponseQuantifiers}) satisfy Requirement~(\ref{Requirements:EntanglementPure}).

\subsubsection{Hilbert-Schmidt distance}
The unitary response measure of QCs $Q_A^{U_2}$ was then defined by resorting to the squared Hilbert-Schmidt distance in~\cite{gharibian2012quantifying,giampaolo2013quantifying} and was shown to be equivalent to the Hilbert-Schmidt geometric measure of QCs $Q_A^{G_2}$ when subsystem $A$ is a qubit, as recalled in Eq.~(\ref{Equation:HSGeometricU2}). Consequently, due to the problem of Eq.~(\ref{Equation:HSAncillaProblem}) highlighted in~\cite{piani2012problem}, the response measure $Q_A^{U_2}$ does not satisfy Requirement~(\ref{Requirements:UnmeasuredParty}) and does not qualify as a valid measure of QCs.

\subsubsection{Trace distance}
The trace distance response measure of QCs $Q_A^{U_1}$ was studied in \cite{nakano2013negativity} and connected both to the trace distance geometric measure of QCs and to the negativity of quantumness defined via the entanglement activation approach, when the subsystem $A$ is a qubit, see Sections~\ref{Section:Measures-Geometric-Trace} and \ref{Section:Measures-NoQ}.

\subsubsection{Bures and Hellinger distance: Discord of response}
In~\cite{roga2014discord} the unitary response measures $Q_A^{U_\delta}$, therein referred to as ``discord of response'', were defined more generally as in Eq.~(\ref{Equation:UnitaryResponseQuantifiers}) by resorting to any contractive distance $D_\delta$, with a particular focus on the squared Bures distance (see Table~\ref{Table:Distances}). Such a study was then complemented in~\cite{roga2015geometric} with a comprehensive comparison between several distance-based measures of QCs, including notable advances for the evaluation of the unitary response measures $Q_A^{U_{\text{Bu}}}$ and $Q_A^{U_{\text{He}}}$ based respectively on the squared Bures and the squared Hellinger distances (see Table~\ref{Table:Distances}). In particular, for pure states $\ket{\psi}_{AB}$, it holds \cite{roga2015geometric}
\begin{eqnarray}
Q_A^{U_{\text{Bu}}}(\ket{\psi}_{AB}) &=& 2\left(1-\sqrt{1-E^{U_{{F}}}(\ket{\psi}_{AB})}\right)\,, \\
Q_A^{U_{\text{He}}}(\ket{\psi}_{AB}) &=& 2 E^{U_{{F}}}(\ket{\psi}_{AB})\,,
\end{eqnarray}
where \begin{equation}
E^{U_{{F}}}(\ket{\psi}_{AB}) = \inf_{{U}_A} \left[1-\big|\bra{\psi}(U_A \otimes \mathbb{I}_B)\ket{\psi}_{AB}\big|^2\right]
\end{equation}
is the entanglement of response defined as in Eq.~(\ref{Equation:EntanglementofResponse}) with the infidelity (see Table~\ref{Table:Distances}) chosen as distance function and with optimisation restricted to root-of-unity local unitaries $U_A$. In the case of mixed states, closed formulae for $Q_A^{U_{\text{Bu}}}$ and $Q_A^{U_{\text{He}}}$ were first derived in \cite{aaronson2013comparative,roga2014discord} for the special case of two-qubit Bell diagonal states. More generally, for arbitrary states $\rho_{AB}$ when $A$ is a qubit, these two measures can be linked  analytically to their geometric counterparts by the simple relation \cite{roga2015geometric}
\begin{equation}\label{Equation:Roga}
Q_A^{U_{\delta}}(\rho_{AB}) = 4 Q_A^{G_\delta}(\rho_{AB}) - \big[Q_A^{G_\delta}(\rho_{AB})\big]^2\,,
\end{equation}
with $\delta=\text{Bu}, \text{He}$. In particular, by chaining this with Eqs.~(\ref{Equation:HellingerGeometric}) and (\ref{Equation:HSGeometricMID}), we get the simple relation
\begin{equation}\label{Equation:HellingerResponse}
Q_A^{U_{\text{He}}}(\rho_{AB}) = 4 Q_A^{G_2}(\sqrt{\rho_{AB}}) = 4 Q_A^{M_2}(\sqrt{\rho_{AB}})\,,
\end{equation}
which shows that the unitary response measure of QCs based on the squared Hellinger distance is equivalent to the remedied Hilbert-Schmidt geometric measure defined in \cite{chang2013remedying}, hence is analytically computable for all states $\rho_{AB}$ when $A$ is a qubit. In this specific case, the Hellinger unitary response measure $Q_A^{U_{\text{He}}}(\rho_{AB})$ is also equivalent to the coherence based measure defined in terms of the Wigner-Yanase skew information $Q_A^{C^{\text{WY}}}(\rho_{AB})$, introduced in \cite{girolami2013characterizing} as ``local quantum uncertainty'' (see Section~\ref{Section:Measures-Asymmetry}). Namely,
\begin{equation}\label{Equation:HellingerResponseLQU}
Q_A^{U_{\text{He}}}(\rho_{AB}) = 2 Q_A^{C^\text{WY}}(\rho_{AB})\,,
\end{equation}
for all states $\rho_{AB}$ when $A$ is a qubit. In this case, such an observation independently provides a closed formula for $Q_A^{U_{\text{He}}}(\rho_{AB})$ \cite{girolami2013characterizing}, equivalent to the one given in \cite{chang2013remedying,roga2015geometric}.

The Bures and Hellinger unitary response measures have been studied for Gaussian states with the restriction to Gaussian unitaries in \cite{roga2015device}, within the operational context of quantum reading (see Section~\ref{Section:Applications-Discrimination}).

\subsubsection{Chernoff distance: Discriminating strength}
Finally, in~\cite{farace2014discriminating} the restriction to a harmonic spectrum was lifted by defining the ``discriminating strength'' as follows
\begin{equation}\label{Equation:DiscriminatingStrength}
Q^{U^\Gamma_{\text{C}}}_A(\rho_{AB}) := 1 - \max_{U_A^\Gamma} \text{C}\big(\rho_{AB}, U_A^\Gamma [\rho_{AB}]\big),
\end{equation}
where
\begin{equation}\label{Equation:ChernoffBound}
\text{C}(\rho,\sigma):=\min_{0\leq s \leq 1} \mbox{Tr}(\rho^s\sigma^{1-s})
\end{equation}
is a measure of indistinguishability between the states $\rho$ and $\sigma$, which coincides with the Uhlmann fidelity $F$ \cite{uhlmann1976transition} (see Table~\ref{Table:Distances}),
\begin{equation}\label{Equation:Fidelity}
F(\rho,\sigma) := \left[\mbox{Tr}\left(\sqrt{\sqrt{\sigma} \rho \sqrt{\sigma}}\right)\right]^2\,,
\end{equation}
when either $\rho$ or $\sigma$ is pure.
The quantity in Eq.~(\ref{Equation:ChernoffBound}) enters in the definition of the quantum Chernoff bound \cite{audenaert2007discriminating}
\begin{equation}\label{Equation:ProperChernoffBound}
\xi(\rho,\sigma):=-\log\text{C}(\rho,\sigma).
\end{equation}
In Eq.~(\ref{Equation:DiscriminatingStrength}), the optimisation is over all local unitaries $U_A^\Gamma$ on subsystem $A$ with non-degenerate spectrum $\Gamma$, as in Eq.~(\ref{Equation:UnitaryResponseQuantifiersGamma}). The discriminating strength, whose defining operational interpretation in the context of channel discrimination will be discussed in Section~\ref{Section:Applications-Discrimination}, can be thus regarded as a unitary response measure of QCs based on the Chernoff distance $D_\text{C}(\rho,\sigma):= 1-\text{C}(\rho,\sigma)$. The choice of the spectrum $\Gamma$ in order to maximise the measure $Q^{U^\Gamma_{\text{C}}}_A$ was analysed as well in \cite{farace2014discriminating}.  When subsystem $A$ is a qubit (in which case any spectrum leads to an equivalent measure up to a normalisation factor, hence the harmonic spectrum can be assumed without any loss of generality \cite{girolami2013characterizing}),  it was further shown in~\cite{farace2014discriminating} that the discriminating strength coincides up to a constant multiplicative factor with the local quantum uncertainty \cite{girolami2013characterizing}, and thus with the Hellinger unitary response measure of QCs, via Eq.~(\ref{Equation:HellingerResponseLQU}). In continuous variable systems, the discriminating strength was investigated in \cite{rigovacca2015gaussian} by restricting to a minimisation over Gaussian unitaries, and a closed expression was derived for a subclass of two-mode Gaussian states, exploiting the analytical formula for the quantum Chernoff distance between a pair of two-mode Gaussian states obtained in \cite{pirandola2008computable}.

At present, the status of Requirement~(\ref{Requirements:LCPO}) for all the above mentioned unitary response measures of QCs remains unknown.

\subsection{Coherence based measures}\label{Section:Measures-Coherence}

\subsubsection{Relative entropy}\label{Section:Measures-Coherence-RE}
As with many other resources in quantum information theory, the relative entropy can be used to define a coherence quantifier $C^{\text{RE}}$ which is a full monotone in any possible resource theory of coherence \cite{aaberg2006quantifying,baumgratz2014quantifying}, and is easily computable for all quantum states as indicated in Table~\ref{Table:Coherence}. Following the approach outlined in Section~\ref{Section:Measures-Types-Coherence}, we can define basis-dependent quantifiers of local coherence with respect to the relative entropy by adopting a geometric approach \cite{yao2015quantum,ma2016converting}, and simplify their expression by resorting to the equivalence with the measurement induced geometric approach \cite{modi2010unified}. We have then
\begin{equation}
\label{Equation:CoherenceREGbasis}
\begin{aligned}
C^{\text{RE}\;{\{\ket{a}_A\}}}_{A}(\rho_{AB}) &:= \inf_{\chi_{AB} \in \mathscr{I}^{\{\ket{a}_A\}}_A} D_{\text{RE}}(\rho_{AB}, \chi_{AB}) = D_{\text{RE}}\big(\rho_{AB}, \pi_A[\rho_{AB}]\big)\,, \\
C^{\text{RE}\;{\{\ket{a}_A, \ket{b}_B\}}}_{AB}(\rho_{AB}) &:= \inf_{\chi_{AB} \in \mathscr{I}^{\{\ket{a}_A,\ket{b}_B\}}_{AB}} D_{\text{RE}}(\rho_{AB}, \chi_{AB}) = D_{\text{RE}}\big(\rho_{AB}, \pi_{AB}[\rho_{AB}]\big)\,,
\end{aligned}\end{equation}
where, as usual, the LPMs $\pi_A$ and $\pi_{AB}$ project the subsystems in the chosen reference bases $\{\ket{a}_A, \ket{b}_B\}$ as in Eqs.~(\ref{Equation:PostLPMA}) and (\ref{Equation:PostLPMAB}).
The coherence based measures of QCs defined in terms of relative entropy are then obtained upon minimisation over said local bases as in Eqs.~(\ref{Equation:CoherenceMeasures}), and as anticipated return the relative entropy based geometric measures of QCs,
\begin{equation}\label{Equation:CoherenceMeasuresRE}\begin{aligned}
Q_{A}^{C^{\text{RE}}}(\rho_{AB}) &= \inf_{\{\ket{a}_A\}} {C^{{\text{RE}}\;{\{\ket{a}_A\}}}_{A}}(\rho_{AB}) = Q_A^{G_{\text{RE}}}(\rho_{AB})\,,\\
Q_{AB}^{C^{\text{RE}}}(\rho_{AB}) &= \inf_{\{\ket{a}_A, \ket{b}_B\}} {C^{{\text{RE}}\;{\{\ket{a}_A, \ket{b}_B\}}}_{AB}}(\rho_{AB})=Q_{AB}^{G_{\text{RE}}}(\rho_{AB})\,,
\end{aligned}\end{equation}
discussed in Section~\ref{Section:Measures-Geometric-RE} together with all their equivalent interpretations in terms of the other types of QCs quantifiers.

\subsubsection{Infidelity}
Exploiting the results of \cite{streltsov2011linking,streltsov2015measuring}, one can also derive coherence based measures of QCs in terms of infidelity (see Table~\ref{Table:Coherence}), which turn out to match their geometric counterparts, analogously to the case of relative entropy. Compactly, we have indeed
\begin{equation}\label{Equation:CoherenceMeasuresRE}\begin{aligned}
Q_{A}^{C^{F}}(\rho_{AB}) &:= 1-\sup_{\{\ket{a}_A\}} \,\,\,\, \sup_{\chi_{AB} \in \mathscr{I}^{\{\ket{a}_A\}}_A} {F}(\rho_{AB}, \chi_{AB}) = Q^{G_F}_A(\rho_{AB})\,,\\
Q_{AB}^{C^{F}}(\rho_{AB}) &:= 1-\sup_{\{\ket{a}_A, \ket{b}_B\}} \,\,\,\, \sup_{\chi_{AB} \in \mathscr{I}^{\{\ket{a}_A,\ket{b}_B\}}_{AB}} D_{F}(\rho_{AB}, \chi_{AB}) = Q^{G_F}_{AB}(\rho_{AB})\,,
\end{aligned}\end{equation}
where the infidelity based geometric measures are monotonically related to the corresponding measures defined in terms of squared Bures distance (see Table~\ref{Table:Distances}) and enjoy accordingly all their properties, discussed in Section~\ref{Section:Measures-Geometric-Bures}.

\subsubsection{$\ell_1$ norm}
The $\ell_1$ norm defines a very intuitive and easy to compute quantifier of coherence (see Table~\ref{Table:Coherence}), which is also a full monotone in all possible resource theories of quantum coherence \cite{baumgratz2014quantifying}. A consistent pathway to obtain QCs measures from the $\ell_1$ norm of coherence has been presented in Section~\ref{Section:Measures-NoQ}, where the ensuing coherence based quantifiers have been shown to reproduce exactly the entanglement activation measures based on negativity. In particular, the two-sided measure $Q_{AB}^{C^{\ell_1}}(\rho_{AB})$, alias the two-sided negativity of quantumness $Q_{AB}^{E^{N}}(\rho_{AB})$ \cite{nakano2013negativity}, exhibits a particularly appealing form, as it amounts to the total of the moduli of the off-diagonal elements of the density matrix $\rho_{AB}$ (traditionally referred to as ``coherences''), minimised over all local product bases. This is valid for arbitrarily large multipartite systems as well, with respect to the product of orthonormal bases on each subsystem \cite{piani2011all}.

\subsubsection{Asymmetry: Local quantum uncertainty and interferometric power}\label{Section:Measures-Asymmetry}
Beyond redefinition of existing measures, the coherence based approach can be used to define original measures of QCs of high operational significance, which do not follow straightforwardly from other approaches. This is the case in particular for two coherence quantifiers defined in the context of the resource theory of asymmetry, namely the Wigner-Yanase skew information and the quantum Fisher information (see Table~\ref{Table:Coherence}). Both quantities are monotones under phase covariant operations \cite{marvian2013the}, but not under more general incoherent operations \cite{baumgratz2014quantifying}, as proven in \cite{marvian2014extending,girolami2014observable} for the skew information, and in \cite{marvian2015quantum} for the quantum Fisher information. These two metrics belong to the family of monotone and Riemannian metrics discussed in Section~\ref{Section-Measures-Geometric-Other} with geodesics corresponding respectively to the Hellinger and the Bures distances (see Table~\ref{Table:Distances}) \cite{bengtsson2007geometry}.
In \cite{girolami2013characterizing} and \cite{girolami2014quantum}, valid measures of QCs have been successfully defined based on these two asymmetry monotones, as we now review within a unified approach.

Given a measure of coherence in the context of quantum asymmetry $C^{\eta\; \Gamma}(\rho,K^\Gamma)$, defined with respect to a fixed reference basis $\{\ket{i}\}$ identified as the eigenbasis of the observable $K^\Gamma = \sum_i  k_i \ket{i}\!\bra{i}$ with non-degenerate spectrum $\Gamma = \{k_i\}$, then  the corresponding one-sided measure of QCs of a bipartite state $\rho_{AB}$ can be defined adapting Eq.~(\ref{Equation:CoherenceMeasures}) as
\begin{equation}\label{Equation:AsymmetryMeasures}
Q_A^{C^{\eta}\; \Gamma}(\rho_{AB}) := \inf_{K_A^\Gamma} C^{\eta\; \Gamma}(\rho_{AB}, K^\Gamma_A \otimes \mathbb{I}_B)\,,
\end{equation}
where the minimisation is over all local observables $K^\Gamma_A$ on subsystem $A$ with non-degenerate spectrum $\Gamma$. Notice the resemblance with the case of unitary response measures in Eq.~(\ref{Equation:UnitaryResponseQuantifiersGamma}). Similarly to such a case, in fact, it is not immediate to extend Eq.~(\ref{Equation:AsymmetryMeasures}) to define a corresponding two-sided measure of QCs, as including an optimisation over observables $K^\Gamma_B$ on $B$ as well might nullify the resulting two-sided quantifier even on maximally entangled states, like in the instance of Eq.~(\ref{Equation:NoABU}).

%It is instructive to recall that the observable $K_A$ may be decomposed as $K^\Gamma_A = V_A \mbox{diag}(\Gamma) V_A^\dagger$, where $V_A$ identifies the reference basis and is varied over the special unitary group on $A$ in the optimisation of Eq.~(\ref{Equation:AsymmetryMeasures}), while the (fixed) spectrum $\Gamma$ rules the fine graining \cite{girolami2013characterizing}. For this reason, as discussed in Section~\ref{Section:Measures-Types-Unitary}, a preferred choice can be to restrict to the harmonic spectrum $\Gamma^{\star}$ given by the $d_A$-th complex roots of unity, in which case the superscript $\Gamma$ will be dropped from Eq.~(\ref{Equation:AsymmetryMeasures}); in particular, if $d_A=2$, this choice is effectively unique (up to an irrelevant rescaling factor). However we need not make such a restriction in general.

Using Eq.~(\ref{Equation:AsymmetryMeasures}), we can now define the ``local quantum uncertainty'' $Q_A^{C^{\text{WY}}\; \Gamma}(\rho_{AB})$ as the coherence based measure of QCs induced by the Wigner-Yanase skew information \cite{girolami2013characterizing}, and the ``interferometric power'' $Q_A^{C^{\text{QF}}\; \Gamma}(\rho_{AB})$ as the coherence based measure of QCs induced by the quantum Fisher information \cite{girolami2014quantum}. More precisely, each of these expressions defines a family of measures, dependent upon the fixed spectrum $\Gamma$ of the observables entering the definition. The original names derive from their physical meaning. As the skew information $C^{\text{WY}}(\rho,K)$ can be regarded as a quantifier of the genuinely quantum share of the uncertainty (out of the total variance) associated to the measurement of $K$ on the state $\rho$, which is due to the fact that $[\rho, K] \neq 0$ \cite{luo2003wigner}, the QCs measure $Q_A^{C^{\text{WY}}}$ consequently quantifies the minimum quantum uncertainty affecting the measurement of a local observable on subsystem $A$ in the state $\rho_{AB}$; in this way, QCs are clearly linked to non-commutativity, as classical-quantum states are all and only the states which can commute with a local observable on $A$ \cite{girolami2013characterizing}. On the other hand, the interferometric power $Q_A^{C^{\text{QF}}\; \Gamma}(\rho_{AB})$  recognises this non-commutativity as an advantageous feature to maintain a guaranteed precision in a metrological task, as we will detail in Section~\ref{Section:Applications-Metrology}.

Beyond the specific interpretations, however, both measures enjoy similar properties, due to their common origin. In particular, they both obey all Requirements~(\ref{Requirements:Faithfulness})--(\ref{Requirements:UnmeasuredParty}) \cite{girolami2013characterizing,girolami2014quantum}, while Requirement~(\ref{Requirements:LCPO}) is currently being tested, with some partial progress achieved for $Q_A^{C^{\text{QF}}\; \Gamma}(\rho_{AB})$ \cite{bromley2016there}. By construction, it holds
\begin{equation}\label{Equation:LQUIP}
Q_A^{C^{\text{WY}}\; \Gamma}(\rho_{AB}) \leq
\frac14 Q_A^{C^{\text{QF}}\; \Gamma}(\rho_{AB})\,,
\end{equation}
with equality on pure states (for this reason, the $\frac14$ factor is usually included in the definition of the interferometric power for normalisation) \cite{girolami2014quantum,adesso2014gaussian}.
Closed analytical formulae are available for both measures on all bipartite states $\rho_{AB}$ when subsystem $A$ is a qubit \cite{girolami2013characterizing,girolami2014quantum}, demonstrating that these measures reconcile the typically contrasting requirements of computability and reliability \cite{tufarelli2013geometric}. In this case (dropping the spectrum superscript without loss of generality as discussed in Section~\ref{Section:Measures-Types-Unitary}), the local quantum uncertainty $Q_A^{C^{\text{WY}}}(\rho_{AB})$ is also equivalent to the unitary response measure based on the Hellinger distance $Q_A^{U_{\text{He}}}(\rho_{AB})$ \cite{girolami2013characterizing} (which in turn reduces to the remedied Hilbert-Schmidt geometric measure $Q_A^{G_2}(\sqrt{\rho_{AB}})$ \cite{chang2013remedying}), and to the discriminating strength $Q^{U^\Gamma_{\text{C}}}_A(\rho_{AB})$ \cite{farace2014discriminating}, as discussed in Section~\ref{Section:Measures-Response}.
Surprisingly, a similar equivalence does not hold between the interferometric power $Q_A^{C^{\text{QF}}}(\rho_{AB})$ and the unitary response measure based on the Bures distance, and it is still unknown whether the former may admit any alternative interpretation in terms of a geometric or a unitary response type measure. The interferometric power has been extended to continuous variable systems, and a closed formula derived for all two-mode Gaussian states restricting to an optimisation over Gaussian observables $K$ with harmonic spectrum \cite{adesso2014gaussian}.

Finally, the interferometric power was measured experimentally in a two-qubit nuclear magnetic resonance ensemble realised in chloroform, as part of an implementation of black box quantum phase estimation with noisy resources (See~\ref{Section:Applications-Metrology}) \cite{girolami2014quantum}.

\subsubsection{Other coherence quantifiers}\label{Section:Measures-Coherence-Others}
Given that the formalism of coherence based measures of QCs has been introduced only in the present paper in its full generality, there have been no further records (to the best of our knowledge) to define other quantifiers of this type. However, the procedure outlined in Section~\ref{Section:Measures-Types-Coherence} could be applied to any valid coherence measure in principle, with the analysis in Section~\ref{Section:Measures-Asymmetry} being specifically useful for monotones arising from the resource theory of asymmetry specialised to coherence.  In this context, let us mention that the authors of \cite{frerot2015quantum} defined the ``quantum variance'', a related quantifier of quantum uncertainty yielding a computable and experimentally accessible lower bound to both Wigner-Yanase skew information and quantum Fisher information, and discussed its applications to measuring QCs and to exploring the associated quantum coherence length in many-body systems.

One suggestion for the future could be to define and study the measures of QCs based on the recently introduced ``robustness of coherence'' \cite{napoli2016robustness} (see Table~\ref{Table:Coherence}), which may be expected to enjoy  some appealing operational interpretations in a channel discrimination setting. This is left open for a keen reader (or co-author). Another possibility could be to study the measures of QCs induced by Tsallis generalisations of the relative entropy of coherence \cite{rastegin2016quantum}, and investigate their relationship with corresponding geometric and informational measures defined via the same entropies.

\subsection{Recoverability measures}\label{Section:Measures-Recoverability}
The recoverability approach to QCs, as outlined in Section~\ref{Section:Measures-Types-Recoverability}, has been explored only quite recently; for this reason, there is only one measure which has been considered in explicit detail so far \cite{seshadreesan2015fidelity}, as we now review.

\subsubsection{Surprisal of measurement recoverability}\label{Section:Measures-Surprisal}

The ``surprisal of measurement recoverability'' $Q_{A}^{R_F}$ has been originally defined by resorting to the (negative logarithm of) Uhlmann fidelity \cite{seshadreesan2015fidelity}, corresponding the one-sided definition in Eqs.~(\ref{Equation:GeometricRecoverability}) with $D_\delta(\rho,\sigma) = - \log F(\rho,\sigma)$, where the fidelity $F(\rho,\sigma)$ is given by Eq.~(\ref{Equation:Fidelity}). Explicitly,
\begin{equation}\label{Equation:SurprisalMeasurementRecoverability}
Q_{A}^{R_F} (\rho_{AB}):= - \log \sup_{\Lambda^{\text{EB}}_A} F\big(\rho_{AB},\Lambda^{\text{EB}}_A[\rho_{AB}]\big),
\end{equation}
where the optimisation is over the convex set of entanglement-breaking channels $\Lambda^{\text{EB}}_A$ acting on subsystem $A$

The motivation behind this definition arose from an alternative angle from which the quantum discord (see Section~\ref{Section:Measures-Informational-MutualInformation}) can be looked at, introduced for the first time in~\cite{piani2012problem}. Indeed, by using the fact that any LGM $\tilde{\Pi}_A$ on subsystem $A$ can be written as a unitary $U_{A\rightarrow A'C}^{\tilde{\Pi}_A}$ from $A$ to a composite system $A'C$ followed by discarding $C$, i.e., $\tilde{\Pi}_A[\rho_{AB}]=\mbox{Tr}_C\big(U_{A\rightarrow A'C}^{\tilde{\Pi}_A}[\rho_{AB}]\big)$, one then gets
\begin{equation}\label{Equation:QuantumDiscordCM}
Q_A^{\tilde{I}_{\mathcal{I}}}(\rho_{AB}) = \inf_{\tilde{\Pi}_A} {\cal I}_{A'}\big(U_{A\rightarrow A'C}^{\tilde{\Pi}_A}[\rho_{AB}]\big),
\end{equation}
where we denote by ${\cal I}_{A}(\rho_{ABC})$ the conditional quantum mutual information of the tripartite state $\rho_{ABC}$ with respect to subsystem $A$, defined as
\begin{equation}
{\cal I}_{A}(\rho_{ABC}) := \mathcal{S}(\rho_{AB}) + \mathcal{S}(\rho_{AC}) - \mathcal{S}(\rho_{A}) - \mathcal{S}(\rho_{ABC}).
\end{equation}
The latter quantity, often denoted as ${\cal I}(B;C|A)_\rho$ in information theory literature \cite{wilde}, captures the correlations present between $B$ and $C$ from the perspective of $A$ in the i.i.d. resource limit, where an asymptotically large number of copies of the tripartite state $\rho_{ABC}$ are shared between the three parties.

In the quest for developing a version of the conditional quantum mutual information which could be relevant for the finite resource regime, in~\cite{seshadreesan2015fidelity} the surprisal of the fidelity of recovery of a tripartite quantum state $\rho_{ABC}$ with respect to subsystem $A$ was defined as follows,
\begin{equation}
F^R_A(\rho_{ABC}):= -\log \sup_{\Lambda^R_{A\rightarrow AC}} F\big(\rho_{ABC},\Lambda^R_{A\rightarrow AC}[\rho_{ABC}]\big),
\end{equation}
where the optimisation is over all recovery maps $\Lambda^R_{A\rightarrow AC}$ from $A$ to the composite system $AC$. The surprisal of measurement recoverability was then introduced in~\cite{seshadreesan2015fidelity} in complete analogy with quantum discord by simply substituting the conditional quantum mutual information with the surprisal of the fidelity of recovery,
\begin{equation}
Q^{R_F}_A(\rho_{AB}):= \inf_{\tilde{\Pi}_A} F^R_{A'}\big(U_{A\rightarrow A'C}^{\tilde{\Pi}_A}[\rho_{AB}]\big),
\end{equation}
and only afterwards proven to be equivalent to Eq.~(\ref{Equation:SurprisalMeasurementRecoverability}).

For bipartite pure states $\ket{\psi}_{AB}$, the surprisal of measurement recoverability is equal to an entanglement measure,
\begin{equation}
Q^{R_F}_A(|\psi_{AB}\rangle)= -\log \mbox{Tr}(\rho_A^2),
\end{equation}
where $\rho_A$ is the marginal state of subsystem $A$, thus satisfying Requirement~(\ref{Requirements:EntanglementPure}). The validity of Requirement~(\ref{Requirements:LCPO}) is still unknown.

In general, it was shown in \cite{seshadreesan2015fidelity} using the results of \cite{fawzi2015quantum} that the surprisal of measurement recoverability is a lower bound to the quantum discord,
\begin{equation}\label{Equation:SMRdiscord}
Q^{R_F}_A(\rho_{AB}) \leq Q_A^{\tilde{I}_{\mathcal I}} (\rho_{AB})\,.
\end{equation}
Furthermore, \cite{piani2015hierarchy} developed a hierarchy of efficiently computable and faithful lower bounds to $Q_A^{\tilde{I}_{\mathcal{I}}}$ that converge exactly to  $Q^{R_F}_A$, thus showing in particular that the surprisal of measurement recoverability can be bounded numerically by a semidefinite program, and admits an operational interpretation in the task of local broadcasting, as we detail in Section~\ref{Section:Applications-Broadcasting}.

It will be interesting in the future to study two-sided versions $Q^{R_F}_{AB}(\rho_{AB})$ of the surprisal of measurement recoverability, as well as other recoverability measures corresponding to different choices of distance $D_\delta$ in Eqs.~(\ref{Equation:GeometricRecoverability}), and explore their interplay with the other types of measures of QCs collected in this review.

Further related measures can be obtained by alternative generalisations of the conditional quantum mutual information, e.g.~in terms of R\'enyi entropies \cite{seshadreesan2015renyi}.

\subsection{Summary of measures}

We close this Section by summarising in Table~\ref{Table:SummaCumLaude} the key characteristics of the majority of QCs measures discussed in this review, including a synopsis of Requirements~(\ref{Requirements:Faithfulness})--(\ref{Requirements:LCPO}) and interdependency relations between measures of different types. When  more than two measures coincide, as in the case of the relative entropy based measures, due to space constraints we only note the equality with the first occurring measure aside each of the further ones; the chain of identities can then be reconstructed scanning the `Notes' column vertically.

Note that similar (and more comprehensive) tables can be found for entanglement measures in \cite{christandl2006thesis}.

\clearpage

\renewcommand{\baselinestretch}{.5}
\begin{table}[!ht]
\centering
\begin{tabular}{cccccccc}
\hline \hline
\!\!\!\! Notation\!\!\!\! &  Nomenclature &  \!\!\!\!\!\! Notes \!\!\!\!\!\! & \!\!(\ref{Requirements:Faithfulness})\!\! &\!\! (\ref{Requirements:UnitaryInvariance})\!\! & \!\!(\ref{Requirements:EntanglementPure})\!\! & \!\!(\ref{Requirements:UnmeasuredParty})\!\! & \!\!(\ref{Requirements:LCPO})\!\!\\ \hline
 $Q_A^{G_{\text{RE}}}$ & relative entropy of discord &  & \checkmark & \checkmark & \checkmark & \checkmark & \checkmark \\
 $Q_{AB}^{G_{\text{RE}}}$ & relative entropy of quantumness &  & \checkmark & \checkmark & \checkmark & -- &  \checkmark \\
 $Q_{A}^{G_{2}}$ & Hilbert-Schmidt geometric discord &   & \checkmark & \checkmark & \checkmark & $\times$ &  ? \\
 $Q_{AB}^{G_{2}}$ & 2-s Hilbert-Schmidt geometric discord &  & \checkmark & \checkmark & \checkmark & -- &  ? \\
 $Q_{A}^{G_{1}}$ & trace distance geometric discord &   & \checkmark & \checkmark & ? & \checkmark & \checkmark \\
 $Q_{AB}^{G_{1}}$ & 2-s trace distance geometric discord &   & \checkmark & \checkmark & ?  & -- & \checkmark \\
 $Q_{A}^{G_{\text{Bu}}}$ & Bures geometric discord &   & \checkmark & \checkmark & \checkmark & \checkmark & \checkmark \\
 $Q_{AB}^{G_{\text{Bu}}}$ & 2-s Bures geometric measure &   & \checkmark & \checkmark & \checkmark & -- & \checkmark \\
 $Q_{A}^{G_{\text{He}}}$  & Hellinger geometric discord &  & \checkmark & \checkmark & \checkmark & \checkmark & \checkmark\\
 $Q_{AB}^{G_{\text{He}}}$  & 2-s Hellinger geometric discord &   & \checkmark & \checkmark & \checkmark & -- & \checkmark \\ \hline

 $Q_{A}^{M_{\text{RE}}}$  & 1-s relative entropy m.i.g.~measure & $=Q_A^{G_{\text{RE}}}$  & \checkmark & \checkmark & \checkmark & \checkmark & \checkmark \\
 $Q_{AB}^{M_{\text{RE}}}$  & 2-s relative entropy m.i.g.~measure& $=Q_{AB}^{G_{\text{RE}}}$  & \checkmark & \checkmark & \checkmark & -- & \checkmark \\
 $Q_{A}^{M_{2}}$  & 1-s Hilbert-Schmidt m.i.g.~measure & $=Q_{A}^{G_{2}}$  & \checkmark & \checkmark & \checkmark & $\times$ & ? \\
 $Q_{AB}^{M_{2}}$  & 2-s Hilbert-Schmidt m.i.g.~measure&   & \checkmark &  \checkmark &  \checkmark & -- & ? \\
 $Q_{A}^{M_{1}}$  & 1-s trace distance m.i.g.~measure& \raisebox{-2pt}{$\stackrel{d_A=2}{=}$} $Q_{A}^{G_{1}}$  & \checkmark & \checkmark & \checkmark & \checkmark & ? \\
 $Q_{AB}^{M_{1}}$  & 2-s trace distance m.i.g.~measure&   & \checkmark & \checkmark & \checkmark & -- & ? \\
 $Q_{A}^{M_{\text{Bu}}}$  & 1-s Bures m.i.g.~measure&   & \checkmark & \checkmark & \checkmark & \checkmark & ? \\
 $Q_{AB}^{M_{\text{Bu}}}$  & 2-s Bures m.i.g.~measure &   & \checkmark & \checkmark & \checkmark  & -- & ? \\
 $Q_{A}^{M_{\text{He}}}$  & 1-s Hellinger m.i.g.~measure&   & \checkmark & \checkmark & \checkmark & \checkmark & ? \\
 $Q_{AB}^{M_{\text{He}}}$  & 2-s Hellinger m.i.g.~measure &   & \checkmark & \checkmark & \checkmark & -- & ? \\ \hline

 $Q_{A}^{I_{\mathcal{S}}}$  & one-way quantum deficit & $=Q_A^{G_{\text{RE}}}$  & \checkmark & \checkmark & \checkmark & \checkmark & \checkmark \\
 $Q_{AB}^{I_{\mathcal{S}}}$  & zero-way quantum deficit & $=Q_{AB}^{G_{\text{RE}}}$  & \checkmark & \checkmark & \checkmark & -- & \checkmark \\
 $Q_{A}^{\tilde{I}_{\mathcal{S}}}$  & 1-s von Neumann (LGM) informational measure &   & \checkmark & \checkmark & \checkmark & ? & ? \\
 $Q_{AB}^{\tilde{I}_{\mathcal{S}}}$  & 2-s von Neumann (LGM) informational measure &   & \checkmark & \checkmark & \checkmark & -- & ?  \\
 $Q_{A}^{I_{\mathcal{I}}}$  & (LPM) quantum discord &   & \checkmark & \checkmark & \checkmark & \checkmark & ? \\
 $Q_{AB}^{I_{\mathcal{I}}}$  & ameliorated measurement induced disturbance&   & \checkmark & \checkmark & \checkmark & -- & ? \\
 $Q_{A}^{\tilde{I}_{\mathcal{I}}}$  & (LGM) quantum discord  &   & \checkmark & \checkmark & \checkmark & \checkmark & ? \\
 $Q_{AB}^{\tilde{I}_{\mathcal{I}}}$  & (LGM) symmetric quantum discord &   & \checkmark & \checkmark & \checkmark & -- & ? \\ \hline

 $Q_{A}^{E^{G_{\text{RE}}}}$  & 1-s relative entropy activation measure & $=Q_A^{G_{\text{RE}}}$  & \checkmark & \checkmark & \checkmark & \checkmark & \checkmark \\
 $Q_{AB}^{E^{G_{\text{RE}}}}$  & 2-s relative entropy activation measure & $=Q_{AB}^{G_{\text{RE}}}$  & \checkmark & \checkmark & \checkmark & -- & \checkmark \\
 $Q_{A}^{E^d}$  & 1-s distillable entanglement activation measure & $=Q_A^{G_{\text{RE}}}$  & \checkmark & \checkmark & \checkmark & \checkmark & \checkmark \\
 $Q_{AB}^{E^d}$  & 2-s distillable entanglement activation measure &  $=Q_{AB}^{G_{\text{RE}}}$ & \checkmark & \checkmark & \checkmark & -- & \checkmark \\
 $Q_{A}^{E^N}$  & 1-s negativity of quantumness & \raisebox{-2pt}{$\stackrel{d_A=2}{=}$} $Q_{A}^{G_{1}}$  & \checkmark & \checkmark & \checkmark & \checkmark & ? \\
 $Q_{AB}^{E^N}$  & 2-s negativity of quantumness &   & \checkmark & \checkmark & \checkmark & -- & ? \\
 \hline

 $Q_A^{U_2}$  & Hilbert-Schmidt discord of response & \raisebox{-2pt}{$\stackrel{d_A=2}{=}$} $4 Q^{G_{2}}_A$  & \checkmark & \checkmark & \checkmark & $\times$ & ?\\
 $Q_A^{U_{1}}$  & Trace distance discord of response & \raisebox{-2pt}{$\stackrel{d_A=2}{=}$} $2 Q_{A}^{G_{1}}$  & \checkmark & \checkmark & \checkmark & \checkmark & ?\\
 $Q_A^{U_{\text{Bu}}}$  & Bures discord of response & \raisebox{-2pt}{$\stackrel{d_A=2}{=}$} $g(Q_A^{G_{\text{Bu}}})$  & \checkmark & \checkmark & \checkmark & \checkmark & ?\\
 $Q_A^{U_{\text{He}}}$  & Hellinger discord of response & \raisebox{-2pt}{$\stackrel{d_A=2}{=}$} $g(Q_A^{G_{\text{He}}})$  & \checkmark & \checkmark & \checkmark & \checkmark & ?\\
 $Q^{U^\Gamma_{\text{C}}}_A$  & discriminating strength & \raisebox{-2pt}{$\stackrel{d_A=2}{=}$} $Q_{A}^{U_{\text{He}}}$  & \checkmark & \checkmark & \checkmark & \checkmark & ?\\ \hline

 $Q_{A}^{C^{G_{\text{RE}}}}$  & 1-s relative entropy coherence measure & $=Q_A^{G_{\text{RE}}}$  & \checkmark & \checkmark & \checkmark & \checkmark & \checkmark \\
 $Q_{AB}^{C^{G_{\text{RE}}}}$  & 2-s relative entropy coherence measure & $=Q_{AB}^{G_{\text{RE}}}$  & \checkmark & \checkmark & \checkmark & -- & \checkmark \\
 $Q_{A}^{C^{\ell_1}}$  & 1-s $\ell_1$ norm coherence based measure & $=Q_{A}^{E^N}$ & \checkmark & \checkmark & \checkmark & \checkmark & ? \\
 $Q_{AB}^{C^{\ell_1}}$  & 2-s $\ell_1$ norm coherence based measure & $=Q_{AB}^{E^N}$  & \checkmark & \checkmark & \checkmark & -- & ? \\
 $Q_{A}^{C^\text{WY}}$  & local quantum uncertainty & \raisebox{-2pt}{$\stackrel{d_A=2}{=}$} $2 Q_{A}^{U_{\text{He}}}$  & \checkmark & \checkmark & \checkmark & \checkmark & ? \\
 $Q_{A}^{C^\text{QF}}$  & interferometric power &   & \checkmark & \checkmark & \checkmark & \checkmark & ? \\ \hline

 $Q_A^{R_F}$  & Surprisal of measurement recoverability  &  & \checkmark & \checkmark & \checkmark & \checkmark & ? \\
\hline\hline
\end{tabular}
\caption{\label{Table:SummaCumLaude} Summary of measures of QCs, categorised by types as in Table~\ref{Table:Types}. Additional legend: `m.i.g.' stands for  `measurement induced geometric'; `1-s' means one-sided, `2-s' means two-sided; the superscript `$d_A=2$' indicates equality when $A$ is a qubit; and $g(x)=4x-x^2$.}
%, and ``$Q^x = Q^y$'' always links to the first occurring $Q^y$.}
\end{table}
\renewcommand{\baselinestretch}{1}

\section{Applications and operational interpretations of quantum correlations}\label{Section:Applications}

After having introduced so many different measures of QCs, even though guided by physical threads inspired by Fig.~\ref{Fig:Defs}, one could honestly wonder how exactly to navigate through the maze of Table~\ref{Table:SummaCumLaude}, or, in other words: {\it ``Which measure should I use?''}

It is desirable for any measure of QCs (or of any other {\em resource}) to be directly linked to the figure of merit in the performance of an operational task. This allows one to justify the quantitative ordering on the set of states imposed by a specific measure of QCs, i.e.~if a given state has more QCs than another, we know that this is precisely because this state performs better in the corresponding operational task.
In the following we provide a selective review of some of the most concrete and physically relevant applications and operational interpretations of measures of QCs.

\subsection{Quantum information and communication}\label{Section:Applications-QC}

Much weight has been placed by the international community behind successfully performing quantum communication, which involves any information transmission task that cannot be achieved perfectly using classical resources, due to the fundamental role of  communication primitives in the landscape of future quantum technologies, including quantum cryptography and computing~\cite{kimble2008quantum}. QCs beyond entanglement have been shown to play a clear role in delineating possibilities and limitations for some well established protocols in quantum information and communication, including broadcasting, distribution of entanglement, quantum state merging and redistribution.

\subsubsection{Local broadcasting}\label{Section:Applications-Broadcasting}

Copying information is a natural ability in our classical realm. Indeed, the classical information you are reading right now is one of (hopefully) many copies. In the quantum realm, however, unconditional copying is not allowed, due to  the no-cloning theorem \cite{wootters1982single,dieks1982communication}. The no-cloning theorem says that it is impossible to create an identical copy of an arbitrary unknown pure quantum state of a system by using a composite unitary acting upon the system and an ancilla. The best that we can do is to reliably clone \emph{orthogonal} pure states, which effectively corresponds to copying of classical information. %, i.e. given a pure state $\ket{\psi}_{A}$ and an ancilla $\ket{0}_{B}$, no unitary $U_{AB}$ exists so that $U_{AB}\ket{\psi}_{A}\otimes \ket{0}_{B} = \ket{\psi}_{A} \otimes \ket{\psi}_{B}$
Such an impossibility was subsequently generalised to mixed states and general transformations leading to the so called no-broadcasting theorem \cite{barnum1996noncommuting}.
%, which states that given a set of quantum states $\{\rho_{A}^{(i)}\}$, and a transformation $\Lambda_{AB}(\rho_{A}^{(i)}\otimes \ket{0}\bra{0}_{B})$ on $\rho_{A}^{(i)}$ and an ancilla $\ket{0}\bra{0}_{B}$, it is only possible for
%\begin{equation}
%\mbox{Tr}_{A}\left[\Lambda_{AB}(\rho_{A}^{(i)}\otimes \ket{0}\bra{0}_{B})\right] = \mbox{Tr}_{B}\left[\Lambda_{AB}(\rho_{A}^{(i)}\otimes \ket{0}\bra{0}_{B})\right] = \rho_{A}^{(i)}
%\end{equation}
%if $\{\rho_{A}^{(i)}\}$ all commute

These no-go theorems enshrine classicality in single systems. For composite systems, the task of ``local broadcasting'' was outlined in \cite{piani2008no}. Consider a composite bipartite state $\rho_{AB}$ shared between laboratory $A$ and $B$, with each laboratory also possessing an ancilla, respectively $A'$ and $B'$, in the composite state $\sigma_{A'B'}=\ket{0}\bra{0}_{A'}\otimes \ket{0}\bra{0}_{B'}$. The objective is for $A$ and $B$ to perform local operations only (and no communication) on the joint state $\rho_{AB}\otimes \sigma_{A'B'}$, in order to produce a joint state $\rho_{AA'BB'}=(\Lambda_{AA'}\otimes \Lambda_{BB'})[\rho_{AB}\otimes \sigma_{A'B'}]$ which contains two (possibly correlated) copies of $\rho_{AB}$,
\begin{equation}
\mbox{Tr}_{A'B'}(\rho_{AA'BB'}) = \mbox{Tr}_{AB}(\rho_{AA'BB'}) = \rho_{AB}.
\end{equation}
It was shown in \cite{piani2008no} that this task can only be performed if $\rho_{AB} \in \mathscr{C}_{AB}$, which is a very intuitive result saying that only classically correlated states can be locally broadcast. More generally, it was shown that, even if the task was relaxed to that of obtaining two (possibly different) reduced states but with the same total correlations as measured by the mutual information between $A$ and $B$ (or $A'$ and $B'$), then the correlations themselves  could only be broadcast if $\rho_{AB}$ is classically correlated,
\begin{equation}
\rho_{AB} \in \mathscr{C}_{AB} \,\,\,\,\,\, \Leftrightarrow \,\,\,\,\,\, \exists \,\, \rho_{AA'BB'}\,\,\,\,\, \mbox{such that} \,\,\,\,\,\, \mathcal{I}\big(\mbox{Tr}_{A'B'}(\rho_{AA'BB'})\big) = \mathcal{I}\big(\mbox{Tr}_{AB}(\rho_{AA'BB'})\big) = \mathcal{I}\left(\rho_{AB}\right),\nonumber
\end{equation}
which is called the no-local-broadcasting theorem \cite{piani2008no}. One-sided local (or unilocal) broadcasting was then considered in \cite{luo2010quantum}. Here, the idea is for only $A$ to have access to an ancilla $\ket{0}\bra{0}_{A'}$ and to perform local operations to obtain states $\rho_{AA'B}=(\Lambda_{AA'}\otimes \mathbb{I}_{B})[\rho_{AB}\otimes \ket{0}\bra{0}_{A'}]$ such that
\begin{equation}
\mathcal{I}\big(\mbox{Tr}_{A'}(\rho_{AA'B})\big) = \mathcal{I}\big(\mbox{Tr}_{A}(\rho_{AA'B})\big) = \mathcal{I}\left(\rho_{AB}\right).
\end{equation}
As intuitively expected, this is shown to be possible if and only if $\rho_{AB} \in \mathscr{C}_{A}$. % Both the local and unilocal broadcasting settings can naturally be viewed within the multipartite setting, but we
Partial broadcasting of QCs was further discussed in \cite{chatterjee2016broadcasting}.

By design it appears straight away that QCs are relevant in the local broadcasting paradigm: indeed, the (im)possibility of local or unilocal broadcasting provides another qualitative characterisation of classical versus quantum states, that could be added to Fig.~\ref{Fig:Defs}. Can such a paradigm provide natural operational interpretations to measures of QCs then? The answer is affirmative.

To begin with, precise quantitative links to one-sided informational measures of QCs and unilocal broadcasting were provided recently by \cite{brandao2015generic}, who consider a bipartite system in the state $\rho_{AB}$ and the process of party $A$ redistributing their correlations with $B$ to $N$ ancillae $\{A_{i}\}_{i=1}^N$ by local operations $\Lambda_{A \rightarrow A_1\ldots A_N}$, with the aim to maintain as many correlations as possible between each $A_i$ and the other party $B$. Let us define the reduced state of  $A_i$ and $B$ after the redistribution as
\begin{equation}\label{Equation:Redist}
\rho_{A_iB} = \mbox{Tr}_{\otimes_{j \neq i} A_j} \big[(\Lambda_{A \rightarrow A_1\ldots A_N} \otimes \mathbb{I}_B)[\rho_{AB}] \big]\,.
\end{equation}
Due the no-unilocal broadcasting theorem \cite{piani2008no,luo2010quantum}, for any $\rho_{AB} \not\in \mathscr{C}_A$ we know that some correlations will be generally lost in the redistribution process, i.e.,~each part $A_i$ will have less correlations with $B$ than the initial correlations between $A$ and $B$, or in formulae, ${\cal I}(\rho_{A_iB}) \leq {\cal I}(\rho_{AB})$, $\forall \ i=1,\ldots,N$.  The loss of correlations when redistributing to the ancilla $A_{i}$ can be averaged over all the ancillae, and minimised over all the local redistributing operations, to arrive at the minimum average loss of correlations,
\begin{equation}\label{Equation:Tuco}
\overline{\Delta}_A^{(N)}(\rho_{AB}) := \min_{\Lambda_{A \rightarrow A_1\ldots A_N}} \frac{1}{N} \sum_{i=1}^N \big[{\cal I}(\rho_{AB}) - {\cal I}(\rho_{A_iB})\big].
\end{equation}
Quite remarkably, in the limit of infinitely many ancillae, it was shown that \cite{brandao2015generic}
\begin{equation}\label{Equation:NoDoze}
\lim_{N \rightarrow \infty} \overline{\Delta}_A^{(N)}(\rho_{AB}) = Q_{A}^{\tilde{I}_{\mathcal{I}}}(\rho_{AB})\,,
\end{equation}
that is, the  asymptotic minimum average loss of correlations after attempting to unilocally broadcast any quantum state $\rho_{AB}$ is given by its QCs content as measured by the quantum discord. Eq.~(\ref{Equation:NoDoze}) yields one of the most prominent operational interpretations of the quantum discord, and generalises the similar work of \cite{streltsov2013quantum} whose result was limited to pure states only. Borrowing the title of \cite{streltsov2013quantum}, this means that {\em quantum discord cannot be shared}, while only the classical correlations quantified by Eq.~(\ref{Equation:ClassicalCorrelationsLGM}) can be arbitrarily redistributed to many parties on one side.

These results have intimate connections to ``quantum Darwinism'' \cite{zurek2009quantum}. The idea behind quantum Darwinism is to explain the emergence of classical reality by observing a quantum system indirectly through its environment. The only information objectively accessible about the quantum system is that proliferated into many sub-parts of the environment. Thus, it is clear that $\overline{\Delta}_A^{(N)}(\rho_{AB})$, hence the quantum discord of $\rho_{AB}$ in the limit of asymptotically many sub-parts of the environment, represents the portion of correlations inaccessible to the classical realm \cite{brandao2015generic}.

Curiously, no rigorous result analogous to Eq.~(\ref{Equation:NoDoze}) has been yet established in the two-sided setting of redistributing correlations in a quantum state $\rho_{AB}$ both on $A \rightarrow \{A_i\}_{i=1}^N$ and on $B \rightarrow \{B_j\}_{j=1}^N$. It is reasonable to expect that in this case one would recover the two-sided informational measure  $Q_{AB}^{\tilde{I}_{\mathcal{I}}}(\rho_{AB})$ as the asymptotic minimum average of the correlation gap  ${\cal I}(\rho_{AB}) - {\cal I}(\rho_{A_iB_i})$. We conjecture this to be the case, following the intuition that $Q_{AB}^{\tilde{I}_{\mathcal{I}}}(\rho_{AB})$ should be related to the amount of correlations lost when performing local broadcasting \cite{piani2008no}, since the maximum share of correlations in a state $\rho_{AB}$ that can be broadcast, i.e., transferred to classical registers from both subsystems, is given by the classical mutual information of $\rho_{AB}$, Eq.~(\ref{Equation:ClassicalCorrelationsAB}). We hope our conjecture will be settled in the near future.

Another way to obtain quantitative results is to look at how similar a locally broadcast copy can be to the original state $\rho_{AB}$~\cite{piani2015hierarchy}. Focusing again on the unilocal broadcasting setting of $A$ attempting to locally copy $\rho_{AB}$ to $N$ ancillae $\{A_{i}\}_{i=1}^{N}$ and arriving at the reduced states $\rho_{A_{i}B}$ in Eq.~(\ref{Equation:Redist}), it is clear that for any $A_{i}$
\begin{equation}
\max_{\Lambda_{A \rightarrow A_1\ldots A_N}}F(\rho_{AB},\rho_{A_{i}B}) \leq 1,
\end{equation}
where $F$ is the Uhlmann fidelity, Eq.~(\ref{Equation:Fidelity}). Equality holds if and only if $\rho_{AB}$ can be locally broadcast, or equivalently if $\rho_{AB} \in \mathcal{C}_{A}$: in this case, for $\rho_{AB} = \sum_{i} p_{i} \ket{i}\bra{i}_{A} \otimes \rho_{B}^{(i)}$, a channel implementing the transformation $\ket{i}_{A} \rightarrow \ket{i}^{\otimes N}_{A}$ would be suitable for broadcasting. This transformation is a type of $N$-Bose-symmetric channel $\Lambda^{B_{N}}_{A \rightarrow A_1\ldots A_N}$ mapping $A$ onto the fully symmetric subspace of $A_1\ldots A_N$, i.e.~returning a density operator invariant under any permutation   $\varpi(A_1\ldots A_N)$ of the ancillae~\cite{hudson2004measurements}. It is then interesting to consider~\cite{piani2015hierarchy}
\begin{equation}
Q^{\text{B}_N}_{A}(\rho_{AB}) := -\log \sup_{\Lambda^{B_{N}}_{A \rightarrow A_1\ldots A_N}} F(\rho_{AB},\rho_{A_{1}B})
\end{equation}
%with $Q^{\text{B}_N}_{A}(\rho_{AB}) \leq Q^{\text{B}_{N+1}}_{A}(\rho_{AB})$, providing a family of
as a figure of merit quantifying the quality of $N$-copy local broacasting (note that, due to the symmetry of  $\Lambda^{B_{N}}_{A \rightarrow A_1\ldots A_N}$, it is enough to just consider $\rho_{A_{1}B}$). In the case $N \rightarrow \infty$, the composition of $\Lambda^{B_{N}}_{A \rightarrow A_1\ldots A_N}$ followed by the partial trace over all but ancilla $A_{1}$ is equivalent to the action of an entanglement breaking channel, Eq.~(\ref{Equation:EntanglementBreaking}), or simply $\rho_{A_{1}B}$ is the output of an entanglement breaking channel acting on $\rho_{AB}$. This means that $Q^{\text{B}_\infty}_{A}(\rho_{AB})=Q^{R_{F}}_{A}(\rho_{AB})$ and we recover the surprisal of measurement recoverability $Q^{R_{F}}_{A}(\rho_{AB})$ from Eq.~(\ref{Equation:SurprisalMeasurementRecoverability}). Hence, $Q^{R_{F}}_{A}(\rho_{AB})$ has an operational interpretation in terms of the minimum distinguishability between $\rho_{AB}$ and its unilocally broadcast copies in the asymptotic setting. Furthermore, it has been suggested that each $Q^{\text{B}_N}_{A}(\rho_{AB})$ is a valid measure of one-sided QCs in its own right~\cite{piani2015hierarchy}, obeying at least Requirements~(\ref{Requirements:Faithfulness}), (\ref{Requirements:UnitaryInvariance}) and (\ref{Requirements:UnmeasuredParty}). These quantities are simple to calculate numerically as the solution of a semidefinite program, and display a natural hierarchy~\cite{piani2015hierarchy,fawzi2015quantum}
\begin{equation}
Q^{\text{B}_2}_{A}(\rho_{AB}) \leq Q^{\text{B}_{N}}_{A}(\rho_{AB}) \leq Q^{R_{F}}_{A}(\rho_{AB}) \leq Q_A^{\tilde{I}_{\mathcal I}} (\rho_{AB}),
\end{equation}
hence providing computable lower bounds to quantum discord.

In summary, the ability to locally broadcast a correlated quantum state is fundamentally (and inversely) linked to the QCs present in the state, in both qualitative and various  quantitative senses.

\subsubsection{Entanglement distribution}

The entanglement activation framework presented in Section~\ref{Section:Measures-Types-Activation} provides already an operational setting to interpret measures of QCs in a state $\rho_{AB}$ in terms of their potential for the creation of entanglement with ancillary systems. Under some conditions, this entanglement can be swapped back to the original system $AB$ \cite{gharibian2011characterizing}. More generally, this means that all the measures $Q^{E^\zeta}_{A}$ and $Q^{E^\zeta}_{AB}$ of QCs of the entanglement activation type, some of which are listed in Table~\ref{Table:SummaCumLaude}, may be regarded as having an intrinsically operational character \cite{piani2011all}.

Here we discuss the role of QCs for a different task, namely  ``entanglement distribution''.
Distributing entanglement between spatially separated laboratories is a key starting point in many quantum information protocols. Consider two laboratories $A$ and $B$, who want to increase their shared entanglement. From the fundamental paradigm of entanglement, it is clear that such a task cannot be achieved using only LOCC. Instead, it is necessary to send a quantum particle $C$ from one laboratory to the other (we will assume the particle is sent from $A$ to $B$), i.e.~to utilise {\it quantum} communication in combination with local operations.

More rigorously, we can look at the state $\rho_{ABC}$ of the composite system consisting of the two parties $A$ and $B$ and the carrier particle $C$. It is then relevant to consider two bipartitions of the three parties. Bipartition $AC:B$ represents the carrier particle at laboratory~$A$, while bipartition $A:BC$ represents the carrier particle after having been sent to laboratory~$B$. We then turn to the entanglement $E_{AC:B}(\rho_{ABC})$ and $E_{A:BC}(\rho_{ABC})$, which represent the entanglement before and after the carrier particle has been sent (through a noiseless quantum channel), where $E_{X:Y}$ is a measure of bipartite entanglement for the bipartition $X:Y$. The objective is to achieve $E_{A:BC}(\rho_{ABC}) > E_{AC:B}(\rho_{ABC})$. Laboratory $B$ can then attempt to perform joint operations on both their system and the carrier particle to localise the entanglement onto $B$ and hence increase the entanglement between $A$ and $B$. %, i.e.~ideally $E_{A:B}(\rho_{AB}') = E_{A:BC}(\rho_{ABC}) > E_{AC:B}(\rho_{ABC}) > E_{A:B}(\rho_{AB})$, where $\rho_{AB}$ and $\rho_{AB}'$ are the states of just $A$ and $B$ before and after the local operations between $B$ and $C$, respectively.??

The question is whether we can place a minimum cost on the increase in entanglement $E_{A:BC}(\rho_{ABC}) - E_{AC:B}(\rho_{ABC})$. One might guess that the cost is related to $E_{AB:C}(\rho_{ABC})$, i.e.~so that the increase in entanglement is limited by the entanglement carried by particle $C$ with the two laboratories. In fact, entanglement across the $AB:C$ split turns out not to be necessary, and the task of entanglement distribution can also be accomplished with a separable carrier, as first shown in \cite{cubitt2003separable}. Instead,~\cite{chuan2012quantum,streltsov2012quantum} suggest that more general QCs between $AB$ and $C$, defined with respect to one-sided measurements on $C$, may represent appropriately the minimum cost.
%One might guess that the difference $E_{A:BC}(\rho_{ABC}) - E_{AC:B}(\rho_{ABC})$ is related to $E_{AB:C}(\rho_{ABC})$, the entanglement between the two parties $A$ and $B$ and the ancilla particle $C$. Indeed, entanglement in the partition $AB:C$ is sufficient for entanglement distribution. It was shown in~\cite{cubitt2003separable}, however, entanglement distribution can be achieved even when the system is separable along this partition. Instead,~\cite{chuan2012quantum,streltsov2012quantum} suggest that the \emph{quantum correlations} between $AB$ and $C$, with one-way measurements on $C$ is the relevant figure if merit.
They showed that, when resorting to the geometric measures of QCs (see Section~\ref{Section:Measures-Types-Geometric}), it holds
\begin{equation}\label{Equation:EntanglementDistribution}
Q_{C}^{G_{\delta}}(\rho_{ABC}) \geq E^{G_{\delta}}_{A:BC}(\rho_{ABC}) - E^{G_{\delta}}_{AC:B}(\rho_{ABC}),
\end{equation}
where $E^{G_{\delta}}_{X:Y}(\rho_{ABC})$ is the corresponding geometric measure of entanglement defined in Eq.~(\ref{Equation:EntanglementGeometricMeasures}) for the bipartition $X:Y$. This inequality, which in the case of measures based on the relative entropy represents a strengthening of the standard subadditivity inequality of von Neumann entropy, tells us that QCs on the $C$ side are in general necessary for entanglement distribution. In fact, whenever QCs are not present, the transmission of the carrier particle $C$ effectively reduces to classical communication. Examples of states $\rho_{ABC}$ were given in~\cite{chuan2012quantum} for when this bound is tight, and it was argued in~\cite{streltsov2012quantum} that entanglement distribution with separable states may be more efficient than using entangled carriers if local generation of entanglement at the sender laboratory is expensive.

However, the usefulness of a state for entanglement distribution is not  determined only by its QCs. In \cite{streltsov2014limits} examples of states with nonzero QCs yet that cannot be used for entanglement distribution were provided, and additional conditions to ensure usefulness for this task  were derived in terms of the dimension of $C$ and the number of product terms in the decomposition of the separable $\rho_{ABC}$ along the partition $AB:C$. This situation was likened to entanglement, whereby only some entangled states are useful for distillation (i.e.~the states being not bound entangled)~\cite{horodecki2009quantum}. It was also suggested in \cite{kay2012using} that the quantity $Q_{C}^{G_{\delta}}(\rho_{ABC})$ may not a relevant resource for entanglement distribution because laboratory $A$ initially holds the carrier $C$ and can thus alter the QCs by local operations, potentially making Eq.~(\ref{Equation:EntanglementDistribution}) a rather loose bound.

Finally, a number of experiments have successfully demonstrated entanglement distribution with separable states in discrete and continuous variable setups \cite{fedrizzi2013experimental,peuntinger2013distributing,vollmer2013experimental}.

\subsubsection{Quantum state merging}\label{Section:Merging}

Imagine that our two laboratories $A$ and $B$ each sample from discrete random variables, with a joint probability distribution $p_{ab}$ where $a \in A$ and $b \in B$ are the possible values found at each laboratory, see Section~\ref{Section:Measures-Informational-MutualInformation} for further details. With $N$ repeated samples, they arrive at a sequence of values $\{(a_{1},b_{1}),(a_{2},b_{2}),\ldots ,(a_{N},b_{N})\}$. However, neither laboratory knows the results of the other. If laboratory $B$ wants to transmit their results to $A$, how much information must they send? Shannon's noiseless coding theorem tells us that for large $N$, $B$ can send as little as ${\cal H}(B)$ bits per sample for $A$ to successfully reconstruct the complete sequence~\cite{shannon1948mathematical,nielsen2010quantum}. Such a scheme, however, does not exploit any correlations between the two random variables. If $A$ has some \emph{prior information} on $B$, i.e.~if $\mathcal{I}(A:B)>0$, then can $B$ exploit this to send fewer than $\mathcal{H}(B)$ bits per sample? The answer is yes, only the \emph{partial information} that $A$ is missing about $B$ needs to be transmitted, which can be achieved with $B$ sending only $\mathcal{H}(B|A)=\mathcal{H}(AB)-\mathcal{H}(A)$ ($\leq \mathcal{H}(B)$) bits of information per sample, provided $N$ is large~\cite{slepian1973noiseless}. Importantly, this may be completed without $B$ having any knowledge of the prior information that laboratory $A$ has.

An analogous problem may be studied in the quantum setting. Now consider a quantum source that emits a bipartite pure state $\ket{\psi_{i}}_{AB}$ with probability $p_{i}$ and distributes this state to laboratories $A$ and $B$. %, who do not know which pure state was emitted.
The laboratories know the statistics of the source and hence the mixed state $\rho_{AB} = \sum_{i} p_{i}\ket{\psi_{i}}\bra{\psi_{i}}_{AB}$, but not the ensemble $\{p_{i},\ket{\psi_{i}}_{AB}\}$ realising it. Suppose that the laboratories sample from this source $N$ times and have the sequence of (unknown) pure states $\{\ket{\psi_{i_{1}}}_{AB},\ket{\psi_{i_{2}}}_{AB},\ldots,\ket{\psi_{i_{N}}}_{AB}\}$. For each sample of the source, how can $B$ transmit their share of the distributed pure state to $A$? Both classical and quantum communication (and local operations) are at $B$'s disposal. However, classical communication is a much simpler task than quantum communication, which requires sending of sensitive quantum information down noisy channels. Hence, we allow for unlimited and free classical communication and concern ourselves with finding the most efficient way in terms of quantum communication for $B$ to transmit their share of the unknown state to $A$. With free classical communication we are able to carry out quantum communication via teleportation \cite{bennett1993teleporting}, using a bank of maximally entangled Bell states $\ket{\Phi}_{AB} = \frac1{\sqrt2}(\ket{0}_A\ket{0}_B + \ket{1}_A \ket{1}_B)$ shared between $A$ and $B$. The question then becomes: what is the minimum entanglement, quantified by shared Bell states (or ``ebits''), required per copy of $\rho_{AB}$ for $B$ to transmit their state to $A$?

This question was tackled in~\cite{horodecki2005partial,horodecki2007quantum}. Here it was shown that for any ensemble $\{p_{i},\ket{\psi_{i}}_{AB}\}$ realising the mixed state $\rho_{AB}$ and for $N \rightarrow \infty$, the rate of ebit consumption is at least $\mbox{max}\{0,\mathcal{S}_{B|A}(\rho_{AB})\}$, with $\mathcal{S}_{B|A}(\rho_{AB}) = \mathcal{S}(\rho_{AB}) - \mathcal{S}(\rho_{A})$ the conditional entropy of Eq.~(\ref{Equation:ConditionalVNEntropies}) (not to be confused with the measurement based quantity in Eq.~(\ref{Equation:ConditionalEntropyLGM})),
%\begin{equation}\label{Equation:ConditionalVNEntropies}
%\mathcal{S}_{B|A}(\rho_{AB}) = \mathcal{S}(\rho_{AB}) - %\mathcal{S}(\rho_{A})\, ,
%\end{equation}
representing the partial information needed for $B$ to transmit their state to $A$. However, it is a curious fact that the quantum conditional entropy can be negative for some $\rho_{AB}$ (consider, for example, any pure entangled state). In that case, $B$ may transmit their state to $A$ using only local operations and classical communication. Even better, it is then possible to \emph{create} shared ebits at a rate no more than $-\mathcal{S}_{B|A}(\rho_{AB})$. Thus, the negative quantum conditional entropy can be understood as the potential for future quantum communication between $A$ and $B$ in the form of teleportation.

The optimal protocol allowing $B$ to transmit their state to $A$ is called ``quantum state merging'', and can be equivalently thought of in the following way \cite{horodecki2007quantum}. Imagine that a third reference laboratory $C$ is present so that the composite state of the three systems $\ket{\psi}_{ABC}$ is pure. State merging is the task of performing LOCC and optimally drawing from (or contributing to) the bank of shared ebits to transform the state $\ket{\psi}_{ABC}^{\otimes N}$ into the state $\ket{\psi}_{ADC}^{\otimes N}$, where $D$ is an ancilla in laboratory $A$ designed to hold the state of $B$. The fidelity between $\ket{\psi}_{ABC}^{\otimes N}$ and $\ket{\psi}_{ADC}^{\otimes N}$ must be high, tending to unity as $N \rightarrow \infty$, which means that the local state of the reference laboratory $C$ is effectively unchanged throughout the process. %Working within this setting provides a rigorous framework that % allowed~\cite{horodecki2007quantum} to identify the quantum conditional entropy as the figure of merit in state merging, and
Working within this setting is useful in our following exposition on the role of QCs in state merging.

%Working again in the many copy setting, the objective is to use LOCC and, if needed, the available entanglement resources to transform the state $\ket{\psi}_{ABC}^{\otimes N}$ into a state $\ket{\psi}_{ADC}^{\otimes N}$, where $D$ is an ancilla in laboratory $A$ designed to hold the state of $B$. The final state $\ket{\psi}_{ADC}^{\otimes N}$ must have high fidelity with $\ket{\psi}_{ABC}^{\otimes N}$ (becoming identical to it in the limit $N \rightarrow \infty$). In the end, $B$ will be completely factorised, and separately $A$ and $B$ will have either used or created a number of ebits based on the value of $\mathcal{S}_{B|A}(\rho_{AB})$.

One link with QCs was provided in \cite{cavalcanti2011operational}. They noted the fact that entanglement between $A$ and $B$ is destroyed during the state merging process and so considered the total entanglement consumption $E^f_{A:B}(\rho_{AB})+\mathcal{S}_{B|A}(\rho_{AB})$, where $E^f_{A:B}(\rho_{AB})$ is the entanglement of formation between $A$ and $B$ (see Table~\ref{Table:Entanglement}). Here, $E^f_{A:B}(\rho_{AB})$ quantifies the entanglement present before merging (which is entirely consumed), and $\mathcal{S}_{B|A}(\rho_{AB})$ quantifies the entanglement gained or lost after merging. By manipulating Eq.~(\ref{Equation:Koashi}) \cite{koashi2004monogamy,fanchini2011conservation}, given that the global state of the tripartite system $ABC$ is pure, one finds that this total entanglement consumption is equal to the mutual information based measure of QCs,  i.e.~the quantum discord, of $\rho_{BC} = \mbox{Tr}_{A}(\ket{\psi}\bra{\psi}_{ABC})$ with one-sided measurements on $C$
\begin{equation}
E^f_{A:B}(\rho_{AB})+\mathcal{S}_{B|A}(\rho_{AB}) = Q_{C}^{\tilde{I}_{\mathcal{I}}}(\rho_{BC}).
\end{equation}
Hence, the QCs between $B$ and $C$, with measurements on $C$, can be quantitatively understood as the total entanglement consumption in quantum state merging. The authors of \cite{cavalcanti2011operational} also considered the corresponding asympotic total entanglement consumption, i.e.~\begin{equation}
 E^f_{A:B}(\rho_{AB})_{\infty}+\mathcal{S}_{B|A}(\rho_{AB}) = Q_{C}^{\tilde{I}_{\mathcal{I}}}(\rho_{BC})_{\infty} ,
\end{equation}
where $E^f_{A:B}(\rho_{AB})_{\infty}$ is the entanglement cost \cite{horodecki2009quantum}, thus providing an interpretation to the regularised measure of QCs given  in Eq.~(\ref{Equation:RegularisedDiscordJ}).

These findings give an operational motivation for the asymmetry present in the one-sided measures of QCs, that is, in general $Q_{C}^{\tilde{I}_{\mathcal{I}}}(\rho_{BC}) \neq Q_{B}^{\tilde{I}_{\mathcal{I}}}(\rho_{BC})$ because the total entanglement consumed for $B$ to merge with $A$ is not necessarily the same as for $C$ to merge with $A$. A link to dense coding was also provided in \cite{cavalcanti2011operational}. In dense coding, one can use quantum communication to transmit classical information at a faster rate than with classical communication, and it was shown that $Q_{C}^{\tilde{I}_{\mathcal{I}}}(\rho_{AC}) - Q_{C}^{\tilde{I}_{\mathcal{I}}}(\rho_{BC})$, with $\rho_{AC} = \mbox{Tr}_{B}(\ket{\psi}\bra{\psi}_{ABC})$, quantifies exactly the difference in the quantum advantage of dense coding for $C$ to communicate with $A$ as opposed to $B$.

An alternative operational view of QCs in terms of state merging was found in \cite{madhok2011interpreting}. Here it was shown that the regularised QCs $Q_{A}^{\tilde{I}_{\mathcal{I}}}(\rho_{AB})_{\infty}$ of Eq.~(\ref{Equation:RegularisedDiscordJ}) represent the minimum increase in the cost of state merging (in terms of the rate of ebit consumption) if one first performs  local measurements on each copy of $A$. Specifically, if $\mathcal{S}_{B|A}(\rho_{AB})$ is the cost of state merging before and $\mathcal{S}_{B|A}(\tilde{\Pi}_{A}[\rho_{AB}])_{\infty}$ is the cost of state merging after each LGM on $A$, then
\begin{equation}
%=Q_{A}^{\tilde{I}_{\mathcal{S}_{B|A}}}(\rho_{AB})_{\infty}=
\inf_{\tilde{\Pi}_{A}}\mathcal{S}_{B|A}(\tilde{\Pi}_{A}[\rho_{AB}])_{\infty}-\mathcal{S}_{B|A}(\rho_{AB}) = Q_{A}^{\tilde{I}_{\mathcal{I}}}(\rho_{AB})_{\infty}.
\end{equation}
This operational link was then extended in \cite{madhok2011quantum,madhok2013quantum} by considering what is known as the fully quantum Slepian-Wolf protocol \cite{abeyesinghe2009mother}, which is the ``mother of all protocols'' since it is so general that it contains as special cases quantum state merging, dense coding, teleportation and entanglement distillation \cite{nielsen2010quantum,wilde}. Again, the regularised QCs $Q_{A}^{\tilde{I}_{\mathcal{I}}}(\rho_{AB})_{\infty}$ represent the minimum drop in performance of the fully quantum Slepian-Wolf protocol after
local measurements on $A$.

Finally, in \cite{streltsov2015concentrating} the primitive of ``information concentration'' was studied, a variation of state merging  in which two parties $A$ and $B$ aim to maximise their mutual information by means of LOCC operations performed by $B$ and a third party $C$. The figure of merit for this protocol was found to be a tripartite generalisation of the quantum discord.

\subsubsection{Quantum state redistribution}\label{Section:QSR}

%Add Wilde2 email

The quantum state redistribution protocol consists of a sender holding two systems $A$ and $C$, a receiver holding instead only one system $B$, and a reference system $R$ that is inaccessible to both sender and receiver, all sharing a four-partite pure state $|\psi_{ACBR}\rangle$ \cite{devetak2008exact}. The objective is to redistribute the quantum information in such a way that is instead the receiver who holds $C$, while retaining the purity of the overall four-partite state. In order to accomplish this task, sender and receiver are only allowed to perform local quantum operations, consume or generate ebits, and the sender can give only qubits to the receiver. More precisely, contrarily to the quantum teleportation protocol \cite{bennett1993teleporting}, they are not allowed to classically communicate beyond what can be encoded in qubits. It turns out that this transfer can occur perfectly in the asymptotic limit of many copies provided that the amount of qubits that the sender gives to the receiver, also called communication cost, is at least half the conditional quantum mutual information $\mathcal{I}_B(\rho_{CBR})$ of the reduced state $\rho_{CBR}=\mbox{Tr}_A(|\psi_{ACBR}\rangle\langle\psi_{ACBR}|)$ between $C$ and $R$ with respect to $B$; see \cite{devetak2008exact} for more details about the optimal protocol. Interestingly, this protocol is self-dual under time reversal, i.e., it can be reversed by generating the same amount of entanglement and spending the same communication cost. In particular, switching $A$ and $B$ has no effect on the communication cost, indeed for any pure state $|\psi_{ACBR}\rangle$ it happens that $\mathcal{I}_B(\rho_{CBR})=\mathcal{I}_A(\rho_{ACR})$, with $\rho_{ACR}=\mbox{Tr}_B(|\psi_{ACBR}\rangle\langle\psi_{ACBR}|)$. We can thus think of the quantity $\mathcal{I}_B(\rho_{CBR})=\mathcal{I}_A(\rho_{ACR})$ as twice the minimal communication cost to transfer $C$ between $A$ and $B$, regardless of the direction of the transfer, while retaining the purity of the global four-partite state $|\psi_{ACBR}\rangle$.

The quantum state redistribution protocol provides us with another operational interpretation for the quantum discord, as first highlighted in \cite{wilde2015multipartite}. Let us consider two laboratories holding, respectively, the quantum systems $A$ and $B$, which share the bipartite quantum state $\rho_{AB}$.
%Let $|\psi_{ABR}\rangle$ be a purification of $\rho_{AB}$, i.e., such that $\rho_{AB}=\mbox{Tr}_R(|\psi_{ABR}\rangle\langle\psi_{ABR}|)$, where $R$ is an environment inaccessible to both $A$ and $B$.
Suppose that a maximally informative LGM $\tilde{\Pi}_A$ is performed on $A$, with $A'$ denoting the corresponding classical output. As already mentioned in Section~\ref{Section:Measures-Surprisal}, such LGM can be written as a unitary $U_{A\rightarrow A'C}^{\tilde{\Pi}_A}$ from $A$ to the composite system $A'C$ followed by discarding $C$, i.e.,
\begin{equation}\label{eq:measurementinterpretation}
\tilde{\Pi}_A[\rho_{AB}]=\mbox{Tr}_C\big(U_{A\rightarrow A'C}^{\tilde{\Pi}_A}[\rho_{AB}]\big).
\end{equation}
%where $C$ is an environment that is inaccessible to both $A$ and $B$.
Moreover, we also know that the quantum discord $Q_A^{\tilde{I}_{\mathcal{I}}}(\rho_{AB})$ between $A$ and $B$, Eq.~(\ref{Equation:QuantumDiscordCM}), is exactly given by the conditional quantum mutual information $\mathcal{I}_{A'}(\rho_{A'CB})$ of the state $\rho_{A'CB}=U_{A\rightarrow A'C}^{\tilde{\Pi}_A}[\rho_{AB}]$ between $C$ and $B$ with respect to $A'$, i.e.,
\begin{equation}\label{eq:wildeoperationalinterpretation}
Q_A^{\tilde{I}_{\mathcal{I}}}(\rho_{AB}) = {\cal I}_{A'}\big(\rho_{A'CB}\big).
\end{equation}

Now, let $R$ be a reference system purifying the state $\rho_{A'CB}$, i.e., such that $\rho_{A'CB}=\mbox{Tr}_R(|\psi_{A'CBR}\rangle\langle\psi_{A'CBR}|)$. By looking at Eq.~(\ref{eq:measurementinterpretation}), we can then think of the measurement $\tilde{\Pi}_A$ as a process whereby the environment $C$ is lost and given to the reference $R$. One can then ask the following question. What is the optimal quantum communication cost needed to send system $C$ from $R$ to laboratory $A$ in such a way that the action of the LGM $\tilde{\Pi}_A$ can be undone? As already mentioned, this is given by half the conditional quantum mutual information ${\cal I}_{A'}\big(\rho_{A'CB}\big)$ between $C$ and $B$ with respect to $A'$, by assuming that laboratory $B$ plays no role in the protocol. Therefore, Eq.~(\ref{eq:wildeoperationalinterpretation}) provides us with the promised operational interpretation of quantum discord as twice the optimal communication cost needed to restore the environment of a measurement so that it is no longer lost, i.e., to restore the coherence lost in a measurement \cite{wilde2015multipartite}. Due to the  symmetry under time reversal of such optimal protocol, the quantum discord between $A$ and $B$ can also be seen as the communication cost needed to send system $C$ back to the reference $R$, i.e., as the amount of quantum information lost in the measurement process \cite{wilde2015multipartite}, which neatly captures the physical motivation behind the original definition of discord \cite{ollivier2001quantum}.

Building upon this operational interpretation, in \cite{berta2016deconstruction} it has been shown that the quantum discord $Q_A^{\tilde{I}_{\mathcal{I}}}(\rho_{AB})$ of a bipartite quantum state $\rho_{AB}$ is equal to the minimal rate of noise that one needs to apply to subsystem $A$ in such a way that: (i) the resulting state $\tilde{\rho}_{AB}$ is locally recoverable after an LGM acts on $A$ and (ii) the post-measurement state corresponding to $\tilde{\rho}_{AB}$ after such an LGM is indistinguishable from the post-measurement state corresponding to $\rho_{AB}$ after a maximally informative LGM. A state satisfying properties (i) and (ii) is also said to be approximately einselected, where ``einselection'' is the abbreviation for environment-induced superselection and is a process whereby the interaction between a quantum system and the environment is such that only the eigenstates corresponding to particular observables of the system, so-called pointer states, persist in the system \cite{zurek2003decoherence}. In other words, quantum discord is equal to the optimal cost of simulating einselection.  This perhaps stands as a physical interpretation of quantum discord that is even more in line with its own original definition as a measure of the decrease of correlations after einselection is complete \cite{ollivier2001quantum}.

\subsubsection{Remote state preparation}\label{Section:RSP}

In quantum teleportation \cite{bennett1993teleporting,pirandola2015advances}, the objective is to transmit an unknown quantum state from laboratory $A$ to laboratory $B$. To achieve such a feat, one must make use of classical communication and some distributed entanglement. In fact, to teleport a qubit it is necessary and sufficient to use two bits of classical communication and one ebit \cite{bennett1993teleporting}. Instead, in ``remote state preparation'', the objective is to transmit a \emph{known} qubit state. It was shown by \cite{bennett2001remote,bennett2005remote} that in the many copy setting, this is achievable with the use of one classical bit of communication and one shared ebit per copy, and that in general it is possible to trade-off these two quantities: one can pay with more classical bits to save on ebits, and vice versa.

One special case is remote state preparation of the qubit states of the form $\ket{\psi} = \frac{1}{\sqrt{2}}(\ket{0}+e^{i \phi} \ket{1})$, which lie on the equator of the Bloch sphere, as discussed in \cite{lo2000classical,pati2000minimum}. In this case, remote state preparation can be performed without resorting to the many copy limit. Consider the shared ebit $\frac{1}{\sqrt{2}}(\ket{01}-\ket{10})$, which can be equivalently written (up to a global phase) as $\frac{1}{\sqrt{2}}(\ket{\psi \psi^{\perp}}-\ket{\psi^{\perp} \psi})$, where $\ket{\psi^{\perp}}= \frac{1}{\sqrt{2}}(\ket{0}-e^{i \phi} \ket{1})$ is the qubit state orthogonal to $\ket{\psi}$. If $A$ performs an LPM on their half of the ebit in the basis $\{\ket{\psi},\ket{\psi}^{\perp}\}$, then it is clear that $B$ will either get the state $\ket{\psi^{\perp}}$, if the LPM by $A$ resulted in $\ket{\psi}$, or the state $\ket{\psi}$, if $A$ found $\ket{\psi^{\perp}}$. If $B$ has $\ket{\psi^{\perp}}$, then $\ket{\psi}$ can be retrieved by simply applying a $\pi$ rotation around the $z$ axis of the Bloch sphere, i.e.~ $\ket{\psi}=\sigma_{3}\ket{\psi^{\perp}}$, where $\sigma_{3}$ is the third Pauli matrix. However, $B$ can only know whether to apply the rotation from the measurement outcome of $A$, and hence remote state preparation of equatorial states in this single copy case requires one shared ebit and one bit of classical communication.

If one has access to an ebit as a resource in the above protocol, then remote state preparation can be achieved perfectly. Instead, if one uses a more general resource state, such as a mixed state of two qubits $\rho_{AB}$, then an error is introduced into the remote state preparation. Consider the state to be prepared with Bloch vector $\vec{m}$ lying on the equatorial plane, and the resultant state after the aforementioned protocol with Bloch vector $\vec{n}$.  In \cite{dakic2012quantum}, a quadratic payoff function $P^2_{\vec{m},\vec{n}} := (\vec{m} \cdot \vec{n})^{2}$, which captures how well the target $\vec{m}$ and the actual $\vec{n}$ overlap and is linked to the fidelity between the corresponding states, was adopted to quantify the performance of the protocol. Precisely, for a given resource state $\rho_{AB}$, a figure of merit for remote state preparation can be defined by first maximising $P_{\vec{m},\vec{n}}$ over all output states $\vec{n}$ (or equivalently all possible LPMs performed by $A$), then averaging the result over all pure states $\vec{m}$ on the equator, and finally minimising the latter quantity over all possible choices of the reference north pole state on the Bloch sphere, since the orientation of the axes can be arbitrary, as long as this is pre-agreed between $A$ and $B$. The resulting quantity, that we denote by $\overline{P}^2_A(\rho_{AB})$, can be interpreted as the remote state preparation fidelity achievable with the shared state $\rho_{AB}$ in a worst case scenario. Interestingly, it was shown in \cite{dakic2010necessary} that, for a class of two-qubit states $\rho_{AB}$ (i.e.~the Bell diagonal states), this quantity coincides with (twice) the Hilbert-Schmidt based geometric measure of QCs,
\begin{equation}\label{Equation:Popoff}
\overline{P}^2_A(\rho_{AB}) = 2 Q_{A}^{G_{2}}(\rho_{AB}),
\end{equation}
thus apparently providing an operational interpretation to the latter, even though in a specialised setting.
The theoretical result was supported experimentally using a photonic implementation, where certain separable states with nonzero QCs were shown to  perform remote state preparation better than some entangled states with smaller QCs, according to the figure of merit given by Eq.~(\ref{Equation:Popoff}).

A refinement of this protocol has been presented in \cite{horodecki2014can}, where it is argued, however, that the figure of merit $\overline{P}^2_A(\rho_{AB})$ can be misleading, since it can be surpassed simply by employing the trivial protocol of $B$ randomly preparing a pure equatorial state, regardless of the communication from $A$ and the shared quantum resource $\rho_{AB}$. Instead, the suggested figure of merit should derive from the (non-quadratic) payoff-function $P_{\vec{m},\vec{n}} = \vec{m} \cdot \vec{n}$. Generalisations of the encoding and decoding strategies of $A$ and $B$ are also outlined in \cite{horodecki2014can}. Instead of an LPM on the shared resource $\rho_{AB}$, $A$ can perform a more general two-outcome LGM, and send the result using one bit of classical communication to $B$. $B$ is then allowed to utilise \emph{any} local operations to recover the best approximation of $\vec{m}$. In this more general setting, it is proved that it is impossible to use a shared separable state to outperform an entangled state in remote state preparation, hence providing a very critical analysis of the role of QCs beyond entanglement in this communication task. However, their role can be resurrected when certain reasonable restrictions are placed upon the local operations that $B$ is able to perform when trying to recover $\vec{m}$. Two important cases are highlighted where it is possible for quantumly correlated separable states to outperform entangled ones: (i) $B$ does not share a local Bloch reference frame with $A$, %and hence can only perform operations with some symmetry about the pole $\ket{p}$
and (ii) $B$ is restricted to unital operations. Even stronger, under condition (ii) and in the case of the shared resource state $\rho_{AB}$ being Bell diagonal, the remote state preparation figure of merit recovers again a link to the Hilbert-Schmidt based geometric measure of QCs.

A further criticism of the role of QCs in remote state preparation was given by \cite{giorgi2013quantum}. Here, they discriminate between the QCs that can be created by local operations and those that cannot, and it is argued that only the latter should be considered. In this sense, it is then shown that there are quantumly correlated states that are useless for remote state preparation, while there are states without QCs that can be made useful. Furthermore, it is discussed in \cite{chaves2011noisy} that an increase in QCs does not necessarily imply an increase in the quality of remote state preparation, even though in some cases a quantitative link is found between the fidelity of remote state preparation using noisy cluster states and the negativity of quantumness $Q_{A}^{E^{N}}$ defined in Section~\ref{Section:Measures-NoQ}.

Finally, in \cite{tufarelli2012quantum} a version of remote state preparation is discussed for a system consisting of a qubit and a harmonic oscillator. It is found that the geometric measure $Q_{A}^{G_{2}}(\rho_{AB})$, where $A$ is the qubit, can provide a lower bound to the payoff of remote state preparation under certain restrictions. These restrictions are on the receiving oscillator $B$: if $B$ is restricted to only local unitary corrections following the measurement by $A$ then $Q_{A}^{G_{2}}(\rho_{AB})$ presents a limit on the performance of remote state preparation, whereas if $B$ is free to perform arbitrary (non-unitary) local operations then $Q_{A}^{G_{2}}(\rho_{AB})$ becomes no longer relevant, and the payoff reduces to an entanglement measure. This lends support to the findings of \cite{horodecki2014can}.

\subsubsection{Quantum cryptography}\label{Section:Applications-Cryptography}
The idea behind ``quantum key distribution'' is to harness quantum mechanics to tackle the classically problematic task of distributing cryptographic keys securely, without a third party eavesdropper discovering what has been communicated \cite{gisin2002quantum}. In fact, the distributors and receivers of a  key encoded in a {\it quantum} system can detect an attempt at eavesdropping (up to a sensitivity threshold) due to the disturbance induced by measurements on their system. From the outset, this problem seems tailored to utilising the resource of QCs and in particular the inherent non-orthogonality of bipartite states with nonzero QCs.
Consequently, it was shown in \cite{pirandola2014quantum} that quantumly correlated states are necessary for quantum key distribution in the device-dependent setting --- which corresponds to the distributing and receiving parties being able to trust their devices.% and the noise present in the environment.

In the practically relevant case of distribution of secure keys over a lossy channel with transmissivity $\eta$, such as a free-space link or an optical fibre, the secret key capacity $K$ has been very recently determined and found to coincide with the maximum quantum discord that can be distributed to the remote parties through such a lossy channel \cite{pirandola2015ultimate}. In the following we briefly explain this result, which connects QCs to the ultimate limits of quantum communication in continuous variable systems. Further details are available in \cite{pirandola2015ultimate}.

In general, a quantum cryptographic protocol \cite{gisin2002quantum} involves two distant laboratories, a sender $A$ and a receiver $B$, separated by a quantum channel. The sender prepares ensembles of (non-orthogonal) input states and transmits them to the receiver, who can measure the outputs. They can resort to unlimited classical communication, which allows them to extract a key through error correction and privacy amplification. The secret key capacity $K$ is defined as the maximum number of secret key bits per channel use which can be distributed over the quantum channel, obtained upon maximising over all possible input states of the sender $A$ and all possible output measurements by the receiver $B$ (in general, the input-output local operations may be adaptive, i.e., assisted by two-way classical communication~\cite{pirandola2015ultimate}). An achievable protocol can be represented in terms of  $N$ ebits of entanglement distributed over the channel, followed by LOCC on both parties. Let us denote by $\rho_{AB}^{\infty}$ the output state of $A$ and $B$, in the limit $N \rightarrow \infty$ (i.e., in the limit of an ideal maximally entangled  distributed state, also known as an EPR state \cite{einstein1935can}). The result in \cite{pirandola2015ultimate} then shows that, in the important practical case of a bosonic lossy channel with transmissivity $\eta$, it holds
\begin{equation}\label{Equation:Pirla}
K(\eta)=Q^{\tilde{I}_{\mathcal{I}}}_A(\rho_{AB}^{\infty})=-\log_2(1-\eta)\,,
\end{equation}
 where in particular Gaussian LGMs are found to be optimal to calculate the quantum discord $Q^{\tilde{I}_{\mathcal{I}}}_A(\rho_{AB}^{\infty})$ of the asymptotic two-mode Gaussian state $\rho_{AB}^{\infty}$, due to the results of \cite{adesso2010quantum,pirandola2014optimality}.

 The further exploration of the role and power of QCs in quantum key distribution protocols certainly has potential  \cite{pati2012quantum,pirandola2014quantum,pirandola2015ultimate}, but its scope exceeds the present review, and we point the interested reader to \cite{gisin2002quantum,weedbrook2012gaussian} for an overview on quantum cryptography.

\subsubsection{Quantum locking of classical correlations}
\label{Section:Applications-Locking}

Let us begin in the classical setting by considering a generic measure of correlations between two random variables $A$ and $B$. %One key property that we might expect this measure to obey %is termed \emph{proportionality}~\cite{divincenzo2004locking}, which requires
We might expect that this measure cannot increase by more than $m$ bits if one were to carry out $m$ bits of classical communication. Indeed, this intuitive property holds in particular for the mutual information $\mathcal{I}(A:B)$ of Eq.~(\ref{Equation:MutualInformationClassical})~\cite{divincenzo2004locking}. How does this feature extends to the quantum setting, where quantum communication is also possible? Suppose we apply a scheme of local operations and either $n$ qubits of quantum communication or $2n$ bits of classical communication, acting on a bipartite state $\rho_{AB}$ shared between laboratory $A$ and $B$. We may now require that a quantifier of correlations cannot increase by more than $2n$ bits after this scheme, since we can view each qubit of quantum communication as effectively two bits of classical communication due to dense coding. This property holds indeed for the quantum mutual information $\mathcal{I}(\rho_{AB})$ of Eq.~(\ref{Equation:MutualInformation})~\cite{terhal2002entanglement}.%, i.e.
%\begin{equation}
%\mathcal{I}(\Lambda^{(m)}_{AB}(\rho_{AB})) - \mathcal{I}(\rho_{AB}) \leq 2m,
%\end{equation}
%where $\Lambda^{(n)}_{AB}$ is our scheme of local operations and quantum or classical communication using $m$ qubits and $2m$ bits.

Interestingly, this property does not hold for the classical correlations $\tilde{\mathcal{J}}_{AB}$ of %either Eq.~(\ref{Equation:ClassicalCorrelations}) (with local measurements on $A$) or
Eq.~(\ref{Equation:ClassicalCorrelationsAB}), representing the correlations available to laboratories $A$ and $B$ through local measurements. Consider the bipartite $(2N+1)$-qubit state $\rho_{AB}$ with $A$ having $N$ qubits and $B$ having $N+1$,
\begin{equation}\label{Equation:LockingState}
\rho_{AB} = \frac{1}{2^{N+1}} \sum_{i=0}^{2^{N}-1} \sum_{j=0}^{1} ( U_{j} \ket{i}\bra{i} U_{j}^{\dagger} )_{A} \otimes (\ket{i}\bra{i} \otimes \ket{j}\bra{j} )_{B}
\end{equation}
where $U_{0} = \mathbb{I}$ and $|\braket{i'|U_{1}|i}|^{2} = 1/2^{N}$, i.e.~with $\{U_{1}\ket{i}\}$ producing a mutually unbiased basis with respect to the basis $\{\ket{i}\}$~\cite{durt2010mutually}. For example, we may fix $\{\ket{i}\}$ to be the computational basis and choose $U_{1} = H^{\otimes N}$, with $H$ the Hadamard gate. This state may be prepared by having laboratory $B$ pick a random $N$-bit string with label $i$ and sending to $A$ either $U_{0}\ket{i}=\ket{i}$ or $U_{1}\ket{i}$, again at random based on the value $j\in \{0,1\}$. The total correlations are here $\mathcal{I}(\rho_{AB})=N$, but the classical correlations are $\tilde{\mathcal{J}}_{AB}(\rho_{AB})=N/2$. Hence, some of the correlations are \emph{locked}, i.e.~inaccessible to $A$ and $B$ through local measurements. However, if $B$ sends the result $j$ to laboratory $A$, then $A$ can reverse $U_{j}$.
% so that they together share
%\begin{equation}
%\Lambda^{(1)}_{AB}(\rho_{AB}) = \frac{1}{2^{N+1}} \sum_{i=0}^{2^{N}-1} \sum_{j=0}^{1} (\ket{i}\bra{i} \otimes \ket{j}\bra{j} )_{A} \otimes ( \ket{i}\bra{i} )_{B}.
%\end{equation}
Laboratory $A$ then measures in the computational basis so that they together share the $N$-bit string $i$, and therefore $N+1$ bits of classical correlations (including the communicated bit). They have thus \emph{unlocked} $N/2$ bits of classical correlations with only one bit of classical communication.

%\begin{equation}
%\tilde{\mathcal{J}}_{AB}(\Lambda^{(1)}_{AB}(\rho_{AB}))-\tilde{\mathcal{J}}_{AB}(\rho_{AB})=N/2,
%\end{equation}
%i.e. the classical correlations have been \emph{unlocked} with only one bit of classical communication.

This curious phenomenon is a truly quantum effect, yet can occur in separable states like in Eq.~(\ref{Equation:LockingState}). Clearly then, entanglement is not a figure of merit in locking of classical correlations, so what about more general QCs? The first hints that QCs play a role were given in~\cite{wu2009correlations,datta2009signatures}. There it was noted that both the two-sided mutual information based measure of QCs $Q_{AB}^{\tilde{I}_{\mathcal{I}}}(\rho_{AB})$ and the measurement induced disturbance (see Section~\ref{Section:Measures-Informational-MutualInformation}) are equal to $N/2$ in the above example, exactly the amount of locked classical correlations.

The link with QCs was placed on a firmer footing in~\cite{boixo2011quantum}. In their setting, laboratory $B$ samples from a random source of numbers $i$ and wants to send the result to $A$, but has to use a noisy quantum channel so that $A$ eventually receives $\rho_{A}^{(i)}$. Together $A$ and $B$ share a quantum-classical state $\rho_{AB} \in \mathcal{C}_{B}$, just like the one in Eq.~(\ref{Equation:LockingState}), and the objective is for $A$ to infer the value of $i$. %The ``accessible information'' of $A$~\cite{nielsen2010quantum} tells us how well this task can be done, and coincides with the classical correlations $\tilde{\mathcal{J}}_{A}(\rho_{AB})$ ($=\tilde{\mathcal{J}}_{AB}(\rho_{AB})$).
The classical correlations tell us how well $A$ can do this. If laboratory $B$ can then send a key to reveal the value of $i$ to $A$, it turns out that the amount of classical correlations unlocked in doing so is exactly the quantum discord $Q_{A}^{\tilde{I}_{\mathcal{I}}}(\rho_{AB})$ \cite{boixo2011quantum}. %(note that this is equal to the two-way measure $Q_{AB}^{\tilde{I}_{\mathcal{I}}}(\rho_{AB})$ for quantum-classical $\rho_{AB}$).
Furthermore, in the asymptotic limit of many copies of $\rho_{AB}$, it was shown that the regularised QCs measure $Q_{A}^{\tilde{I}_{\mathcal{I}}}(\rho_{AB})_{\infty}$ of Eq.~(\ref{Equation:RegularisedDiscordJ}) quantifies the quantum advantage for $A$ to successfully infer the message from $B$ when compared to a corresponding classical protocol, in terms of the key length per copy required to unlock the correlations;  $Q_{A}^{\tilde{I}_{\mathcal{I}}}(\rho_{AB})_{\infty}$ then also gives the amount of classical correlations unlocked per copy in the quantum protocol \cite{boixo2011quantum}.

\subsection{Quantum metrology and discrimination}\label{Section:Applications-Metroid}
Precision measurements are of key relevance in all quantitative sciences and underpin many technological applications, such as navigation, sensing, and medical imaging \cite{giovannetti2011advances}. It is important therefore to identify the resources which can lead to a precision enhancement in suitable implementations. Multipartite entanglement has been extensively investigated as a resource for quantum enhanced measurements, see e.g.~\cite{smerzi2014review,toth2014review,smerzi2016review}. Here we review the role played by QCs in overcoming classical limitations for certain practical tasks of estimation and discrimination.

\subsubsection{Black box quantum metrology}\label{Section:Applications-Metrology}

Quantum metrology studies how to harness quantum mechanical effects to enhance the precision in estimating physical quantities not amenable to a direct observation \cite{giovannetti2004quantum,giovannetti2006quantum,giovannetti2011advances}. A relevant class of problems in quantum metrology can be formalised in terms of phase estimation in an interferometric setup, which is akin to the scheme in Fig.~\ref{Fig:Defs}(c). The estimation procedure then consists of the following steps. An input state $\rho_{AB}$ enters a two-arm channel, in which the idler subsystem $B$ is unaffected while the probe subsystem $A$  undergoes a local unitary transformation $U_A$, so that the output density matrix  can be written as $\rho_{AB}^{\varphi} := (U_A \otimes \mathbb{I}_B) \rho_{AB} (U_A \otimes \mathbb{I}_B)^{\dagger}$, with $U_A=e^{-i \varphi H_A}$, where  $\varphi$ is the parameter one aims to estimate and $H_A$ is a (non-degenerate) Hamiltonian generator of the local unitary dynamics. Finally, the information on  $\varphi$  is recovered by constructing an unbiased estimator  $\hat{\varphi}$ obtained by classical processing of the data resulting from (possibly joint) measurements of suitable observables on the output state $\rho_{AB}^\varphi$.

For any  input state $\rho_{AB}$ and known generator $H_A$, and provided  $n$ i.i.d. iterations of the probing procedure are implemented, the maximum achievable precision is determined theoretically by the  quantum Cram\'er-Rao bound \cite{braunstein1994statistical}, which says that the mean square error $\Delta^2 \hat{\varphi}$ of any unbiased estimator scales as
\begin{equation}\label{Equation:CramerRao}
n \  \Delta^2 \tilde{\varphi}   \geq \frac{1}{C^{\text{QF}}(\rho_{AB}, H_A \otimes \mathbb{I}_B)}\,,
\end{equation}
where $C^{\text{QF}}$ is the quantum Fisher information defined in Table~\ref{Table:Coherence}, which quantifies how sensitive the output state is to an infinitesimal change of the encoded parameter $\varphi \rightarrow \varphi + \delta \varphi$.
There always exists an optimal measurement strategy at the output stage which makes the inequality in Eq.~(\ref{Equation:CramerRao}) asymptotically tight for $n \gg 1$, which means that the quantum Fisher information $C^{\text{QF}}(\rho_{AB}, H_A \otimes \mathbb{I}_B)$ can be regarded as the figure of merit determining the optimal precision achievable in the estimation of a parameter unitarily encoded by the local generator $H_A$ when using an input state $\rho_{AB}$. Therefore, under the assumption of complete prior knowledge of $H_A$, it is clear that coherence in the eigenbasis of $H_A$ (or, more precisely, asymmetry with respect to the group of transformations with unitary representation $U_A=e^{-i \varphi H_A}$) \cite{marvian2015quantum} is the essential resource for the estimation. Since maximal coherence in a known reference basis can be achieved by a superposition state of subsystem $A$ only, there is no need for any correlated idler ancilla $B$ at all in this conventional  setting.

In \cite{girolami2014quantum}, the rules of the game were changed by considering a ``black box'' paradigm in which the eigenbasis of the generator $H_A$ is completely unknown a priori (while only its spectrum is known). We can imagine e.g.~a referee controlling the local dynamics on $A$ (i.e.~like an examiner who is setting an exam), who decides the basis of $H_A$ (i.e.~the question to ask) only after the experimenters selected the input state $\rho_{AB}$ (i.e.~completed their preparation), and then discloses the choice (i.e.~asks the question) so that optimal measurements can still be performed at the output stage (i.e.~the best possible answer can be returned, given the initial preparation). Assuming the referee is fully adversarial (i.e.~the unlucky case that the examinees get asked the question they are least prepared for), the meaningful figure of merit for the protocol has to be defined in a worst case scenario, by considering the minimum achievable precision over all possible bases of $H_A$. In formulae, this corresponds to minimising the quantum Fisher information,
\begin{equation}\label{Equation:InterferometricPower}
\inf_{H_A} C^{\text{QF}}(\rho_{AB}, H_A \otimes \mathbb{I}_B) =: Q_A^{C^{\text{QF}}}(\rho_{AB})\,,
 \end{equation}
 over all local generators $H_A$ of fixed spectral class.
 This is exactly the definition of the interferometric power $Q^{C^{\text{QF}}}_A(\rho_{AB})$ discussed in Section~\ref{Section:Measures-Asymmetry}, whose operational meaning is now very clear: the degree of QCs of any state $\rho_{AB}$ measured by the interferometric power amounts to the minimum precision that the state $\rho_{AB}$ guarantees for the estimation of a parameter $\varphi$ encoded in a local unitary dynamics on $A$, in the worst case scenario in which the  eigenbasis of the generator of said unitary is initially completely unknown.
The more the QCs content of the state $\rho_{AB}$ according to $Q_A^{C^{\text{QF}}}$, the more the state $\rho_{AB}$ will be useful for phase estimation on $A$ with respect to any possible non-trivial generator (i.e., in our analogy, the better the candidates will be confident to respond to any possible exam question, securing a higher base mark).

This connection was investigated experimentally in \cite{girolami2014quantum} in a nuclear magnetic resonance implementation of black box quantum metrology, in which classical-quantum states were shown to be unable to estimate the parameter in case of a most adverse setting of $H_A$, while states with nonzero QCs were found to successfully accomplish the task in all tested settings, with precision bounded from below by their interferometric power.
The paradigm of black box metrology was also extended to the technologically relevant case of optical interferometry with Gaussian states and local Gaussian unitary dynamics in \cite{adesso2014gaussian}, in particular elucidating the roles of QCs, entanglement, and state mixedness in order to maximise the performance in the phase estimation task.

While the above results provide a concrete scenario in which QCs beyond entanglement are found to play centre stage, in practice one may want to assess instead the versatility of input states $\rho_{AB}$ in terms of their average metrological performance, rather than their worst case scenario only. One can then introduce alternative figures of merit e.g.~by replacing the minimum with an average according to the Haar measure in Eqs.~(\ref{Equation:AsymmetryMeasures}). Such a study has been carried out in \cite{farace2016building} by defining the  average local Wigner-Yanase skew information $\overline{Q}_A^{C_{\text{WY}}}(\rho_{AB})$ (see Table~\ref{Table:Coherence}). Unlike the minimum, which defines the local quantum uncertainty $Q_A^{C_{\text{WY}}}(\rho_{AB})$, the average local skew information is found not to be a measure of QCs anymore. In particular, it vanishes only on states of the form $\frac {\mathbb{I}_A}{d_A} \otimes \tau_B$, that is, tensor product states between a maximally mixed state on $A$, and an arbitrary state on $B$ \cite{farace2016building}. This entails that, to ensure a reliable discrimination of local unitaries on average (rather than in the worst case), the input states need to have two resource ingredients: local purity of the probing subsystem $A$, and correlations (of any nature) with the ancilla. The interplay between the average and the minimum performance, as well as a study on the role of entanglement, are detailed in
\cite{farace2016building}.
%A further variation on the black box paradigm has been investigated recently, by considering parameter estimation in the case of tunable, partial prior knowledge of the generator of the unitary dynamics encoded the parameter; however, the role of QCs has not beeen inves

Let us finally mention an earlier study  of QCs in metrology. The authors of \cite{modi2011quantum} investigated unitary phase estimation using $N$-qubit probe states (with phase transformation applied to each qubit) initialised in mixed states with a) no correlations; b) only classical correlations; or c) non-classical correlations (QCs or entanglement); all classes of probe states having the same (tunable) degree of purity for fairness of comparison. They found that uncorrelated and classically correlated probes resulted in a quantum Fisher information scaling linearly with $N$, while quantum strategies allowed for a quadratic scaling $N^2$, as expected in quantum metrology \cite{giovannetti2006quantum}. For the particular classes of probe states considered in their work for part c), they observed that the enhancement, compared to case a), persisted even when the mixedness was so high that entanglement disappeared. Therefore they argued that QCs (which were found to increase with $N$ according to a measure based on the relative entropy distance from the set of $N$-qubit fully classical states, i.e.~the multipartite extension of the relative entropy of quantumness) may be responsible for this enhancement. It is however still unclear whether these conclusions are special to the selected classes of states, or can be further extended to more general settings.

\subsubsection{State discrimination and quantum illumination}\label{Section:Applications-Discrimination}

%- worst case quantum discrimination: Spehner~\cite{spehner2013geometric,spehner2013geometric2}, dissonance~\cite{roa2011dissonance,li2012assisted}, maybe this~\cite{chen2011detecting}. Nice (big) review~\cite{spehner2014quantum}.

%- device independent quantum reading: Inception~\cite{pirandola2011quantum}, showing that it's to do with QCs (of response)~\cite{roga2015device} and done experimentally~\cite{dall2012experimental}.

There exist direct links between measures of QCs and the task of ambiguous quantum state discrimination,
%. The Bures distance-based measure of one-way discord-type correlations is directly linked to the  success  in the ambiguous quantum state discrimination protocol~\cite{spehner2013geometric,spehner2014quantum}
which in turn plays a key role both in quantum communication and cryptography \cite{bae2015quantum}.
In this protocol, a family of $n$ known states $\{\rho_i\}_{i=1}^n$ encodes a message.  A sender randomly selects the states from this family via a probability distribution $\{p_i\}_{i=1}^n$ and gives them one by one to a receiver, whose task is to identify them and thus decode the message. To do this, the receiver performs a generalised measurement $\{\mu_i\}_{i=1}^n$ with $n$ outcomes
%\footnote{A generalised measurement with $n$ outcomes is described by a set of $n$ positive operators $M_i \geq 0$ satisfying $\sum_{i=1}^n M_i = \mathbb{I}$. The probability of getting the $j^{\rm th}$ outcome, after that such a measurement has been performed on a system in the state $\rho$, is given by $\mbox{Tr}(\rho M_i)$.}
 on each of the states given to them by the sender and concludes that the received state is $\rho_j$ when the measurement outcome is the $j^{\rm th}$ one. Since the states $\{\rho_i\}_{i=1}^n$ need not be orthogonal, there is in general no measurement that can perfectly distinguish them, so that the best the receiver can do is to perform a measurement minimising the probability of equivocation. Such optimal measurement is the one maximising the so-called success probability $P_{\text{succ}}=\sum_{i=1}^n p_i P_{i|i}$, where $P_{i|i}=\mbox{Tr}(\mu_i[\rho_i])$ is the probability of getting the $i^{\rm th}$ result provided that the given state is the $i^{\rm th}$ one.

In~\cite{spehner2013geometric,spehner2014quantum} the case of  $\{\rho_i\}_{i=1}^n$ being states of a bipartite system $AB$, with $n$ equal to the dimension of subsystem $A$, has been studied in order to investigate the role played by QCs in ambiguous quantum state discrimination. We note that $A$ and $B$ must not be confused with the sender and the receiver. They are simply the subsystems of the bipartite quantum system that the sender gives to the receiver each time. Therefore, the receiver can then perform a generalised  measurement on the whole bipartite quantum system in order to decode the message. In this case, the maximal fidelity between the state after encoding according to the receiver, $\rho_{AB}=\sum_{i=1}^n p_i\rho_i$,  and the set $\mathscr{C}_A$ of classical-quantum states, is proven to be exactly the maximal success probability in the ambiguous quantum state discrimination of the  states $\{\rho_i\}_{i=1}^n$ with prior probabilities $\{p_i\}_{i=1}^n$, i.e.,
\begin{equation}
\max_{\chi_{AB}\in\mathscr{C}_A}F(\rho_{AB},\chi_{AB}) = P_{\text{succ}},
\end{equation}
that implies
\begin{equation}
Q_{A}^{G_{\text{Bu}}}(\rho_{AB})= \sqrt{2\left(1-\sqrt{P_{\text{succ}}}\right)}.
\end{equation}
In particular, this holds due to the fact that the optimal LGM $\{\mu_i\}_{i=1}^n$ that maximises the success probability is found to be a complete rank one LPM on subsystem $A$.
This provides us with the following direct link between the one-sided Bures geometric measure of QCs $Q_A^{G_{\text{Bu}}}(\rho_{AB})$ and the considered protocol: the more the state after encoding according to the receiver, $\rho_{AB}=\sum_{i=1}^n p_i\rho_i$, is close to be classical-quantum, the more successful the receiver will be in the discrimination task by performing a local von Neumann measurement on subsystem $A$.

``Quantum illumination'' \cite{lloyd2008enhanced,tan2008quantum,barzanjeh2015microwave} stands as a paradigmatic example of a quantum state discrimination task. The conventional illumination protocol consists of a probe qudit, in a pure state $\rho_\phi= |\phi\rangle\langle\phi|$, sent into a distant noisy region to detect the possible presence of a target object. If the target is not there, the probe is completely lost and the sender detects only environmental noise described by the maximally mixed state $\rho_E = \mathbb{I}/d$. Even if the target is actually present, there is only a small probability $\eta$ that the probe is reflected back and then detected by the sender, whereas the majority of time the probe is lost and again only random noise is detected. Overall, by assuming that the presence and absence of the target happen with the same prior probability, such a conventional protocol is equivalent to a quantum state discrimination task between the state $\rho_c^{(0)}:=\rho_E$, corresponding to the absence of the target, and the state $\rho_c^{(1)}:=\eta \rho_\phi + (1-\eta) \rho_E$, corresponding to the presence of the target, with equal a priori probabilities $p_0=p_1=1/2$. More precisely, the better the sender can discriminate between the states $\rho_c^{(0)}$ and $\rho_c^{(1)}$, the more confident they will be in inferring whether or not the target is actually there.

In~\cite{weedbrook2013discord}, the accuracy of the discrimination between two states $\rho_0$ and $\rho_1$ with a priori probabilities $p_0$ and $p_1$, respectively, has been quantified by the corresponding Shannon distinguishability, i.e.,
${D_{\cal H}}(\rho_0,\rho_1):= \max_{\mathscr{M}}\mathcal{J}(X:X_o^{\mathscr{M}})$, where $\mathcal{J}(X:X_o^{\mathscr{M}})$ is the classical mutual information, Eq.~(\ref{Equation:ClassicalJ}),  between the variable $X$ with probability distribution $\{p_0,p_1\}$ and the variable $X_o^{\mathscr{M}}$ obtained as output by performing the measurement $\mathscr{M}$. Therefore, the best case scenario within the conventional approach to illumination can be quantitatively identified by a figure of merit $I_c^{\max}:= \max_\phi {D_{\cal H}}(\rho_c^{(0)},\rho_c^{(1)})$.

In a quantum illumination protocol assisted by non-classical correlations, the sender improves their strategy by maximally entangling the probe $A$ with an idler ancilla $B$ (which is retained at the sending station) by creating the pure global state $\rho_{\psi_{AB}}= \ket{\psi}\!\bra{\psi}_{AB}$, where $|\psi\rangle_{AB}=(1/\sqrt{d})\sum_k|k\rangle_A\otimes|k\rangle_B$ and $\{|k\rangle\}$ is a qudit orthonormal basis ($d$ being the dimension of both probe $A$ and ancilla $B$). Now, the corresponding equivalent quantum state discrimination task is between the state $\rho_{AB}^{(0)}:=\rho_E\otimes\rho_B$ with a priori probability $p_0=1/2$, corresponding to the absence of the target, and the state $\rho_{AB}^{(1)}:=\eta \rho_{\psi_{AB}} + (1-\eta) \rho_E\otimes \rho_B$ with a priori probability $p_1=1/2$, corresponding to the presence of the target, where $\rho_B=\mbox{Tr}_A(\rho_{\psi_{AB}})=\mathbb{I}/d$ represents the reduced state of the ancilla $B$. Such quantum illumination protocol thus outperforms its conventional counterpart when it is easier to distinguish $\rho_{AB}^{(0)}$ from $\rho_{AB}^{(1)}$ than $\rho_c^{(0)}$ from $\rho_c^{(1)}$ for any conventional input $\rho_\phi$, i.e., when $I_q>I_c^{\max}$ with $I_q:= {D_{\cal H}}(\rho_{AB}^{(0)},\rho_{AB}^{(1)})$. The corresponding advantage can be quantified by the difference $\Delta I :=I_q - I_c^{\max}$. It is interesting to pin down the exact origin of such a quantum advantage, since  the entanglement initially present between probe and idler can all be destroyed due to the effects of the noise acting on the probe in the target region \cite{zhang2013entanglement}.

In~\cite{weedbrook2013discord} an answer has been provided by showing a direct equality between the performance gain in quantum illumination over any conventional illumination protocol and the amount of quantum discord between the probe and its ancilla which is consumed to detect the target, thus linking the expenditure of QCs to the resilient enhancement of quantum illumination. More precisely:
\begin{equation}
\Delta I = \Delta Q_A^{\tilde{I}_{\mathcal{I}}},
\end{equation}
where $\Delta Q_A^{\tilde{I}_{\mathcal{I}}}:=Q_A^{\tilde{I}_{\mathcal{I}}}(\overline{\rho}_{\psi_{AB}}) - \overline{Q}_A^{\tilde{I}_{\mathcal{I}}}$, with $\overline{\rho}_{\psi_{AB}}:=p_0 \rho_{\psi_{AB}}^{(0)} + p_1 \rho_{\psi_{AB}}^{(1)}$ being the state of probe and idler ancilla before detecting the target and $\overline{Q}_A^{\tilde{I}_{\mathcal{I}}}:=p_0 Q_A^{\tilde{I}_{\mathcal{I}}}(\rho_{\psi_{AB}}^{(0)}) + p_1 Q_A^{\tilde{I}_{\mathcal{I}}}(\rho_{\psi_{AB}}^{(1)})$ being the average quantum discord between the probe and the idler ancilla after detecting the target.%, can be considered as the amount of discord which is consumed for resolving the target.

Another direct connection between QCs and ambiguous quantum state discrimination has been provided by~\cite{gu2012observing}, where the following protocol has been considered. Two laboratories $A$ and $B$ share some known quantumly correlated state $\rho_{AB}$. Laboratory $A$ privately encodes the value $x_i$ of a random variable $X$ with probability $p_i$ onto their subsystem through the application of a corresponding local unitary $U^{(i)}_A$. Then, $A$ gives their subsystem to $B$, who is asked to decode the encoded value of the random variable $X$ by carrying out some measurement $\mathscr{M}$ on the whole bipartite system. This is clearly an ambiguous quantum discrimination between the states $\rho_i:=U_A^{(i)}[\rho_{AB}]$ with a priori probabilities $p_i$. In analogy with~\cite{weedbrook2013discord}, the accuracy of laboratory $B$'s performance is quantified by resorting to the classical mutual information $\mathcal{J}(X:X_o^{\mathscr{M}})$ between the encoded variable $X$ and the decoded variable $X_o^{\mathscr{M}}$. Now, let use denote by $I_q'$ the  best possible performance that  $B$  can achieve when they can carry out any measurement, and by $I_c'$ the best possible performance that $B$ can achieve when they are restricted only to a single local measurement on each of $A$ and $B$, followed by classical post-processing. The quantity $\Delta I' := I_q' - I_c'$ thus represents the advantage provided by coherent interactions between $A$ and $B$ over no interaction, in performing local measurements within the decoding task.

In~\cite{gu2012observing} it has been shown that there always exists an optimal choice of local unitaries $U_A^{(i)}$ (for instance, the Pauli operators if $\rho_{AB}$ is a two-qubit state) such that
\begin{equation}
\Delta I' = \Delta Q_A^{\tilde{I}_{\mathcal{I}}},
\end{equation}
where $\Delta Q_A^{\tilde{I}_{\mathcal{I}}}:=Q_A^{\tilde{I}_{\mathcal{I}}}(\rho_{AB}) - Q_A^{\tilde{I}_{\mathcal{I}}}(\rho'_{AB})$, with $\rho'_{AB}:=\sum_i p_i U^{(i)}_A[ \rho_{AB}]$ being the state after encoding by party $A$, thus proving that the quantum discord consumed during encoding yields exactly the extra quantum advantage that $B$ can gain by allowing coherent interactions. This was also demonstrated experimentally in \cite{gu2012observing} in a continuous variable optical setup based on Gaussian states and Gaussian encoding  operations.

The last operational interpretation of QCs within ambiguous quantum state discrimination that we review is the one provided in~\cite{farace2014discriminating}. Here, the setting is that of discriminating among $n$ copies of two states $\rho_0$ and $\rho_1$. In particular, by assuming that the a priori probabilities of getting $\rho_0$ and $\rho_1$ are equal, as it happens e.g.~in a quantum illumination protocol, it is known that for $n \gg 1$ the minimal probability of error in distinguishing between these two states, using the optimal discrimination strategy due to Helstrom \cite{helstrom1976quantum}, scales approximately as $e^{-n \xi(\rho_0, \rho_1)} = \text{C}(\rho_0, \rho_1)^n$, where $\xi(\rho_0, \rho_1)$ is the quantum Chernoff bound defined in Eq.~(\ref{Equation:ProperChernoffBound}) while $\text{C}(\rho_0, \rho_1)$ is defined in Eq.~(\ref{Equation:ChernoffBound}).
%$\text{C}(\rho_0, \rho_1)^n$, where $\text{C}(\rho_0, \rho_1)$ is the quantum Chernoff bound defined in Eq.~(\ref{Equation:ChernoffBound}).
Moreover, the complementary Chernoff distance $D_{\text{C}}(\rho_0,\rho_1)=1-\text{C}(\rho_0,\rho_1)$ turns out to be nothing but the measure of distinguishability adopted in the definition of the discriminating strength  in Eq.~(\ref{Equation:DiscriminatingStrength}). When $\rho_0:=\rho_{AB}$ and $\rho_1:=U_A^\Gamma[\rho_{AB}]$, this coincidence entails that such a measure of QCs  quantifies operationally the guaranteed accuracy in the asymptotic discrimination between the state $\rho_{AB}$ and the transformed state $U^\Gamma_A[\rho_{AB}]$, in the worst case scenario in which only the spectrum $\Gamma$ of the unitary operation $U_A^\Gamma$ is known, and a minimisation is considered over all such unitaries with fixed spectrum $\Gamma$. This task bears some analogy with the black box metrology protocol discussed in Section~\ref{Section:Applications-Metrology}, indeed both settings provide direct operational interpretations for QCs quantifiers in worst case scenarios, with the difference that there the goal was to estimate the parameter imprinted by a local unitary $U_A$ on the probe subsystem $A$, while here the goal is just to determine whether a local unitary $U_A$ was applied or not to the probe subsystem $A$.  This task has also been considered in continuous variable systems in \cite{rigovacca2015gaussian} by restricting $\rho_{AB}$ and $U_A$ to Gaussian states and operations, respectively.

An analogous analysis applies to establish the operational role played by the Hellinger  discord of response $Q_A^{U_{\text{He}}}$ in quantum reading of classical digital memories \cite{pirandola2009quantum,pirandola2011quantum}, as demonstrated in~\cite{roga2015device}  in a continuous variable setting restricted to Gaussian states and operations. The authors of \cite{roga2015device} further provided an operational interpretation for the trace distance based Gaussian discord of response in terms of (one minus) the maximum error probability in quantum reading of a classical memory. The study in  \cite{roga2015device}  includes as well a detailed characterisation of the possible enhancements to QCs and to the corresponding quantum advantage due to thermal noise in the considered task.

Finally, we notice that the presence and use of QCs has also been investigated in other tasks related to quantum illumination, such as ghost imaging \cite{ragy2012nature,ragy2014quantifying}.

%\section{Quantum correlations beyond entanglement: applications in thermodynamics and phase transitions}

\subsection{Further physics applications} One of the reasons why entanglement theory has gained so large an acclaim can be traced to its successful cross-fertilisation into other important areas of physics beyond quantum information theory, such as condensed matter, atomic and optical physics, statistical mechanics, and cosmology. As QCs capture all trademark effects of quantum mechanics, including and beyond entanglement, it seems fit to investigate further the role and applicability they can have in different branches of physics, not limited to those where the benefits of entanglement can already be appreciated. Here we will focus on two particular domains.

\subsubsection{Quantum thermodynamics}\label{Section:Thermodynamics}

The concept of QCs arises naturally in the thermodynamics of quantum systems, or ``quantum thermodynamics'' \cite{goold2016role,millen2016perspective}, particularly in the context of work extraction \cite{oppenheim2002thermodynamical,horodecki2013quantumness} from a bipartite quantum system in contact with a thermal reservoir, a fundamental task for which entanglement is not necessary \cite{hovhannisyan2013entanglement}. Specifically, informational measures of QCs exactly quantify the difference between the maximal work that can be extracted by a Maxwell's demon, who is an entity that perfectly knows the state of the composite system and is able to perform any global operation on it, and the maximal work that can be extracted by two goblins,  who are less powerful entities whose knowledge about the state of the composite system can be limited and are able to perform just some subclass of LOCC. A different quantifier of QCs arises depending on the limitations on the goblins' knowledge and the subclass of LOCC that they are able to perform. Such QCs quantifiers are also referred to as ``demon discords'' \cite{lang2011entropic}.

On the one hand, due to the Landauer's principle, the maximal work that a Maxwell's demon can extract from a bipartite quantum system $AB$ in a state $\rho_{AB}$ and in contact with a thermal reservoir at temperature $T$ is given by
\begin{equation}\label{Equation:DemonWork}
W_{\text{demon}}(\rho_{AB}):=k T (\log d_{AB} - {\cal S}(\rho_{AB})),
\end{equation}
where $k$ is  Boltzmann's constant and $d_{AB}=d_A d_B$ is the dimension of the composite system $AB$.

On the other hand, if the goblins perfectly know the state $\rho_{AB}$ and one goblin (operating on $A$) can perform any complete rank one LPM $\Pi_A$ and communicate classically the result to the other goblin (operating on $B$), then the maximal work that they can extract is given by
\begin{equation}
W_{\text{goblins}}(\rho_{AB}) := k T \bigg(\log d_{AB} - \min_{\Pi_A} {\cal S} \big(\Pi_A[\rho_{AB}]\big)\bigg),
\end{equation}
in such a way that in this case
\begin{equation}
\Delta W(\rho_{AB})  := W_{\text{demon}}(\rho_{AB})- W_{\text{goblins}}(\rho_{AB})=k T Q^{I_\mathcal{S}}_A(\rho_{AB}),
 \end{equation}
 where $Q^{I_\mathcal{S}}_A$ is the thermal discord introduced in \cite{zurek2003quantum}, which also coincides with the one-way quantum deficit introduced in \cite{horodecki2005local}, as already mentioned in Section~\ref{Section:Measures-Informational-Entropy}. In fact, the same conclusions are drawn when the goblins are restricted to the so-called closed local operations and classical communication (closed LOCC), i.e., LOCC that cannot change the total number of particles and consist only of local unitary operations and local projective measurements, with one-way communication allowed from $A$ to $B$ \cite{horodecki2005local}. Furthermore, if each of the goblins performs a complete rank one LPM on their subsystem and classical communication is allowed only after the measurements, the zero-way quantum deficit $Q^{I_\mathcal{S}}_{AB}(\rho_{AB})$ is recovered instead \cite{horodecki2005local}.

Another possibility occurs when the goblins do not know the global state $\rho_{AB}$, but rather only their corresponding reduced states $\rho_A$ and $\rho_B$, while one of them can still perform LPMs on $A$ and communicate classically the result to the other \cite{brodutch2010quantum}. The maximal work that can be extracted by these even less powerful goblins is given by
\begin{equation}
W_{\text{goblins}}^{\star} (\rho_{AB}):= k T \left(\log d_{AB} - {\cal S}\big(\Pi_A^{\star}[\rho_{AB}]\big)\right),
\end{equation}
where $\Pi_A^{\star}$ is the LPM onto the eigenbasis of $\rho_A$. In this case we have that
\begin{equation}
\Delta W^{\star}(\rho_{AB}):=W_{\text{demon}}(\rho_{AB})- W_{\text{\text{goblins}}}^{\star}(\rho_{AB})=k T Q^{I^{\star}_\mathcal{S}}_A(\rho_{AB}),
 \end{equation} where $Q^{I^{\star}_\mathcal{S}}_A(\rho_{AB})$ is the increase in the von Neumann entropy of the composite system $AB$ by performing the projective measurement ${\Pi}^{\star}_{A}$ onto the eigenbasis of the reduced state of subsystem $A$, i.e.,
\begin{equation}\label{Equation:DiagonalEntropyBased}
Q_{A}^{I_{\mathcal{S}}^{\star}}(\rho_{AB}) :=  {\cal S}\left({\Pi}^{\star}_{A} [\rho_{AB}]\right) - {\cal S}\left(\rho_{AB}\right).
\end{equation}

The role of QCs, measured specifically by the quantum discord $Q^{{\tilde{I}}_{\cal I}}_A$, has been studied also within the context of quantum thermal machines. In particular, in \cite{leggio2015quantum}, a quantum machine that needs to drive the temperature of a body outside of the range defined by the surrounding thermal reservoirs has been considered, and it has been shown that not only the machine-body quantum discord is necessary to this aim but also that a high machine efficiency at maximum power is strongly connected to the presence of a peak in such a QCs measure, thus showing the role of the latter as a quantum thermal machine resource. In the same spirit, in \cite{park2013heat} it has been shown that QCs drive the Szilard engine containing a diatomic molecule with a semipermiable wall, in the sense that the work that can be extracted from this engine is lower bounded by the quantum discord between the two atoms of the molecule. Furthermore, in \cite{dillenschneider2009energetics} it has been shown that the quantum discord shared by the two atoms constituting the quantum reservoir of a photo-Carnot engine can increase the thermodynamic efficiency of the latter as opposed to the case where these two atoms have no QCs, thus demonstrating once again the role of QCs as a thermodynamic resource.

The presence of quantum discord in the operation of quantum absorption refrigerators modelled as three qubits in contact with three baths was also investigated in \cite{correa2013performance}, where it was found that QCs are always present between the qubit which is being cooled (cold qubit $A$) and the relevant subspace of the other two qubits which is in direct interaction with the cold qubit (machine virtual qubit $B$), although $Q^{I_{\mathcal{I}}}_A$ was found to play no quantitative role in the optimisation of the coefficient of performance or  the cooling power. On the contrary, in \cite{liuzzo2016thermodynamics} in a simple model of quantum feedback cooling of a two-level system using an interacting ancilla, akin to algorithmic cooling, the cooling performance was found to be closely related to the discord $Q^{I_{\mathcal{I}}}_A$ (and not to entanglement or total correlations) built up between system $A$ and ancilla $B$ in the pre-measurement step, before closing the feedback loop; in particular, the performance characteristics of the protocol were shown to be curves of constant quantum discord, with the optimal performance identified by a peak in such a QCs measure.

Another thermodynamic interpretation of QCs can be found when considering quantum predictive processes, wherein a so-called predictive quantum system $S$ is used to predict the future dynamics of a given portion $X$ of its surrounding environment by exploiting the correlations between $S$ and $X$. In \cite{grimsmo2013quantum} it has indeed been shown that the quantum discord $Q_X^{I_{\mathcal{I}}}$ between the predictive system $S$ and the target portion of its surrounding environment $X$ exactly quantifies the advantage in the predictive power over the most efficient classical predictive strategy, which in turn coincides with the corresponding reduction in the dissipated extractable work from the system $S$. This thus stands as another deep connection between efficient use of quantum information and efficient thermodynamical operation as powered by QCs. It may be interesting to further investigate links between predictive processes and the emergence of objective reality in quantum Darwinism, in particular in view of  Eq.~(\ref{Equation:NoDoze}) \cite{brandao2015generic}.

Finally, in \cite{lloyd2015no} it has been shown that any energy transport must necessarily generate QCs. In particular, over a sufficiently short time $\Delta t$, the flow of heat $\Delta E_{A}$ from a  $B$ to another system $A$, both initially prepared in thermal states at different temperatures $T_A$ and $T_B$, $T_B>T_A$, is directly proportional to the corresponding increase in the one-way QCs as measured by the so-called diagonal discord $Q_{A}^{I_{\mathcal{I}}^{\star}}$, i.e.,
\begin{equation}
\Delta E_{A} =  (\beta_A - \beta_B)^{-1}\Delta Q_{A}^{I_{\mathcal{I}}^{\star}},
\end{equation}
where $\beta_A=1/k T_A$, $\beta_B=1/k T_B$, and finally the diagonal discord is the reduction in the mutual information of the composite system $AB$ by performing the LPM ${\Pi}^{\star}_{A}$ onto the eigenbasis of the reduced state of subsystem $A$, i.e.,
\begin{equation}\label{Equation:DiagonalDiscord}
Q_{A}^{I_{\mathcal{I}}^{\star}}(\rho_{AB}) :=  {\cal I}\left(\rho_{AB}\right) - {\cal I}\big({\Pi}^{\star}_{A} [\rho_{AB}]\big).
\end{equation}
The diagonal discord in Eq.~(\ref{Equation:DiagonalDiscord}) is the mutual information based analogue to the von Neumann entropy based informational quantifier in Eq.~(\ref{Equation:DiagonalEntropyBased}), and can be seen as the one-sided counterpart to the measurement induced disturbance discussed in Section~\ref{Section:Measures-Informational-MutualInformation}. Notice that, due to the lack of an optimisation over LPMs, both quantities in Eqs.~(\ref{Equation:DiagonalEntropyBased}) and (\ref{Equation:DiagonalDiscord}) are not, in general, valid measures of QCs, yet they both obey Requirement~(\ref{Requirements:Faithfulness}), thus being able to faithfully discriminate between classically and quantumly correlated states.

\subsubsection{Many-body systems and quantum phase transitions}\label{Section:Applications-QPT}

General QCs represent not only essential ingredients for quantum technologies and thermodynamics,  but are also a very useful tool for the characterisation of the low temperature macroscopic phases of many-body quantum systems \cite{sarandy2013quantum,werlang2013interplay}. In particular, the analysis of the QCs of quantum ground states can be exploited to detect the points in the low temperature macroscopic phase space where the system passes from one phase to another, a process referred to as ``quantum phase transition''.

Quantum phase transitions typically occur at or near absolute zero temperature, where the de Broglie wavelength is greater than the correlation length of the thermal fluctuations, by varying a parameter of the system's Hamiltonian that is known as tuning parameter. This change of phase is only caused by quantum fluctuations stemming from the Heisenberg uncertainty relation, which still survive in the low-temperature regime, as opposed to a classical phase transition that is instead driven by thermal fluctuations. More precisely, when the tuning parameter crosses a particular value, known as critical point, the system's ground state undergoes a dramatic qualitative change.
%, which in turn causes a drastic variation of the macroscopic properties of the system.
The latter can consist of either the breaking of some symmetry of the system's Hamiltonian and ensuing appearance of a local order or the variation of the topological order of the system's phase as characterised by a nonlocal order parameter based on the pattern of the long-range two-body ground state correlations.

After an initial race to detect the occurrence of quantum phase transitions through the analysis of both the two-body and global entanglement properties of quantum ground states \cite{amico2008entanglement}, QCs more general than entanglement have been adopted for this purpose for the first time in \cite{dillenschneider2008quantum}, when considering in particular the one-dimensional spin-$1/2$ transverse Ising and antiferromagnetic XXZ models. Specifically, there the critical points have been identified by looking at the singularities of the derivatives of the zero-temperature nearest- and next-nearest-neighbour two-body quantum discord in the thermodynamic limit. In \cite{sarandy2009classical}, such analysis has been generalised by discussing also the finite size scaling of the two-body nearest-neighbour quantum discord and then extended to the Lipkin-Meshkov-Glick model.

In both \cite{dillenschneider2008quantum} and \cite{sarandy2009classical} only the zero-temperature regime was considered. However, according to the third law of thermodynamics, the absolute zero temperature is unreachable. This makes a low-temperature extension of the aforementioned analyses compelling, thus inevitably motivating the studies that have been carried out  \cite{werlang2010thermal,werlang2010quantum,rong2012quantum}. Quite remarkably, in those works it has been shown that, when considering the one-dimensional spin-$1/2$ XXZ model, the two-body quantum discord spotlights the quantum phase transitions even at finite temperature. On the contrary, the two-body entanglement and other thermodynamic quantities such as entropy, specific heat,  magnetic susceptibility, or the two-body (classical) correlation functions, are not able to detect the corresponding quantum phase transitions at finite temperature, thus making bipartite QCs stand out as particularly useful tools in the quest for the experimental detection of quantum phase transitions in the one-dimensional spin-$1/2$ XXZ model.

However, recently the authors of \cite{malpetti2016quantum} have pointed out that the two-body quantum discord may provide essentially no information on the quantum nature of two-body correlations in a system at finite temperature. A more appropriate and physically motivated definition of QCs is instead proposed in \cite{malpetti2016quantum}, leading to the introduction of the so-called ``quantum correlation function'', which stands in full analogy with the ordinary correlation function in statistical physics, although it only distills the quantum nature of correlations. Interestingly, the quantum variance introduced in \cite{frerot2015quantum} and discussed in Sec.~\ref{Section:Measures-Coherence-Others} is obtained by integrating the quantum correlation function, in the same way as the ordinary variance of a physical quantity is obtained by integrating the ordinary correlation function.

In the same spirit as the finite-temperature analyses, in \cite{maziero2010quantum,maziero2012long,cakmak2012critical,campbell2013criticality,huang2014scaling,karpat2014quantum} the long-range nature of two-body QCs and their scaling with the inter-spin distance have been exploited in order to detect the critical point of the one-dimensional spin-$1/2$ XY model. Again, the analysis of long-ranged correlations becomes instead infeasible when focusing on two-body entanglement, due to its short-ranged nature.

Other studies then complemented the above ones when either taking symmetry breaking into account \cite{tomasello2011ground,tomasello2012quantum} or considering more sophisticated quantum phase transitions such as the ones in the one-dimensional spin-$1/2$ XY model with Dzyaloshinskii-Moriya interaction \cite{liu2011quantum}, the Dicke model \cite{wang2012quantum}, the Castelnovo-Chamon model \cite{chen2010quantum}, cluster-like systems \cite{chen2010quantum2}, as well as spin-$1$ XXZ and bilinear-biquadratic chains \cite{allegra2011quantum,rossignoli2012measurements,malvezzi2016quantum}. Most of the above works adopted the quantum discord or the one-way quantum deficit as measures of QCs, but other valid and more efficiently computable quantifiers such as the local quantum uncertainty were also proven of use  in the characterisation of quantum critical points \cite{karpat2014quantum,coulamy2016scaling,malvezzi2016quantum}.

Finally, in \cite{rulli2011global} the ``global quantum discord'' was defined as a multipartite measure of QCs by extending a relative entropy based definition of the quantum discord to many-body systems. The global discord was shown able to characterise the infinite-order quantum phase transition in the Ashkin-Teller spin chain, which could not be detected by using the two-body discord alone. An extended study of one-dimensional spin models at finite temperature in \cite{campbell2013global} showed that the global discord could detect critical points even in many-body systems out of their ground state; interestingly, it was also found that the global discord scales with universal critical exponents in the proximity of a quantum phase transition in the Ising model.  Quantum and global discord were further analysed in spin-$1$ Heisenberg chains subject to single-ion anisotropy in \cite{power2015nonclassicality}; these measures were shown to detect the quantum phase transitions confining the symmetry-protected Haldane phase, and to exhibit critical scaling with universal exponents.

%\cite{dillenschneider2008quantum,werlang2010quantum,sarandy2013quantum,maziero2010quantum,rong2012quantum,allegra2011quantum,chen2010quantum,sarandy2009classical,werlang2010thermal,tomasello2011ground}

\section{Concluding remarks and outlook}\label{Section:Conclusions}

%\addtolength{\textheight}{.4cm}

% Cite these two papers by Piani that I don't know what he's talking about: http://arxiv.org/pdf/1501.06855v1.pdf and http://arxiv.org/pdf/1405.1640v2.pdf

In this review we  summarised recent progress in the characterisation of the most general forms of QCs in composite quantum systems, focusing in particular on bipartite states. We devoted particular attention in Section~\ref{Section:QCs} to illustrate the physical significance of the concept of QCs, providing a polyhedral description of how quantumly correlated states differ from classically correlated ones (see Fig.~\ref{Fig:Defs}). Such a qualitative analysis then formed the basis to introduce in Section~\ref{Section:Measures} a plethora of different measures of QCs (see Tables~\ref{Table:Types} and \ref{Table:SummaCumLaude}), each capturing quantitatively one or more defining aspects of quantumly correlated states. The operational significance of several of the measures was then explored in Section~\ref{Section:Applications}, where various relevant applications of QCs to quantum information and to physics more broadly were described.

As the topic of QCs is still under development, there remain lots of open questions to investigate in future research. The attentive reader will have found a bunch of them already interspersed in the previous Sections. One of the most pressing matters to address is certainly the development of a consistent resource theory of QCs, more specifically to formulate the appropriate set of free operations under which any {\it bona fide} measure of QCs should be monotonically non-increasing. We have proposed one such set, namely the local commutativity preserving operations, formalised in Eqs.~(\ref{Equation:LCPOA}) and (\ref{Equation:LCPOAB}) of Section~\ref{Section:Requirements}, and used them to postulate a strong monotonicity for measures of QCs as a desideratum, tagged Requirement~(\ref{Requirements:LCPO}). Fruitful next steps could be either to systematically validate existing measures against this Requirement (thus clearing as many question marks as possible from the last column of Table~\ref{Table:SummaCumLaude}), or to tweak such a Requirement if deemed necessary from a physical point of view. Perhaps insights from the resource theories of entanglement \cite{plenio2007an,horodecki2009quantum} and coherence \cite{marvian2013the,baumgratz2014quantifying,winter2016operational,chitambar2016are,streltsov2016review}, and their respective free operations, could guide such an endeavour, in view of the close connections that both of these resources have with QCs  (refer e.g.~to Definitions~\ref{Definition 3:} and \ref{Definition 5:}).

Nevertheless, we have also highlighted some measures of QCs which are already proven to be in full compliance with the Requirements of Section~\ref{Section:Requirements}. Perhaps the most satisfactory ones are the one-sided and two-sided geometric measures $Q^{G_{\text{RE}}}_A$ and $Q^{G_{\text{RE}}}_{AB}$ defined in terms of relative entropy, which are shown to be further equivalent to a variety of other types of measures. These measures enjoy thus multiple fundamental interpretations, namely in terms of: the minimal state distinguishability from any classical state (geometric approach); the minimal state distinguishability following a least perturbing local measurement (measurement induced geometric approach); the minimum loss of global information, or equivalently the minimum added noise, due to a least perturbing local measurement (informational approach in terms of von Neumann entropy); the  minimum distillable entanglement created with an ancillary system at the pre-measurement stage during a least perturbing local measurement (entanglement activation approach); and the minimum quantum coherence in any local basis (coherence based approach in terms of relative entropy); plus, they have a natural operational meaning as quantum deficits in the context of thermodynamical work extraction. To the best of our knowledge, the measures of QCs based on relative entropy have not been studied explicitly in either the unitary response or the recoverability approach. It is tempting to conjecture that both of them might reduce as well to the corresponding geometric quantifiers, hence closing the circle of equivalences, and providing yet further physical interpretations for such measures of QCs. We leave this as an open question to investigate for the interested reader.

In this respect, let us also point out that the surprisal of measurement recoverability looks quite lonely in its partition of Table~\ref{Table:SummaCumLaude}, and it will be worth studying more explicit measures of QCs within the recently formulated recoverability approach \cite{seshadreesan2015fidelity}, in particular in view of possible operational connections with practical tasks, as discussed in Section~\ref{Section:Applications-Broadcasting} for local broadcasting. Within such a setting, let us also recall the fundamental interpretation revealed for the  quantum discord as the share of correlations that cannot be redistributed to infinitely many subsystems in a  one-sided scenario \cite{streltsov2013quantum,brandao2008reversible}. As remarked after Eq.~(\ref{Equation:NoDoze}), a deserving next step could be to formalise a similar connection in the two-sided case, and study its implications for the quantum to classical transition.

Further connections highlighted through the exploration of quantitative measures of QCs are also inspiring. For instance, the equivalence between the two-sided negativity of quantumness  (an entanglement activation type of measure) and the quantum coherence measured by the $\ell_1$ norm minimised over all local product bases (capturing the seminal notion of coherence in terms of interference effects), which bridges in particular the interpretations illustrated by Fig.~\ref{Fig:Defs}(b) and (d), reveals once more that QCs incarnate truly fundamental signatures of quantum mechanics,   blending together other more widely celebrated phenomena like entanglement, superposition, non-commutativity, and the perturbing effect of measurements. This is equally, and maybe even more strongly, represented in the chain of equivalences discussed above for relative entropy based measures.

Let us remark that such basic ingredients as the non-commutativity of a bipartite quantumly correlated state with any local observable may be seen as a negative feature, i.e.~an unavoidable uncertainty affecting the measurement of any single local observable (as captured by the local quantum uncertainty $Q^{C^{\text{WY}}}_A$) \cite{girolami2013characterizing}, but also, crucially, as a positive feature, since all quantumly correlated states are necessarily modified by --- hence sensitive to --- any local dynamics with a fully non-degenerate spectrum. This is exactly the feat that makes QCs a natural resource for metrology and discrimination tasks in worst case scenarios, as  discussed in Section~\ref{Section:Applications-Metroid}. Further study of the concrete implications of these observations for quantum technologies would be welcome.

In this review we mainly spoke about bipartite QCs. Several of the measures introduced extend straightforwardly to multipartite systems, in particular the geometric ones (by defining the set of fully classical states as the states diagonal with respect to a product of orthonormal bases for each subsystem \cite{piani2008no,modi2010unified}, and any set of partially  classical-quantum states with respect to a selection of subsystems treated as classical, as detailed in \cite{piani2012quantumness}), the measurement induced geometric and informational ones (again depending on a selection of subsystems on which local measurements are applied), the entanglement activation ones \cite{streltsov2011linking,piani2011all,piani2012quantumness}, and the recoverability ones \cite{wilde2015recoverability}.  As previously pointed out, it will be an interesting future direction to develop consistent generalisations of the unitary response and coherence based (in the particular case of asymmetry measures) approaches to define faithful two-sided and more general multipartite measures of QCs, especially in view of the operational merits that these types of measures exhibit in the one-sided case for bipartite systems.
Some measures of multipartite QCs have already been defined, such as the global discord \cite{rulli2011global} (which finds applications in the study of quantum critical points in many-body systems, see Section~\ref{Section:Applications-QPT}), the so-called quantum dissension \cite{chakrabarty2011quantum}, and a few quantifiers of genuine QCs
\cite{zhou2008irreducible,giorgi2011genuine}.  More general (measure-independent) features of QCs in multipartite systems have also been explored, demonstrating in particular that, unlike entanglement measures \cite{coffman2000distributed}, no measure of QCs beyond entanglement can satisfy a conventional monogamy inequality on all states of three or more parties \cite{streltsov2012are}. However, alternative monogamy constraints that impose trade-offs between QCs and other resources in tripartite systems, such as entanglement, coherence, and local entropies, can be derived \cite{koashi2004monogamy,fanchini2011conservation,chuan2012quantum,streltsov2012quantum,xi2015quantum,lancien2016should,dhar2016monogamy}. Even more generally, the results reviewed in Section~\ref{Section:Applications-Broadcasting} reveal that QCs beyond entanglement do obey fundamental limits to their shareability, which is another testament to their true quantum nature, as opposed to the case of freely shareable classical correlations. In this context it is interesting to point out that, while we focused on general QCs within the hierarchy of correlations illustrated in Fig.~\ref{Fig:CorrHier}, there also exist finer layers of correlations that can be identified in quantum states, for instance intermediate between QCs and entanglement, as investigated e.g.~in \cite{jevtic2014quantum}.

As announced in the Introduction, we have not discussed any study of the dynamical  properties of QCs in open quantum systems, even though there are lots of  interesting results in such a direction. In particular, due to the geometry of the set of classical states \cite{ferraro2010almost}, QCs cannot vanish at finite time for any typical dynamics \cite{maziero2009classical,werlang2009robustness}, making them by construction more resilient than entanglement. In certain cases, such a resilience can be extreme, as QCs in bipartite and multipartite systems can remain constant (frozen) in time under local decohering maps even though the global state is evolving \cite{mazzola2010sudden}; this happens for particular classes of states and dynamical conditions. Within the geometric approach to QCs (Section~\ref{Section:Measures-Types-Geometric}), it has been shown in \cite{cianciaruso2015universal} that, under the allowed conditions, such features are universal and occur for all measures $Q^{G_\delta}$ independently of the specific choice of the distance $D_\delta$ in their definition. More details can be found in \cite{cianciaruso2015universal} and references therein, as well as \cite{modi2012classical}, to which the reader is referred for a collection of pertinent literature (including a number of experimental demonstrations).
Note that such freezing phenomena also manifest in the study of quantum coherence \cite{bromley2015frozen}, under the same dynamical conditions as for QCs, which provides a  further objective evidence of the intimate relationship between these two quantum effects.

One historical perspective that certainly needs clearing up is the
question of whether QCs beyond entanglement can be responsible for
super-classical advantages in quantum computing. Needless to say, pinning down the exact contribution of QCs to information processing presents an extremely relevant goal to the wider development of quantum technologies. In fact, after half a decade of quiescence since their initial definition \cite{ollivier2001quantum,henderson2001classical}, QCs attracted a flurry of research interest  \cite{merali2011nature} following the suggestion in~\cite{datta2008quantum} that quantum discord might be accountable for the power behind the protocol of quantum computing with one clean qubit (DQC1)~\cite{knill1998power}. This was also supported by experimental implementations of this protocol using separable states of a few qubits~\cite{lanyon2008experimental,passante2011experimental}. However, numerous questions have been raised about the precise role of QCs in DQC1~\cite{ferraro2010almost,dakic2010necessary,ma2016converting}, with no clear operational link developed so far between the quantum speed-up  (for which a figure of merit is not itself available) and any specific quantifier of QCs. While there remains interest in using separable states with nonzero QCs in quantum algorithms~\cite{chaves2011noisy,cui2010correlations,vedral2010elusive}, and one can even formulate precise arguments for the need of mixed state quantum computations to create QCs in order not to be classically simulatable \cite{eastin2010simulating,cable2015exact}, it is overall a contentious topic, that we have not covered in this review;  the interested reader may consult \cite{modi2012classical} for further details. We chose instead to review a wealth of novel and recent results from operational perspectives other than quantum computation, that themselves highlight the importance of QCs within communication, metrology, thermodynamics, and much more. We believe this is just the beginning to the acknowledgment of QCs as resources, and it will be of pivotal importance to pursue further new operational viewpoints of this multi-faceted phenomenon, in order to better appreciate the practical value of QCs, beyond their intrinsic foundational virtues.

To conclude, our goal with this review was to provide a comprehensive and, in some respects, original introduction to the quantum versus classical frontier, as perceived from the viewpoint of correlations between two or more parts of a composite system. In particular, we have reported detailed answers to the three basic questions posed in the Introduction, i.e.: What are the signature traits of QCs? How can we meaningfully quantify QCs? What are the practical applications of QCs? Along the way we have highlighted many possible avenues to further expand on these answers. We hope to have conveyed sufficient motivation for the reader to be enthused about the study of quantum correlations and we look forward to future progress, along some of the lines discussed here, and beyond.

\ack{This work is supported by the European Research Council (ERC) Starting Grant GQCOP ``Genuine Quantumness in Cooperative Phenomena'' (Grant No.~637352), and by the Foundational Questions Institute (fqxi.org) Physics of the Observer Programme (Grant No.~FQXi-RFP-1601). We are grateful to Marco Piani for many fruitful discussions and to an anonymous Referee for very constructive comments. We thank S. Piano, J. Doe, and E. D'Angelo for their patience and support during the writing of this review. We acknowledge useful feedback on an earlier draft of this manuscript by (in alphabetical order): R.~Angelo, S.~Campbell, N.~Canosa, I.~Chakrabarty, F.~Fanchini, Y.~Huang, A.~Iqbal~Singh, F.~Illuminati, S.~Pirandola, A.~Rastegin, T.~Roscilde, A.~Smerzi, C.~Susa-Quintero, M.~Wilde,  and M.~Yurischev.}

\renewcommand{\baselinestretch}{1}
\section*{References}

\bibliographystyle{iopart-num}
\bibliography{ABC_biblio}

\end{document}